%% file: ms.tex
\documentclass[final,journal,twocolumn,twopage]{IEEEtran}

\usepackage{amssymb,amsmath}
\usepackage{cite}
\usepackage{commath}
\usepackage{latexsym}
\usepackage{tabularx}
\usepackage[]{graphicx}
\usepackage[caption=false]{subfig}
\usepackage{mathtools}
\usepackage{siunitx}
\usepackage{listings}
\usepackage{tikz}
\usetikzlibrary{mindmap,trees}
\usepackage{stfloats}
\usepackage{comment}
\usepackage{fontawesome5}
\usepackage{longtable}
\usepackage[colorlinks=true,linkcolor=black,anchorcolor=black,citecolor=black,filecolor=black,menucolor=black,runcolor=black,urlcolor=black]{hyperref}
\usepackage{ragged2e}
\usepackage[]{array}
\usepackage{multirow}
\usepackage{soul}
\usepackage{booktabs}% http://ctan.org/pkg/booktabs

\usepackage{textcomp}
\usepackage{enumitem}
\newcolumntype{L}[1]{>{\raggedright\let\newline\\\arraybackslash\hspace{0pt}}m{#1}}
\newcolumntype{C}[1]{>{\centering\let\newline\\\arraybackslash\hspace{0pt}}m{#1}}
\newcolumntype{R}[1]{>{\raggedleft\let\newline\\\arraybackslash\hspace{0pt}}m{#1}}
\makeatletter
\newcommand\remembertext[2]{% #1 is a key, #2 is the text
  \immediate\write\@auxout{\unexpanded{\global\long\@namedef{mytext@#1}{#2}}}%
   {%
   \hypersetup{linkcolor=black,anchorcolor=black,citecolor=black,filecolor=black,menucolor=black,runcolor=black,urlcolor=black}%
   \color{black} #2%
   \hypersetup{linkcolor=black,anchorcolor=black,citecolor=black,filecolor=black,menucolor=black,runcolor=black,urlcolor=black}%
   }%
}
\newcommand\recalltext[1]{%
  \ifcsname mytext@#1\endcsname
    {%
    \hypersetup{linkcolor=black,anchorcolor=black,citecolor=black,filecolor=black,menucolor=black,runcolor=black,urlcolor=black}%
    \color{black}\@nameuse{mytext@#1}%
    \hypersetup{linkcolor=black,anchorcolor=black,citecolor=black,filecolor=black,menucolor=black,runcolor=black,urlcolor=black}%
    }%
  \else
    ``???''
  \fi
}
\makeatother
\input{acronyms}

\begin{document}

\title{A Review of Millimeter Wave Device-based Localization and Device-free Sensing\\ Technologies and Applications}

% OTHER Titles:
% \begin{itemize}
%     \item ``Indoor Millimeter Wave Localization and Sensing: Algorithms and applications''
%     \item ``A Survey on Indoor Localization and Sensing using Millimeter Wave Devices''
%     \item ``A Comprehensive View of Millimeter Wave Communication- and Radar-Enabled Localization and Sensing''
% \end{itemize}

\author{
Anish~Shastri,~\IEEEmembership{Student~Member,~IEEE},
Neharika~Valecha,~\IEEEmembership{Student~Member,~IEEE},
Enver~Bashirov,~\IEEEmembership{Student~Member,~IEEE},
Harsh~Tataria,~\IEEEmembership{Member,~IEEE},
Michael~Lentmaier,~\IEEEmembership{Senior~Member,~IEEE},
Fredrik~Tufvesson,~\IEEEmembership{Fellow,~IEEE},
Michele~Rossi,~\IEEEmembership{Senior~Member,~IEEE},
and~Paolo~Casari,~\IEEEmembership{Senior~Member,~IEEE}% <-this % stops a space
\thanks{Manuscript received xxxx xx, xxxx \ldots}% 
\thanks{This work received support from the European Commission's Horizon 2020 Framework Programme under the Marie Sk{\l}odowska-Curie Action MINTS (GA no.~861222), and from Italian Ministry for Education, University and Research (MIUR) under the ``Departments of Excellence'' initiative (Law 232/2016).}%
\thanks{A.~Shastri (email: anish.shastri@unitn.it) and P.~Casari (email: paolo.casari@unitn.it) are with the Department of Information Engineering and Computer Science, University of Trento, 38123 Povo (TN), Italy.}% 
\thanks{N.~Valecha (email: neharika.valecha@eit.lth.se), %H.~Tataria (email: harsh.tataria@eit.lth.se), 
M.~Lentmaier (email: michael.lentmaier@eit.lth.se) and F.~Tufvesson (email: fredrik.tufvesson@eit.lth.se) are with the Department of Electrical and Information Technology, Lund University, 22100 Lund, Sweden.}%
\thanks{E.~Bashirov (email: enver.bashirov@dei.unipd.it), and M.~Rossi (email: rossi@dei.unipd.it) are with the Department of Information Engineering, University of Padova, 35131 Padova, Italy.}%
\thanks{H.~Tataria (email: harsh.tataria@ericsson.com) was with Ericcson AB, 22363 Lund, Sweden, when working on this paper.}%
}%

%\markboth{IEEE Communications Surveys and Tutorials}%
%{SOMEOFUS \textit{et al.}: Millimeter-Wave Localization: A Joint Radar and Communications Perspective}



\maketitle

\begin{abstract}
    \remembertext{new_abstract}{
    The commercial availability of low-cost \ac{mmw} communication and radar devices is starting to improve the adoption of such technologies in consumer markets, paving the way for large-scale and dense deployments in \ac{5g}-and-beyond as well as 6G networks. At the same time, pervasive \ac{mmw} access will enable device localization and device-free sensing with unprecedented accuracy, especially with respect to sub-6~GHz commercial-grade devices.
    
    This paper surveys the state of the art in device-based localization and device-free sensing using \ac{mmw} communication and radar devices, with a focus on indoor deployments. We overview key concepts about \ac{mmw} signal propagation and system design, detailing approaches, algorithms and applications for \ac{mmw} localization and sensing. Several dimensions are considered, including the main objectives, techniques, and performance of each work, whether they reached an implementation stage, and which hardware platforms or software tools were used.
    
    We analyze theoretical (including signal processing and machine learning), technological, and implementation (hardware and prototyping) aspects, exposing under-performing or missing techniques and items towards enabling a highly effective sensing of human parameters, such as position, movement, activity and vital signs. Among many interesting findings, we observe that {\it device-based} localization systems would greatly benefit from commercial-grade hardware that exposes channel state information, as well as from a better integration between standard-compliant \ac{mmw} initial access and localization algorithms, especially with multiple \acp{ap}. Moreover, more advanced algorithms requiring zero-initial knowledge of the environment would greatly help improve the adoption of \ac{mmw} \ac{slam}. \Ac{ml}-based algorithms are gaining momentum, but still require the collection of extensive training datasets, and do not yet generalize to any indoor environment, limiting their applicability.
    
    Device-free (i.e., radar-based) sensing systems still have to be improved in terms of: improved accuracy in the detection of vital signs (respiration and heart rate) and enhanced robustness/generalization capabilities across different environments; moreover, improved support is needed for the tracking of multiple users, and for the automatic creation of radar networks to enable large-scale sensing applications. Finally, integrated systems performing joint communications and sensing are still in their infancy: theoretical and practical advancements are required to add sensing functionalities to \ac{mmw}-based channel access protocols based on \ac{ofdm} and multi-antenna technologies. 
    % We conclude by discussing that better algorithms for consumer-grade devices, data fusion methods for dense deployments, as well as an educated application of machine learning methods are promising, relevant and timely research directions.
    }
\end{abstract}

% Note that keywords are not normally used for peer review papers.

% \fi
\IEEEpeerreviewmaketitle

\glsresetall

%% Section 01 %%
\input{01_Introduction}

%% Section 02 %%
\input{02_Influence_mmWave_channels}

%% Section 03 %%
\input{03_Beamforming_architectures}

%% Section 04 %%
\input{04_Standardization_notes}

%% Section 05 %%
\input{05_Localization_algorithm_survey}

%% Section 06 %%
\input{06_RADAR_survey}

%% Section 07 %%
\input{07_Discussion_open_research_dir}

%% Section 08 %%

\input{08_Conclusions}

%\newpage

\printnoidxglossary[type=\acronymtype,title=List of abbreviations]
\label{sec:glossary}

% If we need appendices
% \input{appendices_file}

\ifCLASSOPTIONcaptionsoff
  \newpage
\fi



%\bibliographystyle{IEEEtran}

%\newpage

% \IEEEtriggercmd{\enlargethispage{-7cm}}
% \IEEEtriggeratref{226}

\bibliographystyle{IEEEtran}
\bibliography{IEEEabrv,refLOC,refNeharika_copy,refRADAR,refOther}  


 \begin{IEEEbiography}[{\includegraphics[width=1in,height=1.25in,clip,keepaspectratio]{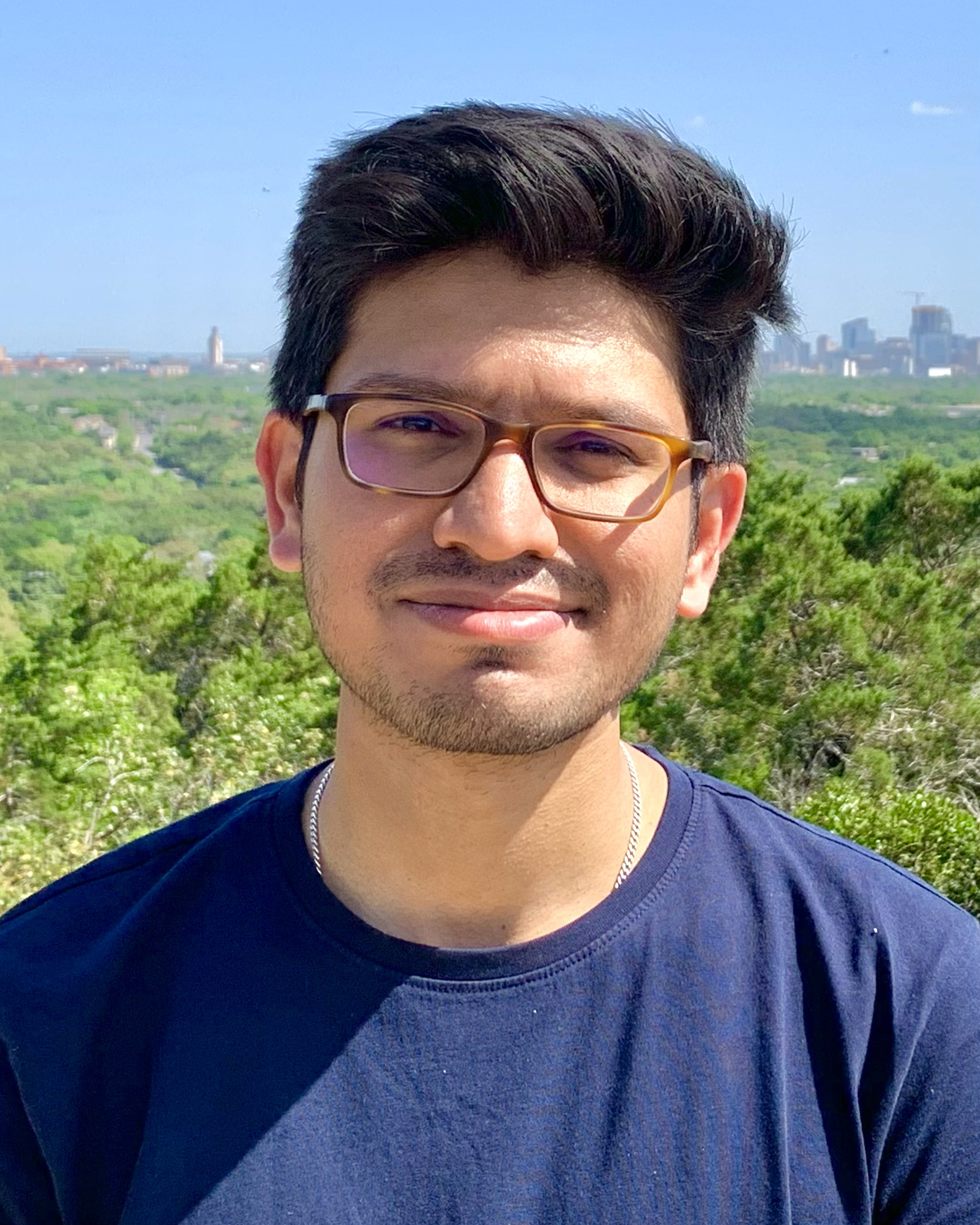}}]{Anish Shastri}
 (S'18) is an Early Stage Researcher with the EU H2020 Marie Skłodowska Curie Actions MINTS ETN, pursuing his Ph.D. at the Department of Information Engineering and Computer Science, University of Trento, Italy. He received his B.Tech in Electronics and Communication Engineering (ECE) from the National Institute of Technology (MANIT) - Bhopal, India, in 2016, and MS by Research in ECE from the International Institute of Information Technology – Hyderabad (IIIT-H), India, in 2019. Before starting his Ph.D., he was a Research Intern in the Indoor Networks Research group at Nokia Bell Labs, Ireland. His research interests include signal processing and machine learning for wireless communications, with his current research focusing on algorithms for mmWave indoor localization, environment mapping, and sensing.
 \end{IEEEbiography}

\begin{IEEEbiography}[{\includegraphics[width=1in,height=1.25in,clip,keepaspectratio]{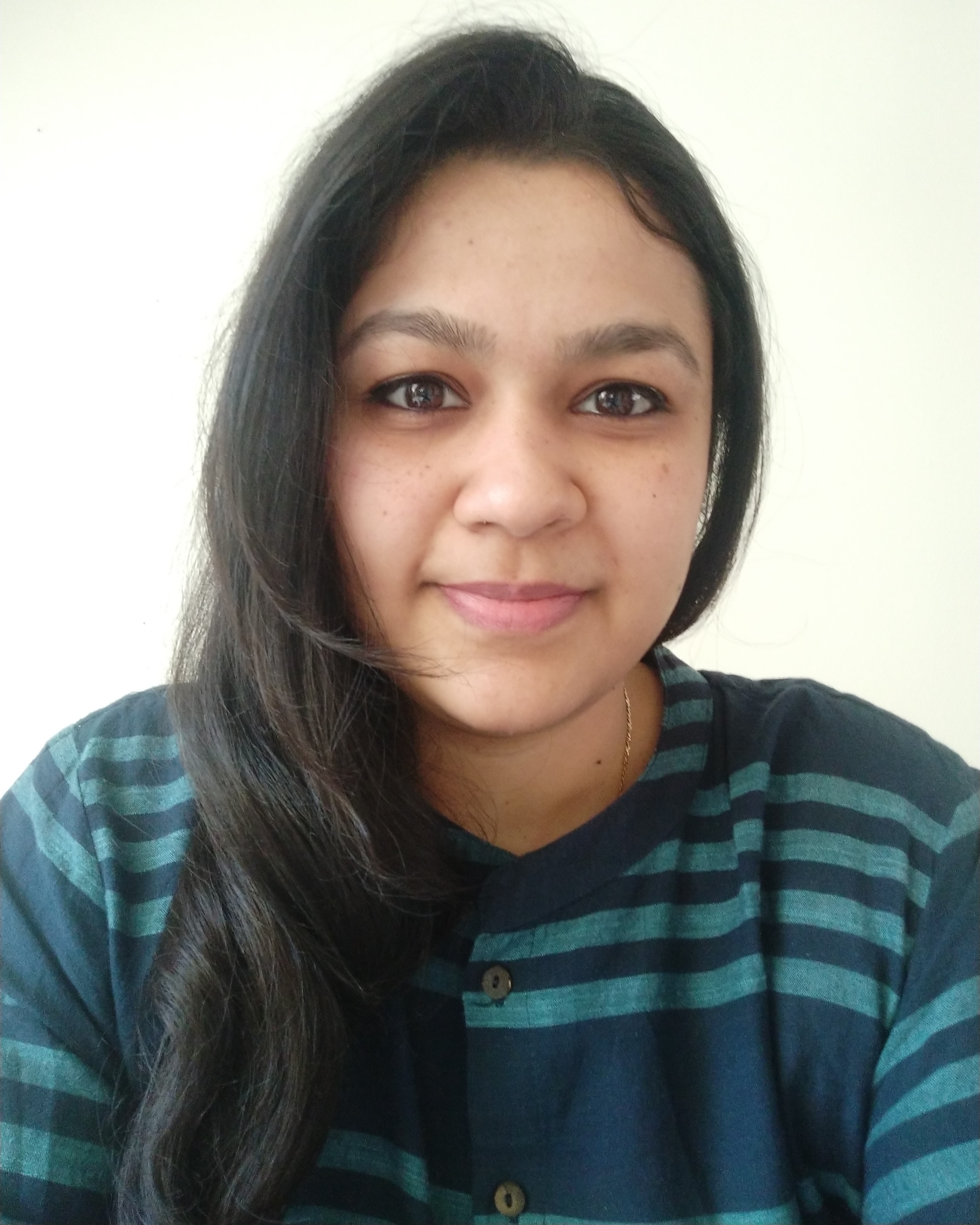}}]{Neharika Valecha}
(S'20) is an Early Stage Researcher with the EU H2020 Marie Skłodowska Curie Actions MINTS ETN, pursuing her Ph.D. at the Department of Electrical and Information Technology, Lund University, Sweden. She received her Masters in Mobile Computing Systems from Institut Eur\'ecom in 2019. Her research interests include signal processing and communication for next generation wireless systems. Currently, she is working with sensing and localization for mmWave indoor scenarios.
\end{IEEEbiography}

\begin{IEEEbiography}[{\includegraphics[width=1in,height=1.25in,clip,keepaspectratio]{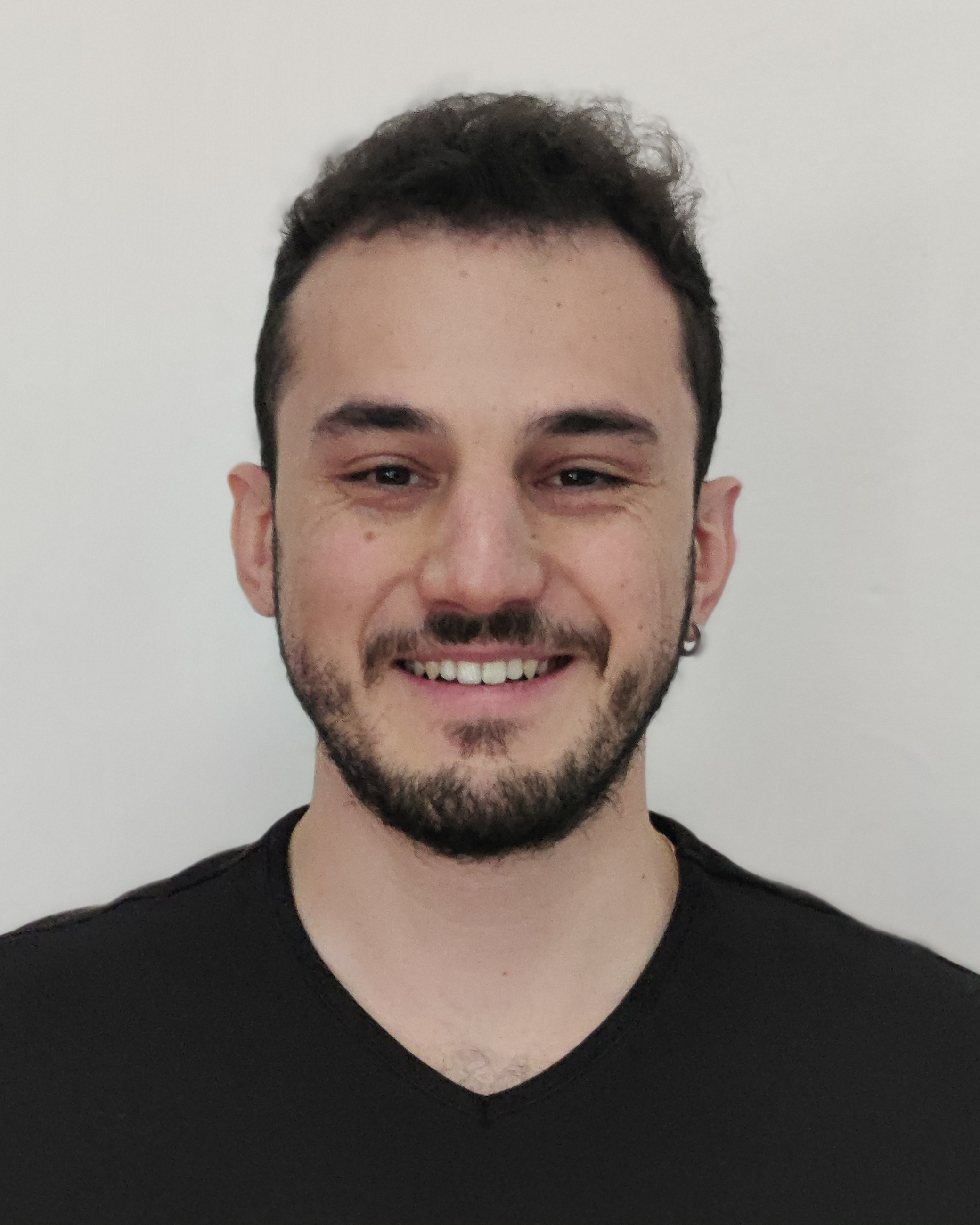}}]{Enver Bashirov}
(S'20) is currently an Early Stage Researcher at EU Horizon 2020 Marie Skłodowska Curie project MINTS, pursuing his Ph.D. degree at the Department of Information Engineering, University of Padova, Italy. He received his B.Sc. degree in Computer Engineering from Bilkent University and M.Sc. degree in Applied Mathematics and Computer Science from Eastern Mediterranean University, North Cyprus. His research interests include sensing applications in mmWave, together with machine learning and signal processing solutions.
\end{IEEEbiography}

\begin{IEEEbiography}[{\includegraphics[width=1in,height=1.25in,clip,keepaspectratio]{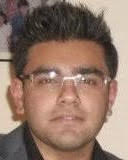}}]{Harsh Tataria} (M'17)
received the B.E. degree (honors) in electronic
and computer systems engineering and the Ph.D.\ degree in communications engineering from the Victoria University of Wellington, New Zealand, in December 2013 and March 2017, respectively. Since then, he has held postdoctoral fellowship positions at Queen's University Belfast, Belfast, U.K., the University of Southern California, Los Angeles, CA, USA, and Lund University, Sweden. His research interests include measurement and modeling of propagation channels, multiple antenna transceiver design, and statistical analysis techniques of multiple antenna systems at centimeter-wave, millimeter-wave, and sub-terahertz frequencies.
\end{IEEEbiography}

\begin{IEEEbiography}[{\includegraphics[width=1in,height=1.25in,clip,keepaspectratio]{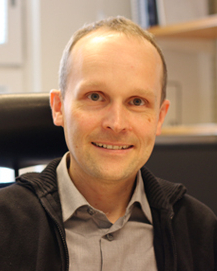}}]{Michael Lentmaier}
(SM'11) is an Associate Professor at the Department of Electrical and Information Technology at Lund University, Sweden, which he joined in January 2013. His research interests include design and analysis of coding systems, graph based iterative algorithms and Bayesian methods applied to decoding, detection and estimation in communication systems. He received the Dipl.-Ing. degree in electrical engineering from University of Ulm, Germany in 1998, and the Ph.D. degree in telecommunication theory from Lund University, in 2003. He then worked as a Post-Doctoral Research Associate at University of Notre Dame, Indiana and at University of Ulm. 
From 2005 to 2007 he was with the Institute of Communications and Navigation of the German Aerospace Center (DLR) in Oberpfaffenhofen, where he worked on signal processing techniques in satellite navigation receivers. From 2008 to 2012 he was a senior researcher and lecturer at the Vodafone Chair Mobile Communications Systems at TU Dresden, where he was heading the Algorithms and Coding research group.  He is a senior member of the IEEE and served as an editor for the \textsc{IEEE Communications Letters} (2010-2013), \textsc{IEEE Transactions on Communications} (2014-2017), and \textsc{IEEE Transactions on Information Theory} (2017-2020).  He was awarded the Communications Society \& Information Theory Society Joint Paper Award (2012) for his paper ``Iterative decoding threshold analysis for LDPC convolutional codes.''
\end{IEEEbiography}

\begin{IEEEbiography}[{\includegraphics[width=1in,height=1.25in,clip,keepaspectratio]{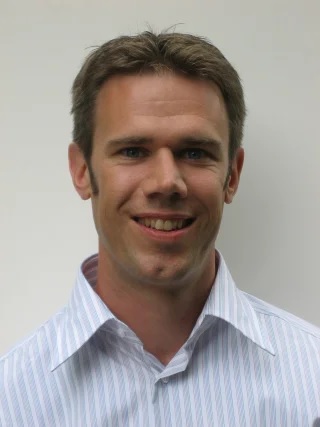}}]{Fredrik Tufvesson}
(F'17) received his Ph.D. degree from Lund University, Lund, Sweden, in 2000. After two years at a startup company, he joined the Department of Electrical and
Information Technology, Lund University, where he is currently a Professor of radio systems. His main research interest is the interplay between the radio channel and the rest of the communication system with various applications in 5G/B5G systems such as massive MIMO, mm wave communication, vehicular communication and radio based positioning.
Fredrik has authored around 100 journal papers and 150 conference papers, he is an IEEE Fellow, and his research has been awarded with the Neal Shepherd Memorial Award (2015) for the best propagation paper in the \textsc{IEEE Transactions on Vehicular Technology}, and with the IEEE Communications Society best tutorial paper award (2018, 2021).
\end{IEEEbiography}

\begin{IEEEbiography}[{\includegraphics[width=1in,height=1.25in,clip,keepaspectratio]{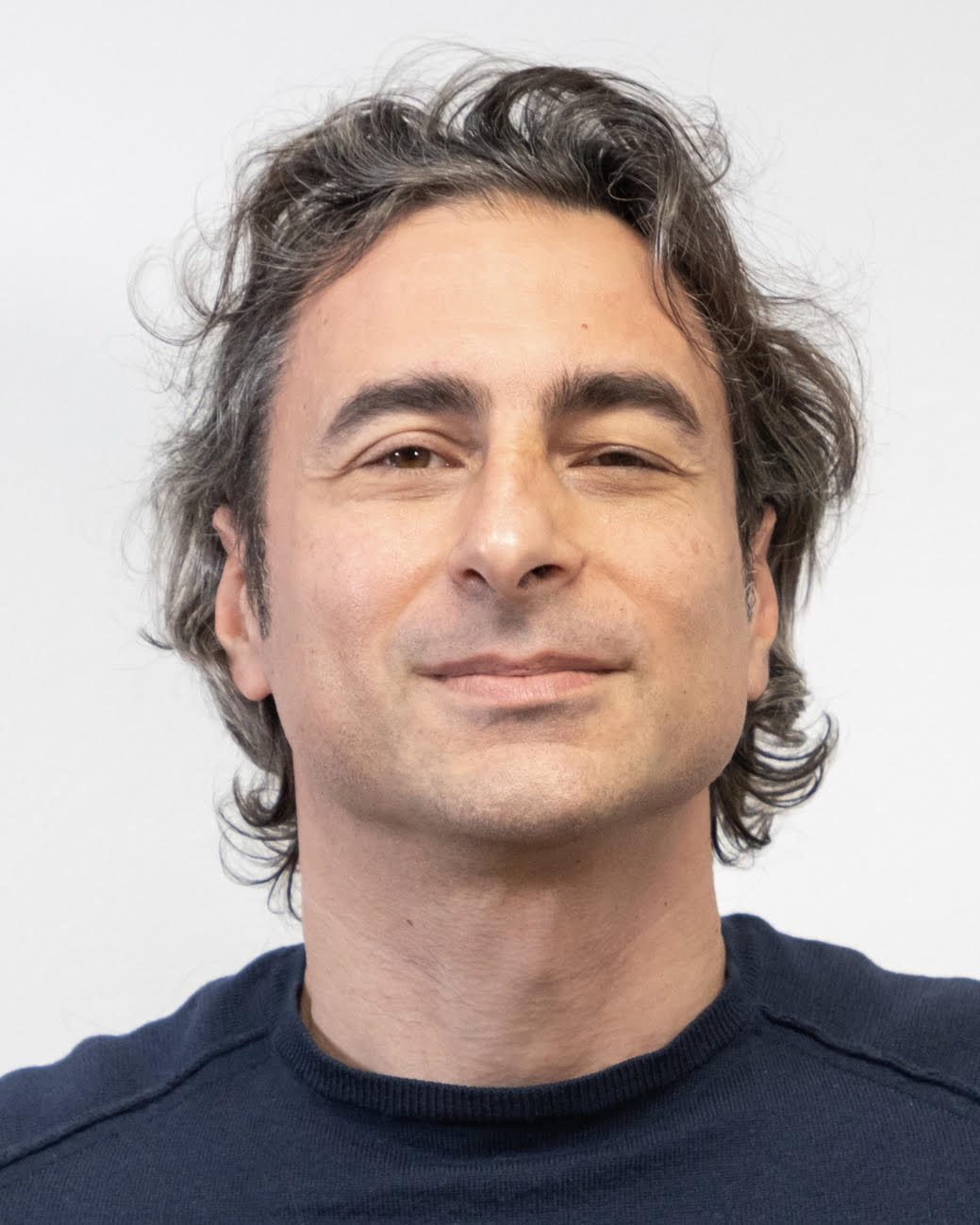}}]{Michele Rossi}
(SM'13) is a Professor of Wireless Networks in the Department of Information Engineering (DEI) at the University of Padova (UNIPD), Italy, where is the head of the Master's Degree in ICT for internet and Multimedia (\url{http://mime.dei.unipd.it/}). He also teaches Human Data Analysis at the Data Science Master's degree at the Department of Mathematics (DM) at UNIPD (\url{https://datascience.math.unipd.it/}). Since 2017, he has been the Director of the DEI/IEEE Summer School of Information Engineering (\url{http://ssie.dei.unipd.it/}). His research interests broadly embrace wireless sensing systems, green mobile networks, edge and wearable computing. Over the years, he has been involved in several EU projects on wireless sensing and IoT and has collaborated with major companies such as Ericsson, DOCOMO, Samsung and Intel. His research is currently supported by the European Commission through the H2020 projects MINTS (no. 861222) on ``mmWave networking and sensing'' and GREENEDGE (no. 953775) on ``green edge computing for mobile networks'' (project coordinator). Prof.~Rossi has been the recipient of seven best paper awards from the IEEE and currently serves on the Editorial Boards of the \textsc{IEEE Transactions on Mobile Computing}, and of the \textsc{Open Journal of the Communications Society}.
\end{IEEEbiography}

\begin{IEEEbiography}[{\includegraphics[width=1in,height=1.25in,clip,keepaspectratio]{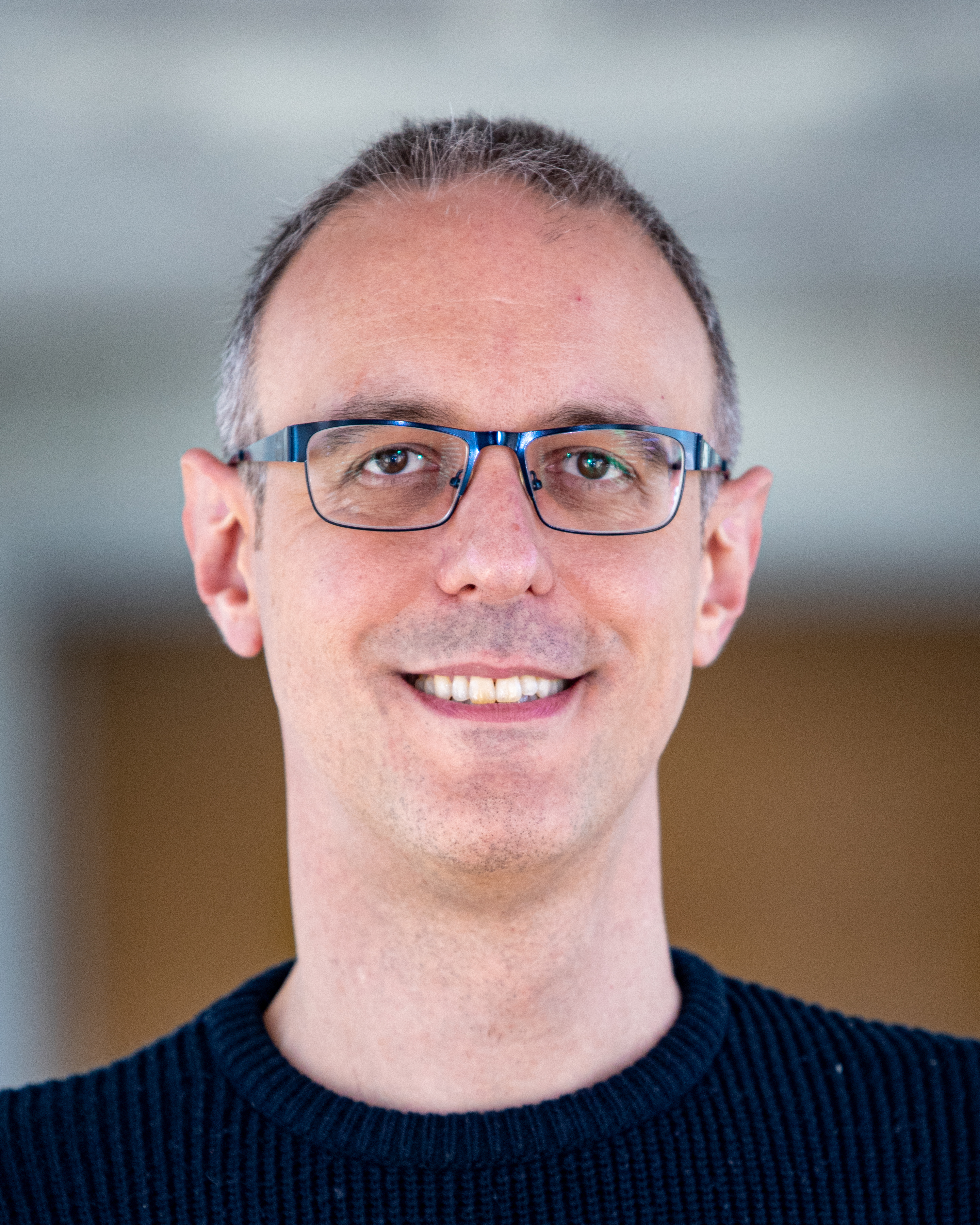}}]{Paolo Casari}
received the PhD in Information Engineering in 2008 from the University of Padova, Italy. He was on leave at the Massachusetts Institute of Technology in 2007, working on underwater communications and networks. He collaborated to several funded projects including EU FP7 and H2020 efforts, EDA projects, as well as US ARO, ONR and NSF initiatives, and is currently the PI of the NATO SPS project SAFE-UComm. In 2015, he joined the IMDEA Networks Institute, Madrid, Spain, where he led the Ubiquitous Wireless Networks group. In October 2019, he joined the faculty of the University of Trento, Italy, as a tenure-tracked assistant professor.

Dr. Casari is currently on the editorial boards of the \textsc{IEEE Transactions on Mobile Computing} and of the \textsc{IEEE Transactions on Wireless Communications}, and regularly serves in the organizing committee of several international conferences. Previously, he has been guest editor of a special issue of \textsc{IEEE Access} on ``Underwater Acoustic Communications and Networking.'' He received two best paper awards. His research interests include diverse aspects of networked communications and computing, such as channel modeling, network protocol design, localization, resource allocation, simulation, and experimental evaluation.
\end{IEEEbiography}

% \begin{IEEEbiographynophoto}{Name Surname}


\end{document}

%% file: acronyms.tex
\usepackage[acronym,shortcuts,nonumberlist,nopostdot]{glossaries}
\usepackage{glossary-mcols}

\makenoidxglossaries

\newacronym{4g}{4G}{fourth-generation}
\newacronym{5g}{5G}{fifth-generation}
\newacronym{3gpp}{3GPP}{Third-generation partnership project}
\newacronym{abft}{A-BFT}{association beamforming training}
\newacronym{abt}{ABT}{asymmetric beamforming training}
\newacronym{ad}{AD}{angle-Doppler}
\newacronym{adc}{ADC}{analog-to-digital converter}
\newacronym{adoa}{ADoA}{angle difference-of-arrival}
\newacronym{ae}{AE}{autoencoder}
\newacronym{aoa}{AoA}{angle of arrival}
\newacronym{aod}{AoD}{angle of departure}
\newacronym{ap}{AP}{access point}
\newacronym{api}{API}{application program interface}
\newacronym{bi}{BI}{beacon interval}
\newacronym{brp}{BRP}{beam refinement protocol}
\newacronym{bs}{BS}{base station}
\newacronym{bti}{BTI}{beacon transmission interval}
\newacronym{cacfar}{CA-CFAR}{cell-averaging constant false alarm rate}
\newacronym{cbap}{CBAP}{contention based access period}
\newacronym{cdf}{CDF}{cumulative distribution function}
\newacronym{cir}{CIR}{channel impulse response}
\newacronym{cnn}{CNN}{convolutional NN}
\newacronym{com}{COM}{center of mass}
\newacronym{cots}{COTS}{commercial off-the-shelf}
\newacronym{csi}{CSI}{channel state information}
\newacronym{dac}{DAC}{digital-to-analog converter}
\newacronym{dbscan}{DBSCAN}{density-based spatial clustering of applications with noise}
\newacronym{dft}{DFT}{discrete Fourier transform}
\newacronym{dkf}{DKF}{discrete KF}
\newacronym{dl}{DL}{deep learning}
\newacronym{dsp}{DSP}{digital signal processor}
\newacronym{dti}{DTI}{data transmission interval}
\newacronym{ecd}{ECD}{expand-contract dilation}
\newacronym{ekf}{EKF}{extended Kalman filter}
\newacronym{em}{EM}{expectation maximization}
\newacronym{endc}{EN-DC}{enhanced UTRA-dual connectivity}
\newacronym{esprit}{ESPRIT}{estimation of signal parameters via rotational invariance techniques}
\newacronym{fcos}{FCOS}{fully connected \mbox{one-stage}}
\newacronym{fft}{FFT}{fast Fourier transform}
\newacronym{fmcw}{FMCW}{frequency-modulated continuous wave}
\newacronym{fpga}{FPGA}{field-programmable gate array}
\newacronym{fscn}{FSCN}{fractionally strided convolutional network}
\newacronym{ftm}{FTM}{fine time measurement}
\newacronym{gan}{GAN}{generative adversarial network}
\newacronym{gru}{GRU}{gated recurrent unit}
\newacronym{gscm}{GSCM}{geometry-based stochastic channel model}
\newacronym{hdc}{HDC}{hybrid dilated convolution}
\newacronym{hr}{HR}{heart rate}
\newacronym{hvrae}{HVRAE}{hybrid variational RNN autoencoder}
\newacronym{if}{IF}{intermediate-frequency}
\newacronym{ifft}{IFFT}{inverse FFT}
\newacronym{iir}{IIR}{infinite impulse response}
\newacronym{iot}{IoT}{Internet of things}
\newacronym{itur}{ITU-R}{International telecommunication union -- radiocommunication Sector}
\newacronym{kf}{KF}{Kalman filter}
\newacronym{lm}{LM}{Levenberg-Marquardt}
\newacronym{lms}{LMS}{least mean squares}
\newacronym{los}{LoS}{line-of-sight}
\newacronym{lstm}{LSTM}{long-short term memory}
\newacronym{mac}{MAC}{medium access control}
\newacronym{mcu}{MCU}{micro-controller unit}
\newacronym{mf}{MF}{matched filter}
\newacronym{mimo}{MIMO}{multiple-input multiple-output}
\newacronym{miso}{MISO}{multiple-input single-output}
\newacronym{ml}{ML}{machine learning}
\newacronym{mmse}{MMSE}{minimum mean-square error}
\newacronym{mmw}{mmWave}{millimeter-wave}
\newacronym{mpc}{MPC}{multipath component}
\newacronym{mse}{MSE}{mean-square error}
\newacronym{music}{MUSIC}{multiple signal classification}
\newacronym{nlos}{NLoS}{non-line-of-sight}
\newacronym{nn}{NN}{neural network}
\newacronym{noma}{NOMA}{non-orthogonal multiple access}
\newacronym{ofdm}{OFDM}{orthogonal frequency-division multiplexing}
\newacronym{oks}{OKS}{object keypoints similarity}
\newacronym{pa}{PA}{power amplifier}
\newacronym{pdp}{PDP}{power-delay profile}
\newacronym{phy}{PHY}{physical layer}
\newacronym{prs}{PRS}{positioning reference signal}
\newacronym{psd}{PSD}{power spectral density}
\newacronym{pw}{PW}{pulsed wave}
\newacronym{ra}{RA}{range-azimuth}
\newacronym{rd}{RD}{range-Doppler}
\newacronym{rda}{RDA}{range-Doppler-azimuth}
\newacronym{resnet}{ResNet}{residual network}
\newacronym{rf}{RF}{radio frequency}
\newacronym{rgb}{RGB}{red-green-blue}
\newacronym{rms}{RMS}{root mean square}
\newacronym{rmse}{RMSE}{root mean square error}
\newacronym{rnn}{RNN}{recursive NN}
\newacronym{rss}{RSS}{received signal strength}
\newacronym{rssi}{RSSI}{received signal strength indicator}
\newacronym{rtt}{RTT}{round-trip time}
\newacronym{rx}{RX}{receive/receiver}
\newacronym{rzf}{RZF}{regularized zero-forcing}
\newacronym{saf}{SAF}{spatial attention fusion}
\newacronym{sage}{SAGE}{space-alternating generalized expectation maximization}
\newacronym{sdr}{SDR}{software-defined radio}
\newacronym{slam}{SLAM}{simultaneous localization and mapping}
\newacronym{sls}{SLS}{sector-level sweep}
\newacronym{snr}{SNR}{signal-to-noise ratio}
\newacronym{soc}{SoC}{system-on-chip}
\newacronym{sp}{SP}{service period}
\newacronym{spi}{SPI}{subsample peak interpolation}
\newacronym{srs}{SRS}{sounding reference signal}
\newacronym{ssw}{SSW}{sector sweep}
\newacronym{sta}{STA}{station}
\newacronym{stft}{STFT}{short-time Fourier transform}
\newacronym{sv}{SV}{Saleh-Valenzuela}
\newacronym{svm}{SVM}{support vector machine}
\newacronym{tdoa}{TDoA}{time difference-of-arrival}
\newacronym{toa}{ToA}{time of arrival}
\newacronym{tof}{ToF}{time of flight}
\newacronym{tx}{TX}{transmit/transmitter}
\newacronym{txss}{TXSS}{transmit sweep}
\newacronym{uav}{UAV}{unmanned aerial vehicle}
\newacronym{ue}{UE}{user equipment}
\newacronym{ukf}{UKF}{unscented Kalman filter}
\newacronym{urllc}{URLLC}{ultra-reliable low-latency communications}
\newacronym{usrp}{USRP}{universal software radio peripheral}
\newacronym{v2x}{V2X}{vehicle-to-everything}
\newacronym{va}{VA}{virtual anchor}
\newacronym{vco}{VCO}{voltage-controlled oscillator}
\newacronym{wlan}{WLAN}{wireless LAN}
\newacronym{zf}{ZF}{zero-forcing}

\setglossarystyle{alttree}
\glssetwidest{URLL-NMP}

%% file: 01_Introduction.tex
\section{Introduction}  \label{sec:intro}

%\IEEEPARstart{T}{his} demo file is intended to serve as a ``starter file''

\Ac{mmw} communications in the 28--300~GHz band are looked at with great interest, as they may be able to quench --at least temporarily-- the ever-increasing bandwidth requirements of such applications as massive \ac{iot}, virtual/augmented reality, mobile cloud services and ubiquitous ultra-high definition multimedia streaming~\cite{xiao_mmwave_comm_future_mobile_networks_JSAC_2017,wang_mmwave_comms_survey_COMST_2018,uwaechia_comprehensive_survey_mmwave_5g_feasib_chall_Access_2020}. This would cover the shortcomings of sub-6~GHz technologies such as WiFi and \ac{4g} cellular networks, which currently cannot support the massive bandwidth and number of users the above applications imply.

The potential of \ac{mmw} technology, however, is not limited to higher-rate communications: rather, \ac{mmw} devices can become a proxy for high-resolution device-based localization as well as device-free sensing. These capabilities follow from the physics of \ac{mmw} propagation. First, the shorter wavelength of \acp{mmw} (compared to sub-6~GHz signals) enables accurate location estimates and lower location error bounds~\cite{henk_error_bounds_3D_loc_mmwave_TWC_2018,henk_leb_mmw_iq_imbalance_TVT_2020}. Second, \acp{mmw} have well-known and peculiar propagation characteristics~\cite{rappaport2013mmw,rappaport_bband_mmw_prop_meas_AntProp_2013} which yield higher spatial scanning resolution. For example, \acp{mmw} propagate quasi-optically, meaning that a \ac{los} \ac{mpc} is predominant over \ac{nlos} contributions to the received signal~\cite{peinecke_phong}. Scattering also has a limited impact off typical non-rough reflecting surfaces such as walls, furniture, metal plates as well as glass layers~\cite{subrt_scattering_indoor},~\cite{rappaport_pathloss_models_UMi_GLOBECOM_2013}. 

Another consequence of \ac{mmw} propagation is that \ac{mmw} signals undergo much higher path loss with respect to microwaves. To compensate for this attenuation, and still enable long-reach wireless links, \ac{mmw} devices resort to large or massive antenna arrays. Via beamforming, they can focus their transmitted energy towards a confined portion of the 3D space, and thus achieve greater directionality. While this requires specific protocols for initial access~\cite{giordani_initial_access_magazine_2016,perf_analysis_mmwave_cell_two_stage_2016,capone_initial_access_Access_2016} and beam training such as the IEEE 802.11ad~\cite{802.11ad,ieee80211adStandard} and 802.11ay~\cite{802.11ayWLANdesign2018,802.11ay_claudioSilva} standards, it also means that a reduced amount of power is typically directed towards secondary multipath components. In addition with the quasi-optical propagation patterns discussed above, the main consequence is that the received angular spectrum of a \ac{mmw} signal is sparse: in typical conditions, one can identify one \ac{los} \ac{mpc} along with a number of \ac{nlos} \acp{mpc} corresponding to signal reflections off the surrounding environment. 
The above features of \ac{mmw} communications have significant implications for localization and sensing~\cite{6gLocSensingOverview}. For example, being able to separate \acp{mpc} in the angular domain enables angle-based localization schemes that are not normally used in sub-6 GHz systems due to limited angular resolution when using small antenna arrays.
Fingerprinting-based algorithms can also be enhanced by incorporating angle-based features to improve location discrimination. From the point of view of device-free sensing, \ac{mmw} propagation also implies typically clearer reflections off sensed targets and parts thereof. For example, a) quasi-optical \ac{mmw} propagation along with b) the large \ac{mmw} bandwidth available at typical \ac{mmw} radar frequencies respectively imply that reflections off targets are usually separate in the a) angle and b) time domains. This makes it possible to measure features that point to each reflection's movement velocity (e.g., the Doppler shift) and use this data to precisely localize and identify different targets. 

In this paper, we focus on indoor \ac{mmw} device-based localization and device-free sensing, and provide a comprehensive review of approaches, technologies, schemes and algorithms to estimate a device or object's location in an indoor environment. The objective of our survey is to shed light on indoor applications of localization and sensing using \ac{mmw} signals. 
Location information can be extremely useful in different indoor setups~\cite{xiao2020overview6G,fettweis_loc_feature_mmwave_IWCMC_2016}. 
For example, in factories and industrial environments, location information can be exploited to enhance \ac{urllc} for industrial \ac{iot} and smart manufacturing~\cite{kfloc2021vashisht,6gLocSensWhitePaper}. Accurate localization and sensing can benefit healthcare scenarios for patient tracking and lifesign/behavior monitoring, help people navigate in indoor areas, provide trajectory suggestions through relevant waypoints in museums, malls, and company headquarters, as well as support mission-critical applications such as disaster relief and indoor security. 
Location systems are also crucial for network performance optimization. Accurate location information can support the fast alignment of transmit and receive antenna arrays, optimize the association between clients and \acp{ap}, and prevent blockage of high-power \ac{los} paths via predictive handovers to provide seamless coverage. This can result in low-latency communications as needed for augmented reality, virtual reality, and tactile Internet applications.

In the following, we start with an overview that touches on \ac{mmw} signal structure and propagation characteristics that make this domain unique with respect to other radio communication and sensing technologies. We consider practical constraints that define the applicability of algorithms and processing schemes to \ac{mmw} devices operating indoors. We then delve into a detailed description of device-based indoor localization algorithms, explaining the main localization techniques employed in the literature, and how they are practically implemented in real \ac{mmw} hardware whenever available. For device-free sensing, we list a number of relevant applications and technologies that leverage \ac{mmw} hardware and signals to detect, localize and track targets indoors, as well as to specifically identify features related to sub-sections of a target (e.g., a part of the human body). Because these device-free approaches are mainly based on \ac{mmw} radar devices, we will briefly discuss how \ac{mmw} radar bands are being standardized for different applications.

\subsection{Differences with respect to previous surveys}
\label{sec:intro.diffsurv}

Localization and sensing are topics of great interest for both current and future-generation wireless communication system engineering. The research on these topics has proceeded at a steady pace, considering aspects as diverse as localization techniques, heterogeneous technologies, different scenarios, and different kinds interactions between the device to be localized and the location server, among others. Several surveys cover these aspects, typically for sub-6~GHz technologies. 
For example, Zafari et~al.~\cite{surv7_indoorlocsystech} and Geok et~al.~\cite{surv3_indoorloc} focus on localization techniques for wireless systems in general, and cover heterogeneous technologies. These works only tangentially consider \acp{mmw}, and instead survey geometric and signal processing-based localization methods for sub-6 GHz systems.
Ngamakeur et~al.~\cite{surv1_devicefree} delve into device-free sensing of different human signatures using sub-6~GHz technologies indoors. Here, the focus is on the localization, tracking and identification of multiple subjects using Wi-Fi and other kinds of wireless sensors.

By leveraging similar technologies, Singh et~al.~\cite{surv5_ML_indoor_WiFi_fingerprinting} consider techniques and algorithms to localize \ac{iot} devices indoors. In this case, the focus of the survey is on a specific source of location information (received WiFi signal strength fingerprints) and on how machine learning works when applied to such datasets.
By expanding into the concept of smart world, the work in~\cite{surv6_ubiquitous} also surveys how sub-6~GHz technologies can help improve a variety of services via data collection and system automation using active and passive sensing techniques.
Finally, the work in~\cite{5gmimo2019survey} touches on aspects related to the modeling and estimation of wireless channels in \ac{5g} cellular systems. While the work touches on localization, the covered techniques apply to outdoor cellular systems, and can thus leverage the density and much higher computational power of their hardware. 

Unlike our survey, none of the above works targets millimeter wave device-based and device-free indoor localization. This area is characterized by several interesting research works to date, but remains a very hot topic due to the inception of \ac{mmw} coverage for future \ac{5g}-and-beyond networks as well as wireless (indoor) local-area networks. The objective of our survey is to cover the most significant work in this area, while giving a comprehensive view of unsolved challenges and open research avenues.

Note that, in our survey, we are \emph{not} seeking an analysis of the limits of \ac{mmw} localization and sensing technology based on purely theoretical arguments, or an operational description of well-known geometric localization algorithms, or even a coverage of the integration between \ac{mmw} communications and \ac{5g}, beyond-\ac{5g}, and future 6G networks. These are related yet tangential topics for which we rather refer the interested reader to one of the several excellent surveys that touch on these aspects, e.g.,~\cite{xiao2020overview6G, 6gLocSensingOverview,5gmimo2019survey,carlo2018CommSurvTut, yang2021integratedLocComm, 6gLocSensWhitePaper, towards6G2020Giordani,wild2021joint,bourdoux20206g,sanusi2021review }.

\subsection{Outline and organization of the manuscript}
\label{sec:intro.outline}

The remainder of this paper expresses three purposes:
to cover the characteristics of \ac{mmw} propagation and communication/sensing hardware that impacts localization and sensing performance, including standardization efforts (Sections~\ref{sec:chan} through~\ref{sec:std}); to detail the state of the art in device-based \ac{mmw} localization (Section~\ref{sec:localgo}) and in device-free \ac{mmw} sensing (Section~\ref{sec:radarsurv}); and finally to discuss our findings, discuss promising research avenues, and draw concluding remarks (Sections~\ref{sec:disc} and~\ref{sec:concl}).

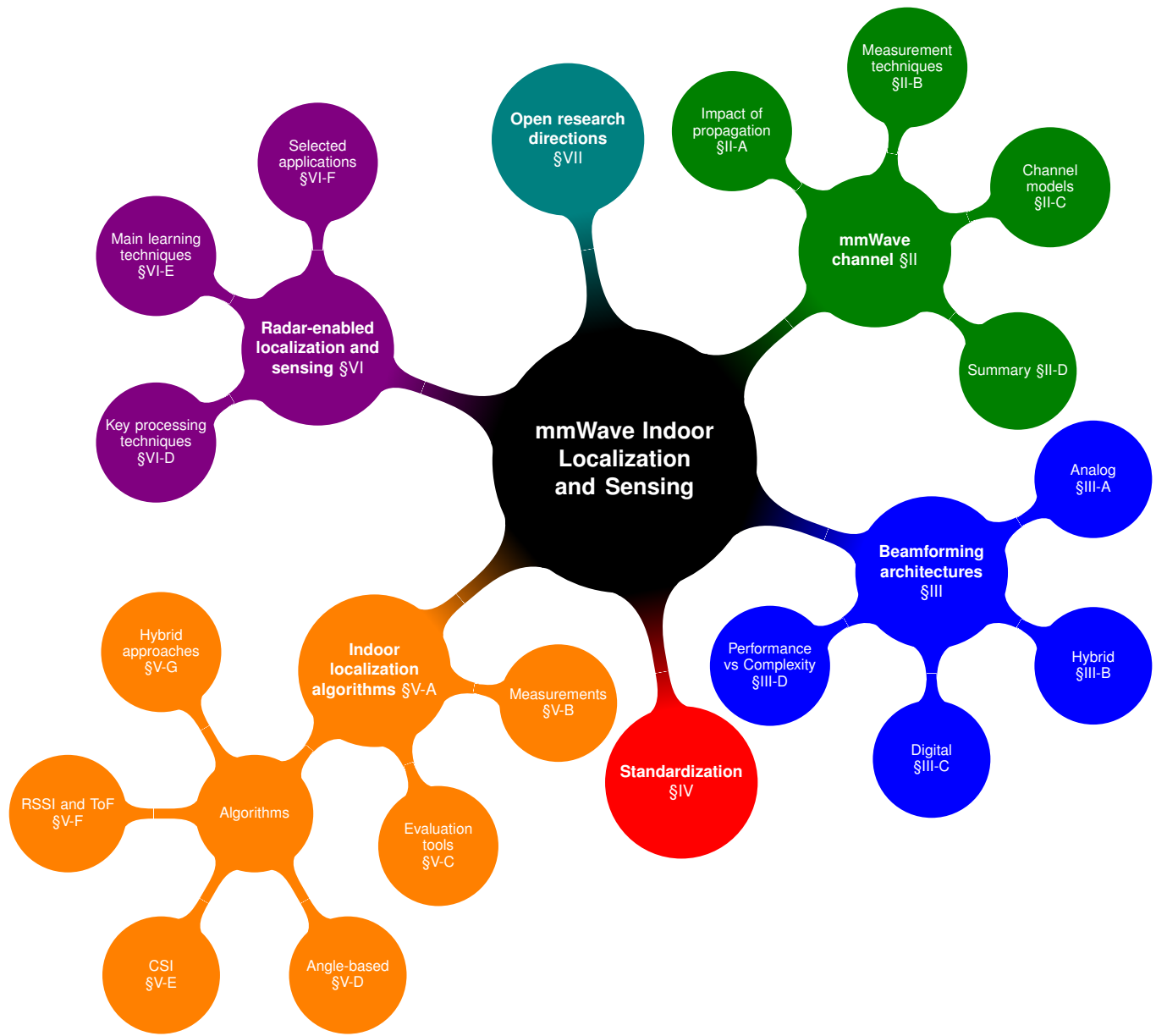
\begin{figure*}[t]
    \centering
    \input{summaryMap}
    \caption{Mind map showing the organization of this survey.}
    \label{fig:surveyorg}
\end{figure*}

In particular, Sections~\ref{sec:localgo} and~\ref{sec:radarsurv} constitute the core of our technological survey. Section~\ref{sec:localgo} discusses device-based localization algorithms for indoor environments, whereas Section~\ref{sec:radarsurv} presents several approaches for radar-based device-free localization. 
Each section is organized to first present the section topic, and then to add progressively more details related to the typical techniques appropriate for each section, the hardware typically used in testbeds, and the description of each surveyed approach. We also include summary tables to help the reader navigate the contents and extract key information. Both Sections~\ref{sec:localgo} and~\ref{sec:radarsurv} end with a summary of the most relevant aspects and findings.

Fig.~\ref{fig:surveyorg} represents the organization of the survey as a mind map, starting from Section~\ref{sec:chan} (top right), proceeding clockwise, and concluding with Section~\ref{sec:disc}.

% Finally, Section~\ref{sec:disc} summarizes our findings and discusses open research avenues. Here, we also argue about which enhancements to \ac{mmw} hardware can help improve the quality and accuracy of localization and sensing. We conclude our paper in Section~\ref{sec:concl}.

%% file: summaryMap.tex
% Author: Till Tantau
% Source: The PGF/TikZ manual
\hypersetup{linkcolor=white,anchorcolor=white,citecolor=white,filecolor=white,menucolor=white,runcolor=white,urlcolor=white}
\resizebox{1\linewidth}{!}{
\begin{tikzpicture}
\sffamily 
\centering
  \path[large mindmap,concept color=black,text=white]
    node[concept] {\textbf{mmWave Indoor Localization and Sensing}}
    [clockwise from=40]
    child[concept color=green!50!black] {
      node[concept] {\textbf{mmWave channel}  \S\ref{sec:chan}}
      [clockwise from=140]
      child { node[concept] {Impact of propagation \S\ref{sec:chan.pec}} }
      child { node[concept] {Measurement techniques \S\ref{sec:chan.meastech} } }
      child { node[concept] {Channel models\\ \S\ref{sec:chan.models}} }
      child { node[concept] {Summary \S\ref{sec:summary}} }
    }  
    child[concept color=blue] {
      node[concept] {\textbf{Beamforming architectures} \S\ref{sec:bfarch}}
      [clockwise from=30]
      child { node[concept] {Analog \\ \S\ref{sec:bfarch.analog}} }
      child { node[concept] {Hybrid \\ \S\ref{sec:bfarch.hybrid}} }
      child { node[concept] {Digital \\ \S\ref{sec:bfarch.digital}} }
      child { node[concept] {Performance vs Complexity \S\ref{sec:bfarch.perfcomp}} }
    }
    child[concept color=red] { node[concept] {\textbf{Standardization} \\ \S\ref{sec:std}} }
    child[concept color=orange] { node[concept] {\textbf{Indoor localization algorithms}  \S\ref{sec:localgo.intro}} 
    % add all the subsections
    [clockwise from=-10]
      child { node[concept] {Measurements \\ \S\ref{sec:localgo.locdepmeas}}}
      child { node[concept] {Evaluation tools \\ \S\ref{sec:localgo.evtools}} }
      child { node[concept] {Algorithms}}
      [clockwise from=-60]
      child { node[concept] {Angle-based \\ \S\ref{sec:localgo.aoa}}}
      child { node[concept] {CSI \\ \S\ref{sec:localgo.csi}} }
      child { node[concept] {RSSI and ToF \\ \S\ref{sec:localgo.rssi-tof}} }
      child { node[concept] {Hybrid approaches \\ \S\ref{sec:localgo.hybrid}} }
    }
    child[concept color=violet]{node[concept] {\textbf{Radar-enabled localization and sensing} \S\ref{sec:radarsurv}}
    [clockwise from=-150]
      child { node[concept] {Key processing \\ techniques  \\ \S\ref{sec:radarsurv.proc}} }
      child { node[concept] {Main learning \\ techniques  \\ \S\ref{sec:radarsurv.learning}} }
      child { node[concept] {Selected applications  \\ \S\ref{sec:radarsurv.detloc}} }
    }
    child[concept color=teal]{
    node[concept] {\textbf{Open research directions} \\ \S\ref{sec:disc}}
    };
\end{tikzpicture}
}
\hypersetup{linkcolor=black,anchorcolor=black,citecolor=black,filecolor=black,menucolor=black,runcolor=black,urlcolor=black}

%% file: 02_Influence_mmWave_channels.tex
\section{Influence of mmWave channels} \label{sec:chan}

\subsection{Impact of mmWave frequencies on propagation conditions} \label{sec:chan.pec}
The propagation of a wave through any medium depends on its frequency: this basic property helps us predict the behavior of the channel for diffeangularrent carrier frequencies. When it comes to \acp{mmw}, considering the Friis equation under the assumption that the antenna gain $G$ at both link ends is frequency-independent (by reducing the antenna aperture), the free space path loss increases with the square of the carrier frequency $f$. On the contrary, assuming a constant physical area $A$ at both the \glsunset{tx}\glsunset{rx}transmitter (\ac{tx}) and the receiver (\ac{rx}), the antenna gains $G = A ({4 \pi}/{\lambda})^2$ increase on both sides, and thus the overall path loss \emph{decreases} quadratically with increasing frequency $f$~\cite{tataria_standardization_2021}. Specular reflections for dielelectric halfspaces (e.g., ground reflections) depend on frequency as long as the dielectric constant is itself a function of frequency. For reflections at a dielectric layer (e.g., building walls) the specular reflections depend on the electrical thickness of the wall, which in turn is also a function of frequency. Interestingly, we have no evidence that the reflection coefficient varies with frequency, although the transmission power decreases uniformly with increasing frequency due to the skin effect in lossy media~\cite{shafi_microwave_2018}.

Two effects that have gained spotlight with the increased interest in the \ac{mmw} band are diffraction and diffuse scattering. The former reduces noticeably at high frequencies, and larger objects lead to ``sharp'' shadows. The latter effect is more significant as the surface roughness becomes comparable to \ac{mmw} wavelengths. As the surface roughness increases, the objects behave like a Lambertian radiator, which scatters the radiation. Foliage has a similar effect as scattering; with the decreasing wavelength relative to the size of the leaves, we observe more diffused scattering and less penetration. Another factor is atmospheric attenuation due to fog or rain\cite{liebe_updated_1985} and may affect the \ac{mmw} frequencies in case of extreme weather.

Channel models used for localization need to account for the above mentioned phenomena, and are often based on ray tracing or cluster-based modeling with some \ac{gscm}\cite{kunisch_ultra-wideband_2003,poutanen_multi-link_2012, maltsev_channel_2016}. Moreover, for ray tracing approaches, high-resolution environment information is needed to account for such surface roughness, as different materials have different properties (e.g., glass windows vs.\ concrete walls). These effects also depend on the environment: the high concrete walls and glass surfaces of the urban areas lead to different propagation conditions, compared to the greener suburban areas with, e.g., stucco exteriors and shorter walls.

\subsection{Measurement techniques and results} \label{sec:chan.meastech}

To model the properties of a channel, we need to perform the measurements for different propagation scenarios. A channel sounder, that helps to measure these properties is not only an expensive piece of equipment but as we move towards higher frequencies, the susceptibility to \emph{phase noise} as well as \emph{antenna spacing} errors start to increase. Similarly, the cost and energy consumption of up/down-conversion chains, in particular of the front-end mixed signal circuitry in \glsunset{adc}\glsunset{dac}analog-to-digital and digital-to-analog converters (\acsp{adc}/\acsp{dac}) as well as \acp{pa} becomes of paramount importance. For up-to-Gbit/s sampling rates (as often required by best-in-class channel sounding), 12-15~bit resolution is required. To penetrate larger distances (and thus to maximize the forward link gain), \acp{pa} typically need to operate with 6-10~dB backoff power efficiency and need to be continuously driven close to their 1~dB compression point limits.

Consequently, the channel sounders used often for measurements at high frequencies use omnidirectional antennas~\cite{keusgen_propagation_2014} or if directional~\cite{nguyen_empirical_2016}, then the angular resolution is not taken into account. Directionality is achieved by mechanically rotating horn antennas in most cases and the angular resolution corresponds to the beamwidth,  e.g.,~\cite{rappaport_wideband_2015,hur_wideband_2015,weiler_quasi-deterministic_2016}. For indoor measurement scenarios, the directional information though can be enhanced by using switched antenna arrays along with super-resolution algorithms like \ac{sage}\cite{gustafson_directional_2011} and RIMAX~\cite{schneider_large_2010}. It is possible to use  electronically-switched horn arrays~\cite{papazian_radio_2016} as well, which additionally lets us evaluate the \ac{mpc} and intra-cluster information.

\subsubsection{Key outdoor results}

When it comes to outdoor measurements, path loss is a key parameter. For channel modelling, we need to measure the pathloss coefficient, its mean and its variance. The pathloss coefficient for \ac{mmw} frequencies is close to that of microwaves, i.e., often there is no strong frequency dependence beyond the $f^2$ dependence of free-space path loss~\cite{haneda_frequency-agile_2016}. In \ac{los} scenarios, the path loss coefficient lies between 1.6-2.1 (2 for pure free-space propagation) and in  \ac{nlos} scenarios the value increases to 2.5 and 5 (e.g.,~\cite{rappaport_wideband_2015,hur_wideband_2015,hur_proposal_2016}).

On the other hand, the \textit{variance of the path loss around the distance-dependent mean} is higher at \ac{mmw} frequencies, which in turn increases the probability of outage~\cite{jaeckel_correlation_nodate}. The standard deviation as well is strongly dependent on the distance and its values increases from 5-10~dB to more than 20~dB as the distance increases from 30~m to 200~m~\cite{hur_proposal_2016}. This is due to the variation in power levels caused by location and orientation of a street in an urban macro cell~\cite{molisch_spatially_2016} and not due to shadowing as one may expect.

Another parameter important for channel modelling is the \ac{rms} delay spread. But it changes with frequency and thus it may not be the best parameter to model the delay dispersion. Instead, delay windows may be a better alternative as they define the time interval containing part of the energy of \ac{pdp}. Delay spreads in an outdoor environment are measured or simulated by ray tracing ~\cite{rappaport_wideband_2015,hur_wideband_2015,weiler_quasi-deterministic_2016,raschkowski_metis}. Beamforming can help with minimizing the delay spread~\cite{maccartney_jr_exploiting_2015}. The type of beamforming to be used depends on the angular dispersion properties. Angular spreads measured at the \ac{bs} are more accurate than those measured at the \ac{ue} as the ray tracers used often do not include scattering objects such as street signs, parked cars, etc. in their geographic database ~\cite{hur_proposal_2016,thomas_3d_2014}. As observed in ~\cite{hur_wideband_2015,ko_millimeter-wave_2017}, the \ac{rms} angular spread at the \ac{bs} is
of the order of 10$^\circ$ with one cluster only while at the \ac{ue}, the angular spreads are in the range 30-70$^{\circ}$~\cite{rappaport_wideband_2015,hur_wideband_2015,ko_millimeter-wave_2017,kim_directional_2016}.

More information related to fixed wireless scenarios can be found in~\cite{shafi_5g_2017}.

\input{channel_models}

\subsubsection{Key indoor results}
Measurements for indoor environments have picked up in recent years as we look at localization applications for \ac{5g}. The results are often from office and industrial environments, where different material densities can be studied. The path loss coefficient in this case ranges from 1.2-2 in \ac{los} to 2-3 in \ac{nlos} scenarios~\cite{smulders_statistical_2009,erden202028}. The frequency dependence of the path loss is more significant for indoor than outdoors, $f^k$ with $k \approx 2.5$ was observed in~\cite{xing_indoor_2019}. Overall though, the values are closer to those at sub-6 GHz, with an increased probability of outage. Path loss in some cases is shown to follow a dual-slope model and is the same for both \ac{mmw} and sub-6 GHz. The floating intercept model is another alternative used in \ac{3gpp} standards for indoor modelling at high frequencies. Human blockage can cause upto 10-20dB attenuation regardless of one or two people\cite{huang2020multi} and similar values in case of trucks in outdoor scenarios\cite{weiler_environment_2016}. In~\cite{medbo_60_2015}, \ac{fft} based beamforming is used in conjunction with a very large virtual array (25$\times$25$\times$25 elements). It highlights the scattering caused by small objects specifically in \ac{nlos} case and the importance of small scale characterization. Further, it is shown that the indoor environment leads to enhanced diffused \ac{mpc} energy.

Delay spread measured in office scenarios is usually less than 5~ns in \ac{los} conditions, and 10--20~ns in \ac{nlos} conditions~\cite{fu_frequency-domain_2013,haneda_statistical_2015,maccartney_indoor_2015, erden202028,tariq2020mmwave}. Though these measurements were limited to under 100 GHz, recently \cite{xing2021millimeter} performed measurements at 142 GHz and observed delay spread values of 3~ns in \ac{los} and 9~ns for \ac{nlos}. Further, the observed channels are much sparser at frequencies over 100~GHz and we notice higher partition loss compared to 28~GHz. It is worth noting for indoor measurements, the number of \acp{mpc} is higher with more clusters than measured for outdoor with rotating horn antennas~\cite{cano2020channel}. Here the angular spreads are often measured for clusters, with the intra-cluster azimuth and elevation angles are described as having a Laplacian distribution with a spread of 5$^{\circ}$~\cite{gustafson_mm-wave_2014}.

\subsection{Models for mmWave channels} \label{sec:chan.models}

Because \ac{mmw} propagation channels differ from microwave channels, we need to redefine or rather add certain parameters for \ac{mmw} channel modeling. 
As mentioned in~\cite{shafi_5g_2017}, \ac{mmw} channels require 3D modeling of azimuth as well as elevation spreads, inclusion of temporal/spatial/frequency consistency and multipath cluster based modeling. These have further impact when we consider positioning and localization. Prevalent models for \ac{mmw} are \acp{gscm} that imitate the propagation environment with stochastic processes, and create a 3D map. To correctly reproduce the wireless environment, parameter values need to be extracted from the channel impulse response of real time measurements done using a channel sounder. An extensive review of propagation chacteristics at \ac{mmw} frequencies is available in~\cite{salous_millimeter-wave_2016}, which also provides a summary of channel sounder measurements and relevant channel models. The \ac{3gpp} defined different environments for \ac{mmw} channel modeling, these include Urban Macro, Urban Micro, Indoor Office and Rural Macro. Several outdoor and indoor measurements are available, but for this paper we compare large-scale parameter values for an indoor office scenario listed in Table~\ref{models}.

Prominent channel models have been developed for the above mentioned scenarios based on measurements done in each of them. Some key results have already been discussed, but we also observe that cluster-based multipath channel components have been modelled, in order to specifically account for an indoor office environment. Also, as can be seen from the table, the angular spread is no longer limited to the azimuth plane.

\subsubsection{Static vs.~dynamic modeling} Due to the high frequency and thus higher path loss, there is significant deterioration when the \ac{ue} is stationary and more so when the \ac{ue} is moving or is in a high movement zone and transitions from a \ac{los} to \ac{nlos} scenario. This requires the dynamic modeling of the communication channel, as the moving objects in the vicinity also act as random blocking obstacles. The \ac{bs} needs to transmit training beams more frequently so as to update the \ac{aod}/\ac{aoa} estimates, since the location of \ac{ue} changes over time, and slight errors in the orientation of the beams can lead to significant performance loss~\cite{hur_feasibility_2018}.
So far, we have considered a fixed \ac{bs} and slow moving \ac{ue}, but with \ac{5g} and \ac{v2x} communications we expect high mobility scenarios\cite{dupleich_multi-band_2018}. Most \ac{mmw} channel models are still defined only for a fixed \ac{bs}, but have added support for dynamic modeling scenarios for \ac{v2x}.

\begin{figure}[t]
     \centering
         \includegraphics[width=1\columnwidth,trim={2cm 1cm 2.5cm 0},clip]{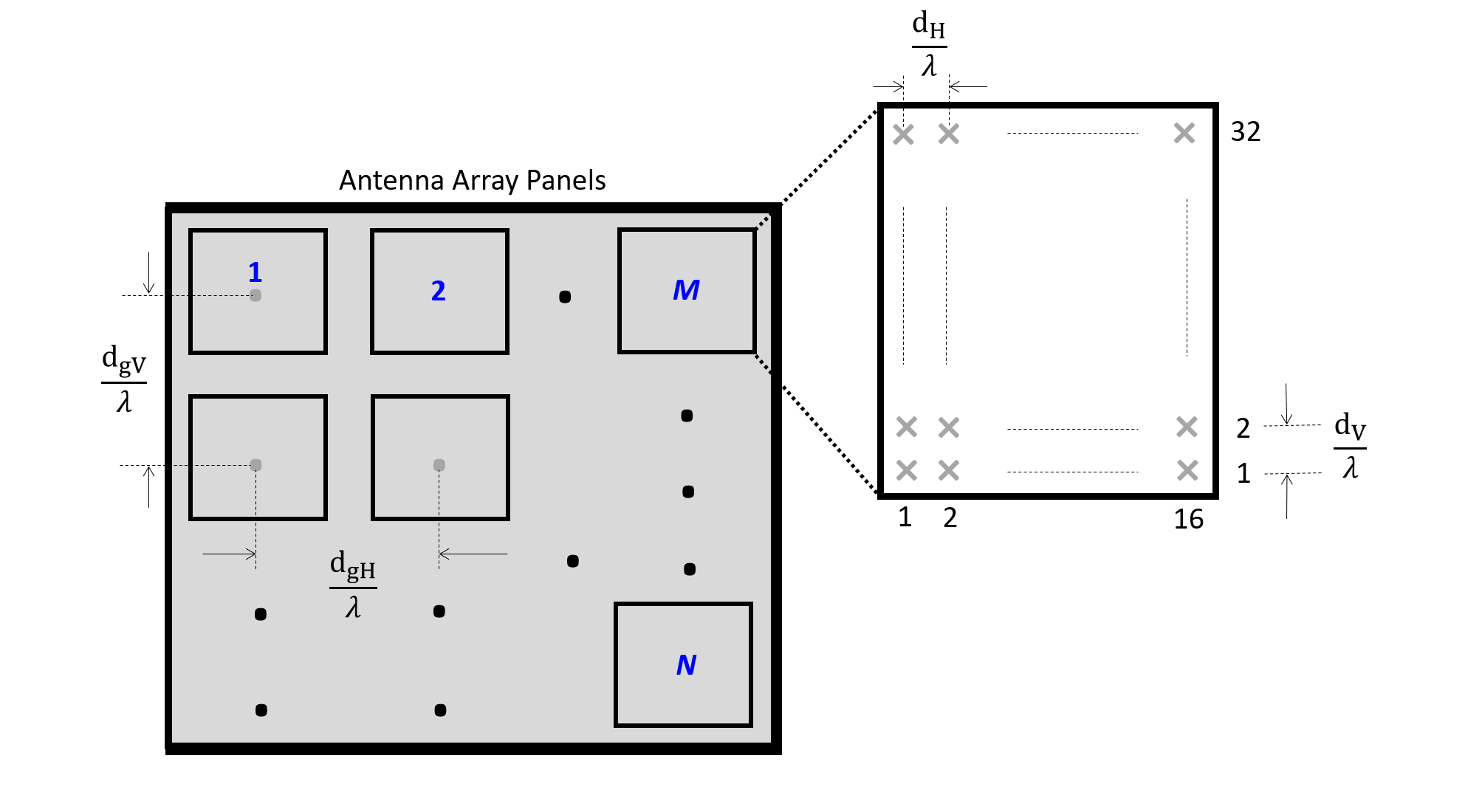}
         \vspace{-5mm}
         \caption{Cross-polarized antenna array panel~\cite{3gpprelease14}.}
    \label{arraymodel}
\end{figure}

\subsubsection{Blockage} \acp{mmw} cannot penetrate obstacles such as human bodies, walls, foliage, etc. Thus, these blockage sources need to be modelled in the link budget itself. One such characterization study is found in~\cite{fuschini2016item}, which measured power loss (in dB) when 70-GHz \ac{mmw} signals propagate through a brick wall, a PC monitor, and book shelves. Blockage does not affect just the total received power but also the angle or power of multipath signal components, due to varying sizes, positions and directions of the blocking object/human. Localizing the position of the \ac{ue} with respect to these blockage sources becomes onerous, especially in a dynamic setting. 

\subsubsection{Spatial consistency and clusters} A new, previously unexplored requirement was added to \ac{3gpp} Release 14~\cite{3gpprelease14}. When \ac{mmw} communications take place through narrow antenna radiation beams, the channel characteristics become highly correlated, especially when two \acp{ue} are close and see the same \ac{bs}. Also, for applications related to \ac{v2x} communications, it is paramount that the channel evolves smoothly without discontinuities during mobility~\cite{tataria_impact_2020}.

\subsubsection{Polarization} The radiation pattern of each antenna element of an array extends over both the azimuthal plane and the elevation plane, and should be separately modelled for directional performance gains. Moreover, as we consider indoor scenarios with higher number of reflections, the polarization properties of the multipath components also come into play.

\subsubsection{Large bandwidth and large antenna arrays}
\begin{figure}[t]
     \centering
         \includegraphics[width=1\columnwidth]{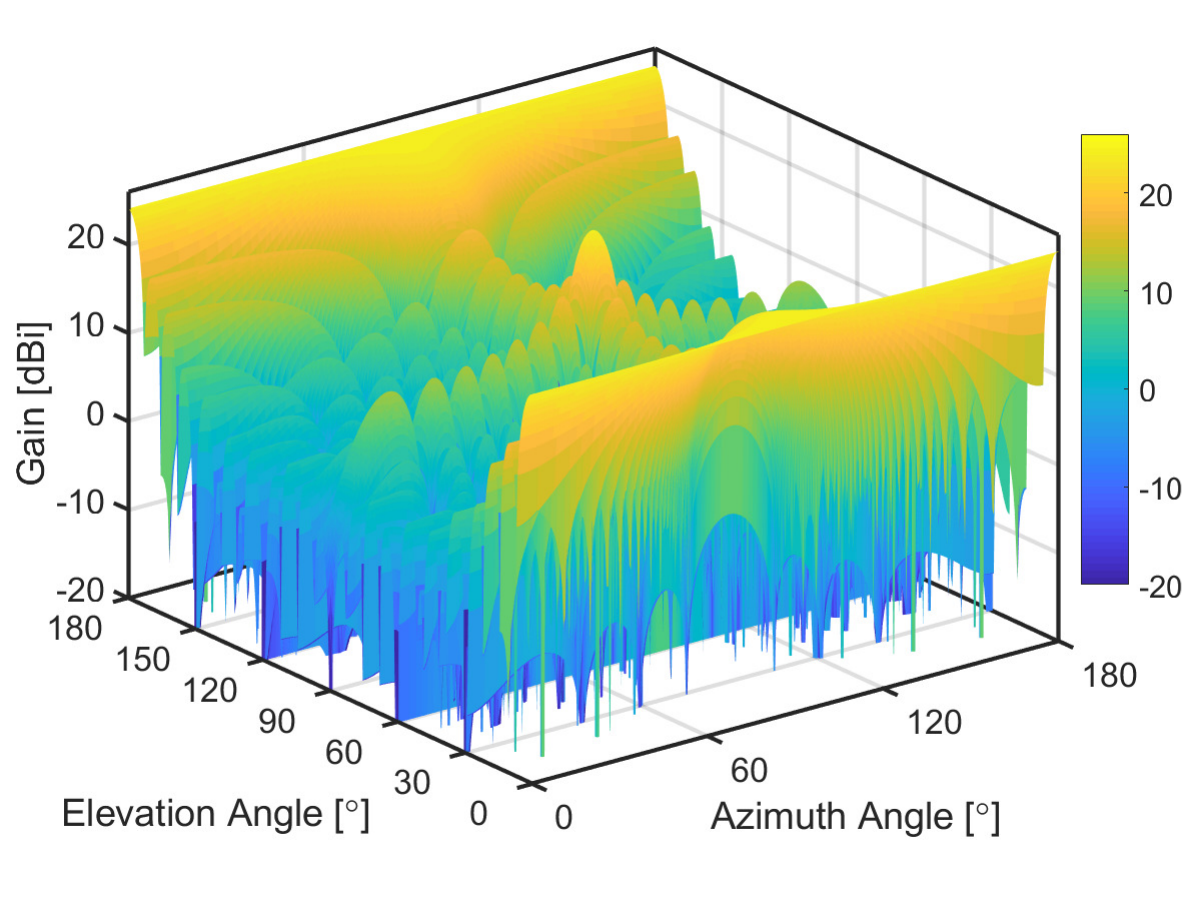}
         \vspace{-5mm}
         \caption{BS antenna array pattern as a function of azimuth and elevation scan angles~\cite{tataria_impact_2020}.}
    \label{sc1}
     \end{figure}

\begin{figure}[t]
         \centering
         \includegraphics[width=1\columnwidth]{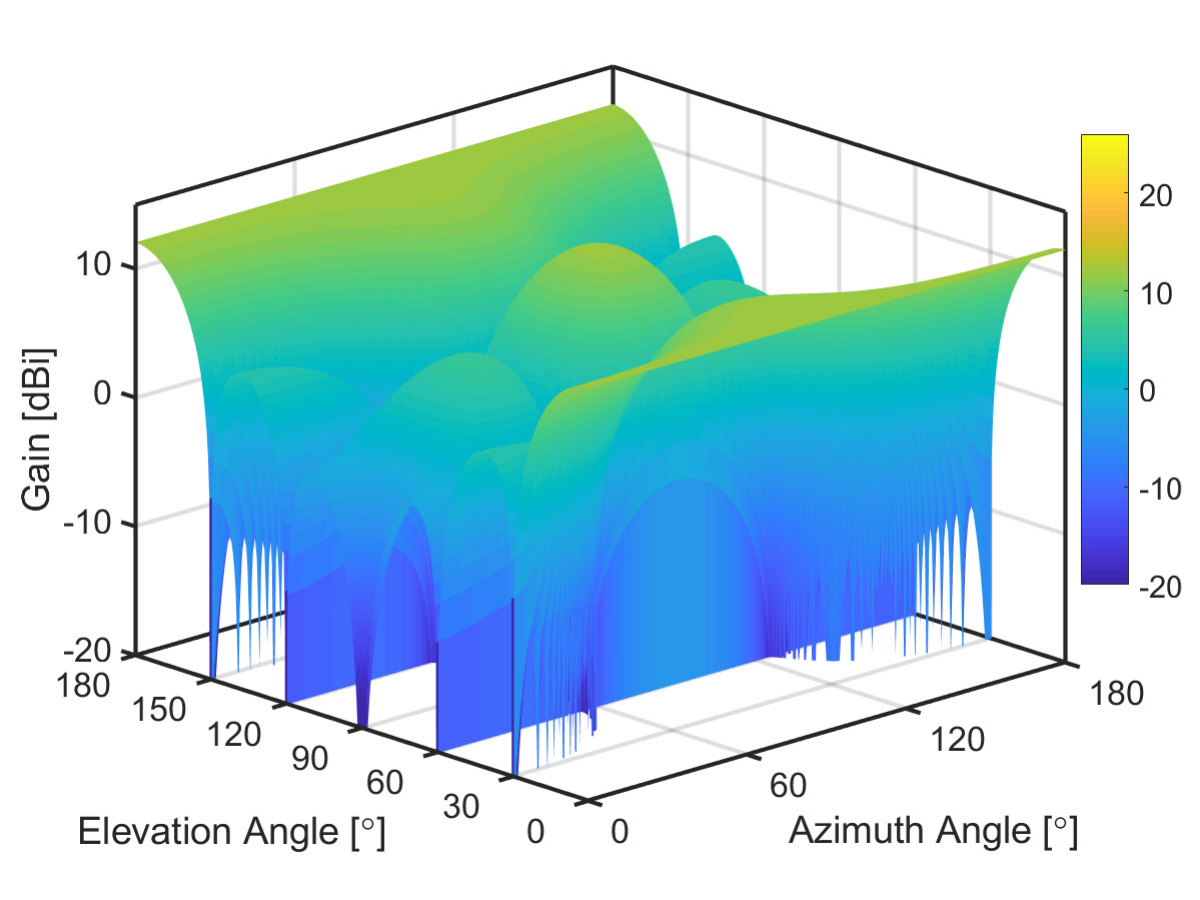}
         \vspace{-5mm}
         \caption{UE antenna array pattern as a function of azimuth and elevation scan angles~\cite{tataria_impact_2020}.}
    \label{sc2}
\end{figure}

Antenna arrays that are larger in size and also massive in the number of antenna elements are needed at \ac{mmw}, thus high resolution channel modeling includes propagation patterns both in the angular domain and in the delay domain. Massive MIMO channel models\cite{steinbauer_double-directional_2001} have previously not considered these exceptions but at \ac{mmw}, accurately modeling of the higher number of multipath components and their \ac{aoa}/\ac{aod} is paramount. Antenna elements in azimuth and elevation plane both need to be evaluated to consider all possible array structures (planar array, rectangular array, cylindrical array). Fig.~\ref{arraymodel} depicts an antenna array panel used for \ac{3gpp}/\ac{itur} antenna modeling~\cite{3gpprelease14, ITUguidelines}. Figs.~\ref{sc1} and~\ref{sc2} show the \ac{bs} and \ac{ue} array radiation pattern based on parameters as defined in~\cite[Table~7.3-1, page~22]{3gpprelease14}.

\subsection{Summary} \label{sec:summary}

The \ac{mmw} channel when considered for indoor applications differs from the microwave channel in key aspects such as free space path loss, diffraction, and penetration loss with respect to different surfaces. This required the need for different measurements to be done for channel characterization. Some key results are presented in Section~\ref{sec:chan.meastech}. Path loss equations and penetration loss for indoor scenarios can be found in~\cite[Tables~7.4.1-1 and~7.4.3-1]{3gpprelease14}.
Various channel models have been developed, these include those by \ac{3gpp}~\cite{3gpprelease14}, \ac{itur}~\cite{ITUguidelines}, METIS~\cite{raschkowski_metis}, MiWEBA~\cite{weiler_quasi-deterministic_2016}, Fraunhofer HHI's QuaDRiGa~\cite{jaeckel_quadriga}, COST2100~\cite{liu_cost_2012}, NYUSIM~\cite{haneda_5g_2016} which still has ongoing measurements for indoor scenarios. The channel models are all \ac{gscm}-based with added cluster based modeling. Small-scale parameter values are further available when considering indoor scenarios found in the documentations mentioned for corresponding models. 

Several measurements have been done in the \ac{mmw} band for outdoor (urban macro and urban micro) scenarios but the indoor measurements are limited to the sub-6~Ghz band for the channel models developed with the exception of~\cite{ju2021millimeter}, where the authors propose an extension for an indoor channel model based on extensive measurements carried out at~28 and 140~GHz. We observe that indoor channel models are an extension of outdoor ones, and can be adapted easily based on the delay and angular spreads of any environment, as well as by adapting path loss modeling.

%% file: channel_models.tex
%%%%%%%%%%%%%%%%%%%%%%%%%%%%%%%%%%%%%%%%%%%%%%%%%%%%%%%%%
%%%% Table of mmWave channel models with parameter specifications
%%%%%%%%%%%%%%%%%%%%%%%%%%%%%%%%%%%%%%%%%%%%%%%%%%%%%%%%%

% \usepackage{multirow}
\begin{table*}[t!]

\centering
\caption{Summary of channel models and their spatial parameter values}
\label{models}
\begin{tabular}{|c|c|c|c|c|c|c|}
\hline
\multicolumn{2}{|c|}{\multirow{3}{*}{Parameter}} & \multicolumn{5}{c|}{mmWave Channel Models}\\ \cline{3-7} 
\multicolumn{2}{|c|}{}                           & \multirow{2}{*}{\begin{tabular}[c]{@{}c@{}}3GPP~\cite{3gpprelease14} /\\  ITU-R~\cite{ITUguidelines}\end{tabular}} & \multirow{2}{*}{\begin{tabular}[c]{@{}c@{}}COST \\ IRACON~\cite{cost_indoor}\end{tabular}} & \multirow{2}{*}{\begin{tabular}[c]{@{}c@{}}METIS\\  \cite{raschkowski_metis}\end{tabular}} & \multirow{2}{*}{\begin{tabular}[c]{@{}c@{}}QuaDRiGa\\ \cite{jaeckel_quadriga}\end{tabular}} & \multirow{2}{*}{\begin{tabular}[c]{@{}c@{}}NYUSIM\\  \cite{nyu_indoor}\end{tabular}} \\
\multicolumn{2}{|c|}{}& &  &   &  &    \\ \hline
\multicolumn{2}{|c|}{$f$ (GHz)}  & 6                                                                                                                                                  & 2.6                                                                                                          & 0.45-63                                                                                                      & 5.4                                                                 & 28                                                                                                 \\ \hline
\multicolumn{2}{|c|}{Type}                       & 2D GSCM                                                                                                                                            & GSCM                                                                                                         & 3D Map-based~\&~GSCM                                                                                           & 3D GSCM                                                             & TCSL                                                                                                        \\ \hline
K- factor                       & $\mu_{K}$      & 7                                                                                                                                                  & N/A                                                                                                          & 7.9                                                                                                          & -1.6                                                                & N/A                                                                                                         \\ \hline
                                & $\sigma_{K}$   & 4                                                                                                                                                  & N/A                                                                                                          & 6                                                                                                            & 2.7                                                                 & N/A                                                                                                         \\ \hline
Delay Spread                    & $\mu_{DS}$     & -7.7                                                                                                                                               & 1.07                                                                                                         & -7.42                                                                                                        & -7.22                                                               & 2.7                                                                                                         \\ \hline
                                & $\sigma_{DS}$  & 0.18                                                                                                                                               & 0.93                                                                                                         & 0.32                                                                                                         & 0.08                                                                & 1.4                                                                                                         \\ \hline
\multirow{2}{*}{AOA Spread}     & $\mu_{ASA}$    & 1.62                                                                                                                                               & 3.94                                                                                                         & 1.65                                                                                                         & 1.67                                                                & 19.3                                                                                                        \\ \cline{2-7} 
                                & $\sigma_{ASA}$ & 0.22                                                                                                                                               & 3.91                                                                                                         & 0.47                                                                                                         & 0.15                                                                & 14.5                                                                                                        \\ \hline
\multirow{2}{*}{AOD Spread}     & $\mu_{ASD}$    & 1.60                                                                                                                                               & 0.71                                                                                                         & 1.64                                                                                                         & 1.54                                                                & 23.5                                                                                                        \\ \cline{2-7} 
                                & $\sigma_{ASD}$ & 0.18                                                                                                                                               & 0.59                                                                                                         & 0.43                                                                                                         & 0.1                                                                 & 16.0                                                                                                        \\ \hline
\multirow{2}{*}{ZOA Spread}     & $\mu_{ZSA}$    & 1.22                                                                                                                                               & 3.73                                                                                                         & 1.28                                                                                                         & 1.61                                                                & 7.4                                                                                                         \\ \cline{2-7} 
                                & $\sigma_{ZSA}$ & 0.297                                                                                                                                              & 2.11                                                                                                         & 0.26                                                                                                         & 0.07                                                                & 3.8                                                                                                         \\ \hline
\multirow{2}{*}{ZOD Spread}     & $\mu_{ZSD}$    & N/A                                                                                                                                                & 1.95                                                                                                         & 1.31                                                                                                         & 1.17                                                                & -7.3                                                                                                        \\ \cline{2-7} 
                                & $\sigma_{ZSD}$ & N/A                                                                                                                                                & 1.80                                                                                                         & 0.31                                                                                                         & 0.07                                                                & 3.8                                                                                                         \\ \hline
\multirow{2}{*}{XPR (dB)}       & $\mu_{XPR}$    & 11                                                                                                                                                 & 15.59                                                                                                        & 29                                                                                                           & 13                                                                  & N/A                                                                                                         \\ \cline{2-7} 
                                & $\sigma_{XPR}$ & 4                                                                                                                                                  & 10.39                                                                                                        & 6.5                                                                                                          & 1.6                                                                 & N/A                                                                                                         \\ \hline
\multirow{2}{*}{Shadow fading}  & $\mu_{PL}$     & 47.9                                                                                                                                               & N/A                                                                                                          & N/A                                                                                                          & 36.1                                                                & N/A                                                                                                         \\ \cline{2-7} 
                                & $\sigma_{PL}$  & 3                                                                                                                                                  & N/A                                                                                                          & 3                                                                                                            & 1.6                                                                 & N/A                                                                                                         \\ \hline
\end{tabular}
\end{table*}

%% file: 03_Beamforming_architectures.tex
\section{Implications of beamforming architectures for mmWave localization} \label{sec:bfarch}

\begin{figure*}
    \centering
    \includegraphics[width=0.7\linewidth]{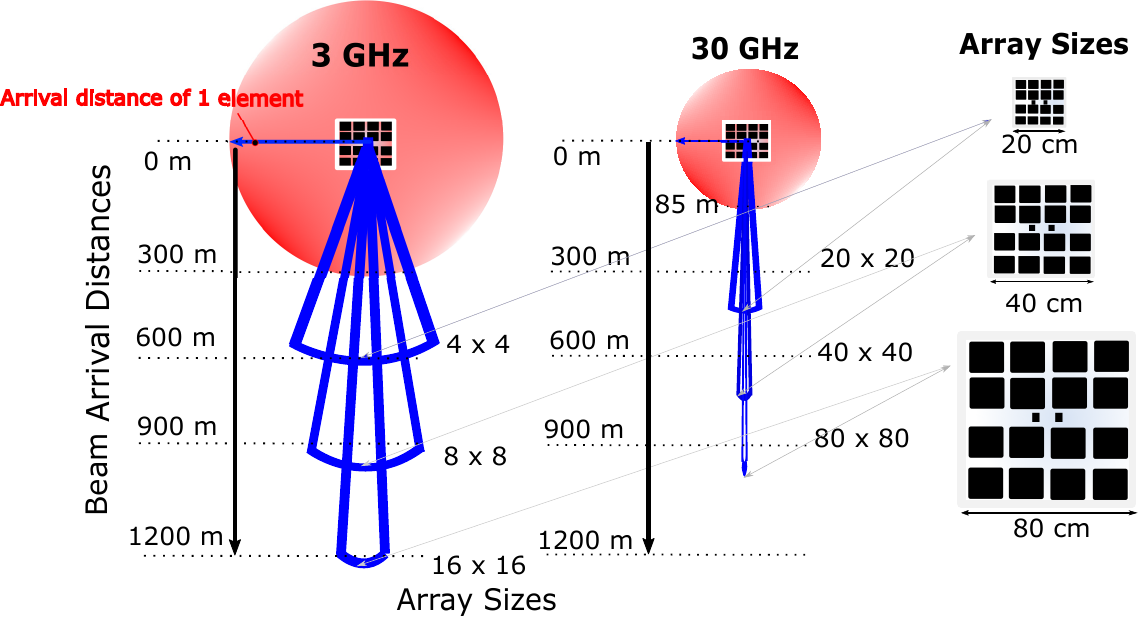}
    \caption{Effect of beamwidth relative to operating frequency and array sizes~\cite{tataria_standardization_2021}.}
    \label{beamwidth}
    \vspace{-10pt}
\end{figure*}

It is a common misconception that for higher frequencies the free space propagation loss is higher. As explained in~\cite{khan_mmwave_2011,molisch2012wireless}, for given aperture area of the antennas used, shorter wavelengths propagate farther due to the narrow directive beams. This is further verified in~\cite{roh_millimeter-wave_2014} with a patch antenna operated at 3~GHz and an antenna array operated at 30~GHz of the same physical size. We observe equal amounts of propagation loss irrespective of the operating frequency. Thus, \ac{mmw} frequencies enable the use of antenna arrays that produce highly directional beams which lead to large array gains. This can be observed from Fig.~\ref{beamwidth}, which shows not only the increase in array size with respect to the beam penetration distance, but also how the larger array size increases the coverage area~\cite{tataria_standardization_2021}.

\subsection{Analog beamforming} \label{sec:bfarch.analog}

Analog beamforming, sometimes also referred to as beam steering, is done by connecting a single \ac{rf} chain to a string of phase shifters that are both energy- and cost- efficient. Each phase shifter multiplies its input by ${\rm e}^{j\frac{2\pi k}{2^N}}$, where $j=\sqrt{-1}$, $N$ is the number of bits, and $k=0,\ldots,2^N -1$ is used to control the phase shifters. Most commonly, codebook-based schemes are used to steer the beams in the direction of the \ac{ue}\slash{}receiver. At the receiver, the \ac{rssi} is the most commonly used parameter to estimate the direction of arrival and delay, and thus localize the device.
However, phase shifters have a constant amplitude constraint and limited phase resolution. It is also worth noting that analog beamforming converges to a single beam for multiple data transmissions, and in multi-user case the inter-user interference is very high. This is a drawback for localization applications, as the phase resolution for analog beamforming is low.

The popularity of analog beamforming systems comes from the availibility of \ac{cots} devices, that are being used for research on \ac{mmw} positioning. The devices come with a pre-programmed codebook to generate beam patterns and with support for retrieving the \ac{rssi} and \ac{csi} which can be used to isolate the position of the \ac{ue}.
One such hardware front-end is available from TMYTEK, an analog correlator with beamformer chips and smart-antenna arrays~\cite{wane2019cognitive}. Another company that provides beamformer integrated circuits and scalable antennas for \ac{mmw} is Anokiwave~\cite{menon2018active}. Siver Semiconductors provides transceiver modules for \ac{mmw} frequencies, i.e., 28~GHz and 60~GHz~\cite{ojefors201857}. National Instruments (NI) also has the PXIe-5831, a \ac{mmw} vector signal transceiver that has beamforming capabilities and phased antenna arrays~\cite{bengtsson2019massive}. It has been used for channel measurements as mentioned above as well~\cite{tataria_channel_2019}. We discuss the hardware devices used in more detail in Section~\ref{sec:localgo.evtools.hw}.

\subsection{Hybrid beamforming} \label{sec:bfarch.hybrid}
Hybrid beamforming is by far the most researched form of beamforming, as it provides a middle ground between complexity and cost. Here, the analog beamformer is used in the \ac{rf} domain, along with a digital precoder at baseband. This can be either a fully connected structure or a partially connected one.
Hybrid analog/digital beamforming structures provide balance between the beam resolution and cost and power consumption. By using multiple \ac{rf} chains concurrently, beam sweeping can be done in a short time leading to shorter beam training time which leads to higher effective data rate.  At \ac{mmw} frequencies the sparse channel behaviour is useful for beam training and higher array gains. Multiple hybrid beamforming techniques for \ac{mmw} have been proposed in the last ten years which broadly fall under codebook dependent, spatially sparse precoding, antenna selection and beam selection~\cite{molisch_hybrid_2017}.
\cite{ayach_low_2012} first gave the idea of what we call hybrid beamforming today. It was a combination of a digital baseband precoder and an \ac{rf} precoder which falls under spatially sparse precoding. The work in~\cite{taeyoung_kim_tens_2013} first proposed the idea of baseband beamforming, or ``hybrid beamforming'' as the authors named it, that chooses the best \ac{rf} beam based on a capacity maximization criterion, and then derives a \ac{zf}-based weighing matrix for digital precoding. Also, both~\cite{song_codebook_2015} and~\cite{payami_effective_2015} suggest codebook-based precoding solutions. Recent works have proposed compressive sensing, least squares- and \ac{dft}-based solutions for hybrid beamforming with use cases in car-to-car scenarios and high speed trains. In most cases, hybrid beamforming is seen to perform as well as fully-digital beamforming, and as being both cost-effective and spectrally efficient.

\subsection{Digital beamforming} \label{sec:bfarch.digital}

Digital beamforming adjusts the amplitude and phase of the transmitted signals using precoding. Linear precoding algorithms such as \ac{mf}, \ac{zf}, and \ac{rzf} methods were classically used for single-antenna user systems. For multiple-antenna users, block diagonalization is a feasible approach. Digital beamforming can be considered as the best option for \ac{mmw} positioning. With the possibility of huge antenna arrays ($256 \times 128$ upwards) a beam resolution of the order of centimeters can be achieved. The calibration accuracy of digital systems allows us to use high-resolution parameter estimation algorithms that can estimate not only the \ac{toa} and \ac{aoa} but also the Doppler frequency offset in case of mobility, making it possible to update the position of a \ac{ue} in real-time. The issue here arises from the use of a \ac{rf} chain per antenna, which leads to a complex, non-cost-effective hardware system for massive \ac{mimo} structures.

As digital beamforming offers higher beam resolution, it is a viable candidate where multi user \ac{mmw} or rather \ac{mmw} massive \ac{mimo} systems are considered. However, commercial hardware for a fully digital system is still in its infancy, and only laboratory results exist. Several authors have proposed alternative techniques for the realization of a digital system that is power efficient. For instance~\cite{dutta_case_2020} gives an option for digital beamforming that employs switches to bypass the hardware constraint of using multiple \ac{rf} chains. In~\cite{yang_digital_2018,hu_digital_2018,xingdong_design_2014}, the authors propose different ways to form an antenna array using waveguides and printed circuit boards that support digital beamforming. Alternatively,~\cite{yu_full-angle_2019, dutta_case_2020} propose  novel frameworks to do digital beamforming for a \ac{mmw} setup using linearization to help with power amplifier loss and improved quantization.

\subsection{Performance vs.\ complexity overview} \label{sec:bfarch.perfcomp}

In localization applications, the requirement for \ac{mmw} indoor systems is to isolate the position of the receiver inside a room, while taking into account blockage caused by humans and objects alike, with \ac{los} being the dominant component. The presence of pillars, metal and glass surfaces affects the channel impulse response and thus make it difficult to extract position information. Presence of antenna arrays greatly enhances the accuracy of the position coordinates. Whereas digital systems have cleaner isolated beams and can potentially yield centimeter-level pointing accuracy, analog setups have a limit to the number of beam patterns they can generate: when trying to increase the resolution, these beam patterns eventually start to overlap. As stated above, the number of \emph{beams} is proportional to the number of available \ac{rf} chains, thus increasing the complexity hundred-fold for digital systems. Calibration issues also prevent analog systems from performing high-resolution parameter estimation which could improve the localization accuracy.
Hybrid beamforming seems a promising tradeoff as of now, due to the easier availability of \ac{cots} devices, and to a performance almost as good as that of fully digital systems.

%% file: 04_Standardization_notes.tex
\section{Progress in standardization of cellular mmWave systems}
\label{sec:std}

\begin{figure}[t]
    \centering
    
    \includegraphics[width=1\columnwidth]{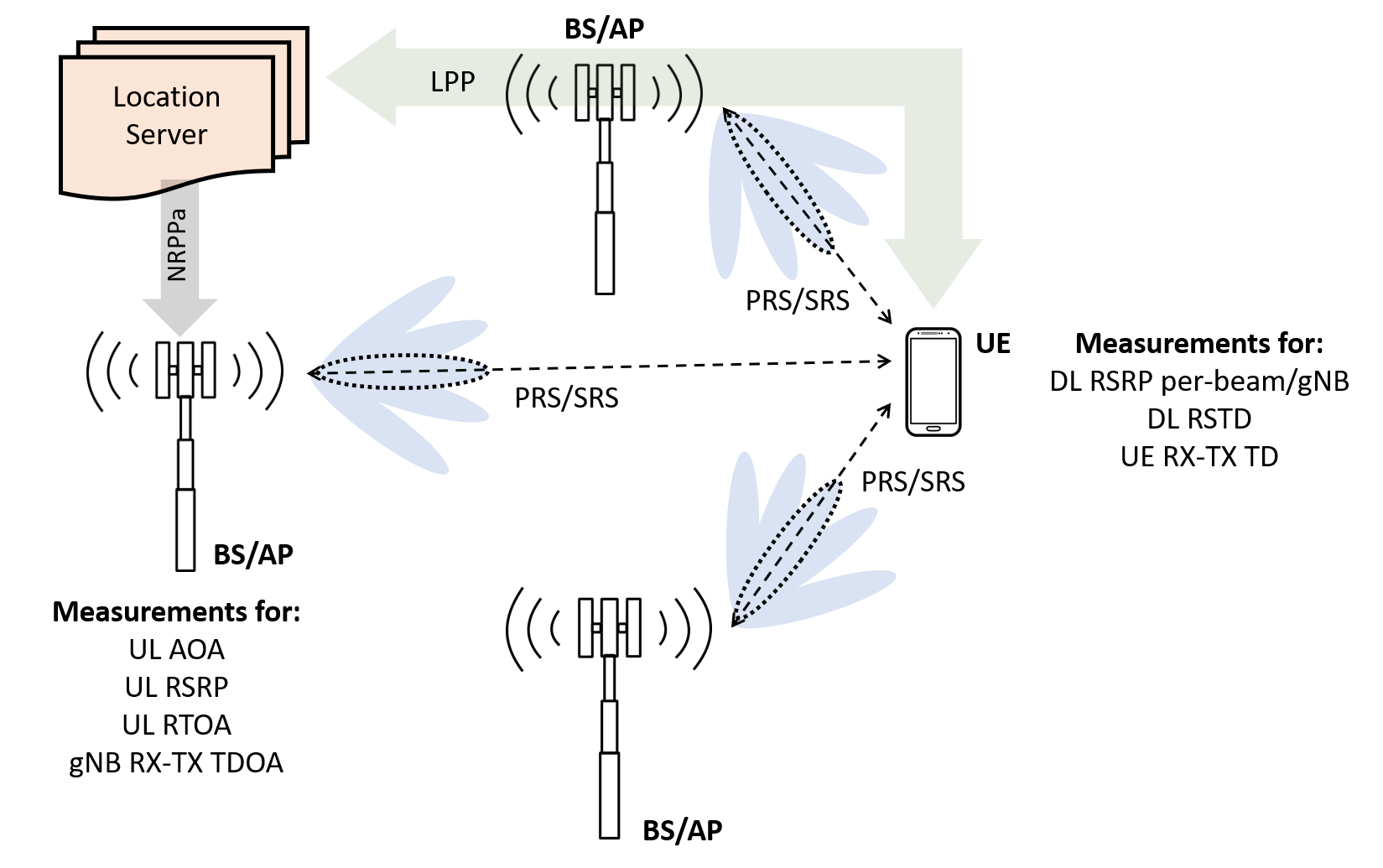}
    \caption{\ac{3gpp} Release 16 radio access type-dependent architecture standardized for \ac{ue} localization in \ac{urllc} scenarios. All \acp{bs}/\acp{ap} are interfaced with a centralized unit enroute to a \ac{urllc} core network.}
    \label{fig:LocatizationArchitecture3GPP}
\end{figure}

The frequency bands used for \ac{5g} systems were proposed at the 2015 World Radio Conference (WRC) by \ac{itur} and approved during WRC 2019. The frequency bands standardized by \ac{3gpp} in Release 15-17~\cite{3GPPrelease15, 3GPPrelease16, 3GPPrelease17} for \ac{5g} systems are classified as FR-I region (below 7.125~GHz) and FR-II region (between 7.25~GHz and 71~GHz). The approved FR-II bands are (in GHz): 24.25--27.5; 31.8--43.5; 45.5--50.2; 50.4--52.6; 66--71. FR-I bands act as the key bands for cellular communications, while the FR-II are more suited to short-range communications. The FR-II bands also provide increased bandwidths compared to FR-I, and are managed via licensed access mechanisms such as \ac{endc}. As some bands overlap with other services, coexistence management is needed for terrestrial access in overlapping satellite communication channels and for fronthaul and backhaul in fixed wireless systems. \ac{5g} commercial deployments have already been taking place since the end of last year, and some spectrum congestion was observed initially amongst multiple operators. Since then, some novel forms of spectrum access/coordination mechanisms have been implemented.\footnote{We note that these are operator- and vendor-specific, since frequency band combinations vary depending on the specific country.} When it comes to localization of \acp{ue}, it was the focus of \ac{3gpp} Release 16~\cite{3GPPrelease16} especially for the use case of \ac{urllc}. In the past, Global Navigation Satellite Systems assisted by cellular networks have been mostly used for \ac{ue} positioning, but their accuracy is high only in outdoor environments, as they rely on satellites to localize \acp{ue}. As we move towards higher frequencies, we require localization indoors as well, and we can accomplish it in \ac{5g} networks using the location server, as it was for long-term evolution-advanced (LTE-A) systems. The location server collects and provides position estimates and assistance data and measurements to the other devices. Various localization methods are used, based on downlink or uplink communications, either separately or in combination, to meet the accuracy requirements for different scenarios. The overall architecture is as depicted in Fig.~\ref{fig:LocatizationArchitecture3GPP}.

\begin{figure*}[b]
    \centering
    \subfloat[Downlink based, \ac{ue}-centric \label{fig:std.ue}]{\includegraphics[width=0.48\linewidth]{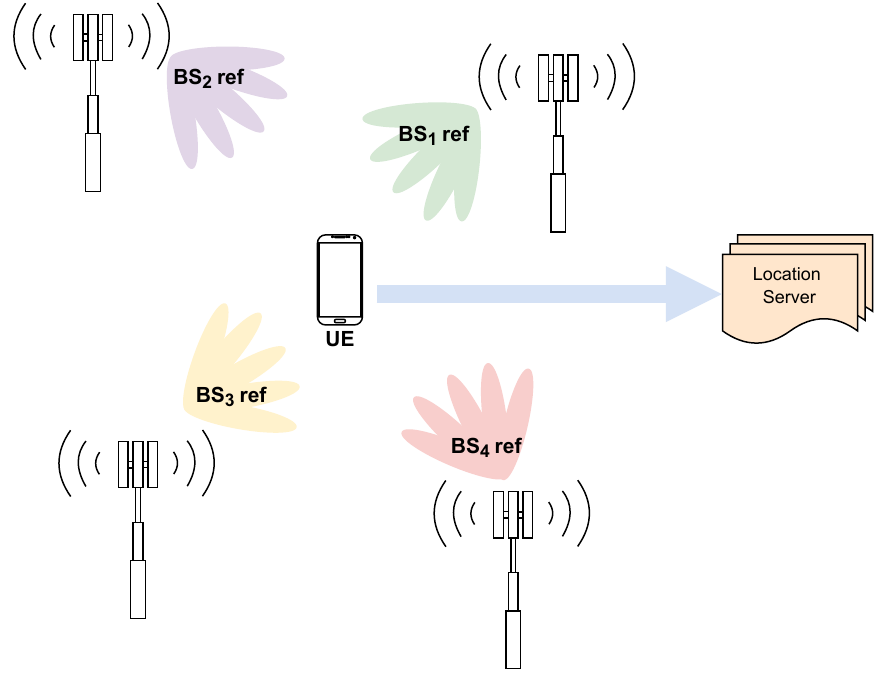}}
    %\hspace{3mm}
    \hfill
    \subfloat[Uplink based, \ac{bs}/\ac{ap}-centric\label{fig:std.bs}]{\includegraphics[width=0.48\linewidth]{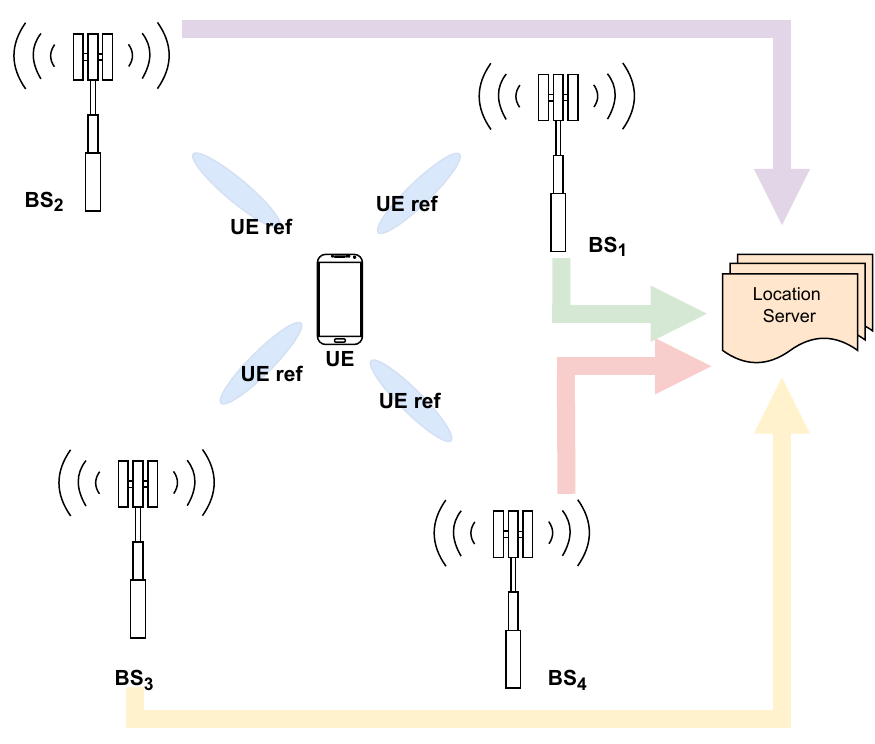}}
    %\vspace{2mm}
    \caption{Architecture of \ac{bs}/\ac{ap}-centric vs \ac{ue}-centric localization.}
    \label{fig:uevsbscentric}
\end{figure*}

As shown in Fig. \ref{fig:std.ue}, downlink-based localization is performed when each of multiple \acp{bs}/\acp{ap} send a different reference signal, known as the \ac{prs}. The \ac{ue} receives the different \acp{prs} and reports the \ac{toa} difference for \acp{prs} received from multiple distinct \acp{bs}/\acp{ap} to the location server. The location server can use the reports to determine the position of the \ac{ue}. Compared to LTE-Advanced, the \ac{prs} has a more regular structure and a much larger bandwidth, which enables a more precise correlation and \ac{toa} estimation.

The canonical \ac{3gpp} Release 15 \acp{srs} with Release 16 extensions added uplink-based localization or \ac{bs}/\ac{ap} centric localization as shown in Fig.~\ref{fig:std.bs}. In this case, the \ac{ue} sends the reference signal. Based on the received \acp{srs}, the \acp{bs}/\acp{ap} can measure and report (to the location server) the arrival time, the received power and the \acp{aoa} from which the position of the \ac{ue} is estimated. The time difference between downlink reception and uplink transmission can also be reported, and used in \ac{rtt}-based positioning schemes, where the distance between a \ac{bs}/\ac{ap} and a \ac{ue} can be determined based on the estimated \ac{rtt}. By combining several such \ac{rtt} measurements, involving different \acp{bs}/\ac{ap} anchors, it becomes possible to estimate the location of the \ac{ue}.

We note that these methods do not utilize the full-dimensional nature of the propagation channel (azimuth and elevation domains), and do not fully take into account the phase information needed to estimate the underlying \acp{mpc} with high resolution. While this is an ongoing topic for research in many study items of \ac{3gpp} Releases~17 and~18, we refer the reader to~\cite{3GPPrelease17,3gpprel18} for further details. Along this same line, a steady stream of work is also conducted in academia, see e.g.,~\cite{li_massive_2019}.

%% file: 05_Localization_algorithm_survey.tex
\section{Device-based mmWave localization algorithms for indoor communication systems}  \label{sec:localgo}

\begin{figure}[t]
    \centering
    \includegraphics[width=\columnwidth]{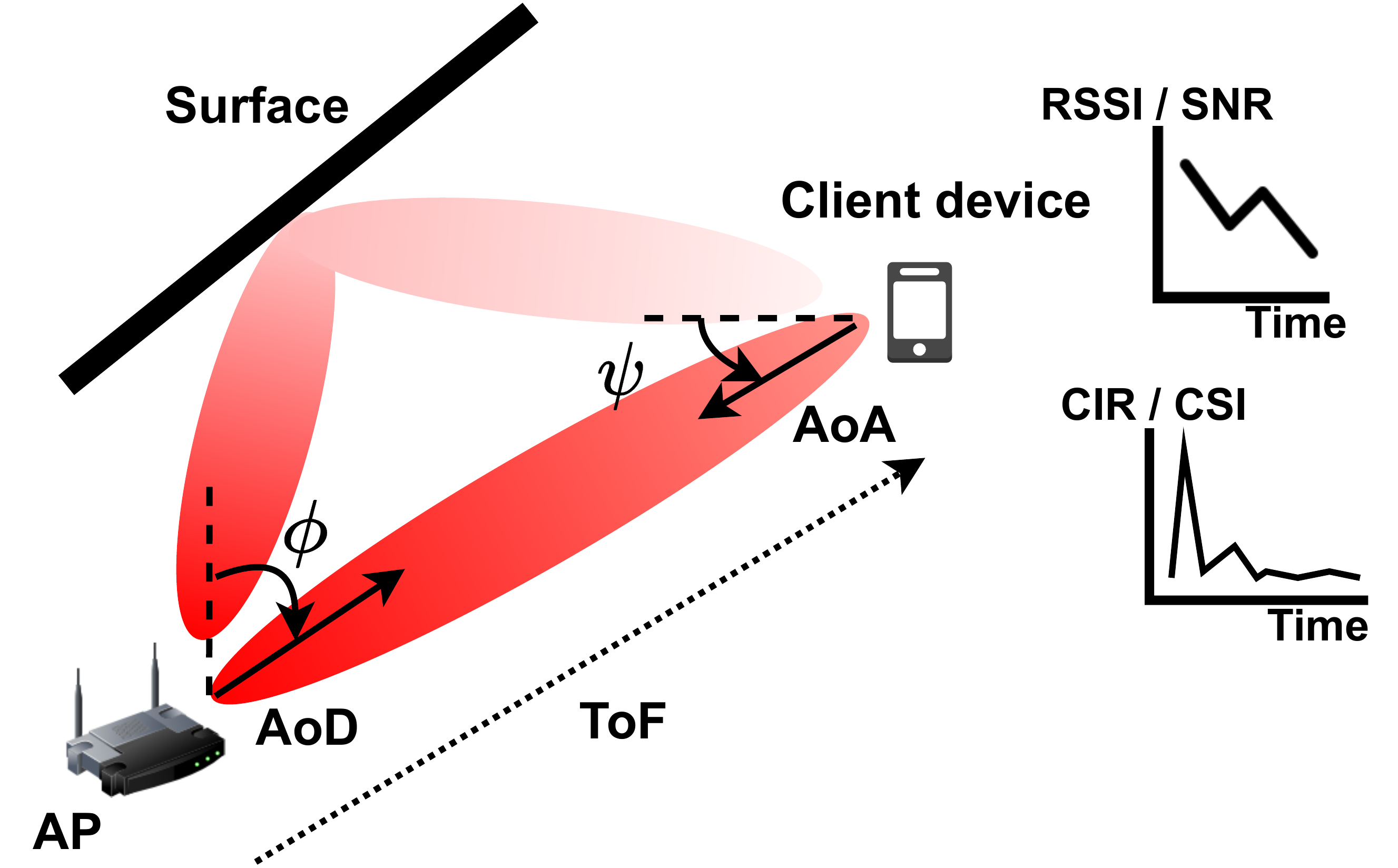}
    \caption{Illustration of the signal measurements obtained from mmWave propagation. The color gradient of the beam represents the decreasing signal strength due to path loss.}
    \label{fig:loc.aoaTofRSSICIR}
\end{figure}

\begin{figure*}[t]
    \centering
    \includegraphics[width=1\linewidth]{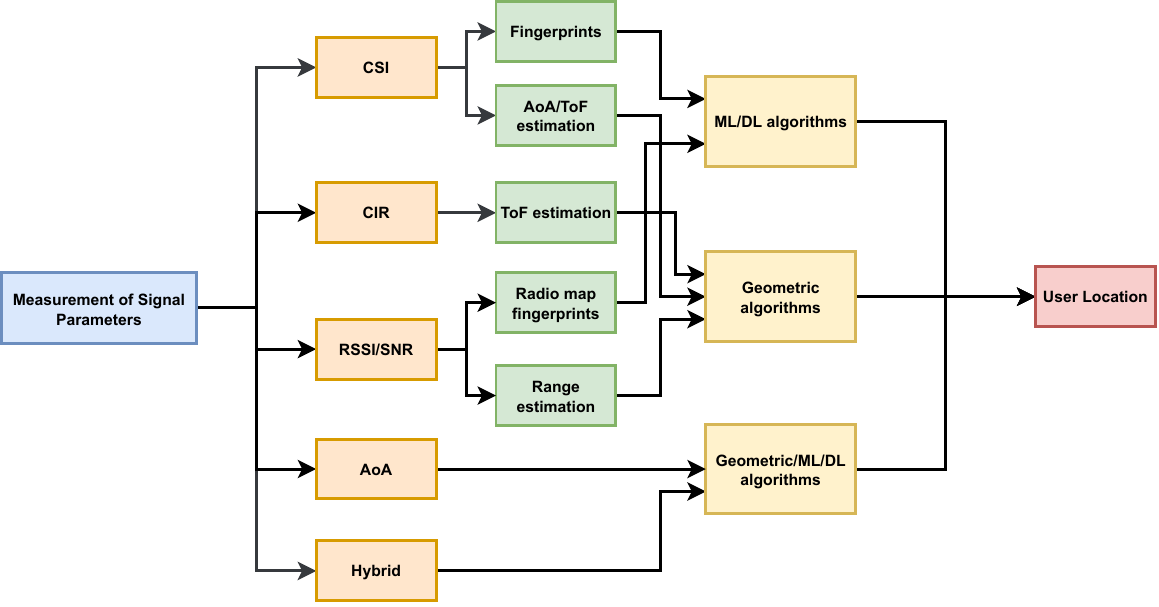}
   
    \caption{General flow chart of the steps of a mmWave localization algorithm from the surveyed literature.}
    
    \label{fig:loc.flowdiag}
\end{figure*}

\subsection{Introduction}
\label{sec:localgo.intro}

In this section, we introduce algorithms and methods that leverage lab-grade and commercial-grade \ac{mmw} hardware to localize devices indoors. 
We start with a brief recap on classical methods for indoor radio localization. The standard techniques designed for localization involve exploiting the parameters of radio signals from existing wireless infrastructure. These have been well explored and surveyed in, e.g.,~\cite{carlo2018CommSurvTut, LiuIndoor2007Survey,indoorLocSurvey2019zafari,xiao2016survey, collaborativeLoc2018Beuhrer, loc5g2021win, 5gPos2015Shahmansoori}. 
With reference to Fig.~\ref{fig:loc.aoaTofRSSICIR}, localization algorithms typically make use of signal parameters related to received signal power (\ac{rssi} and \glsunset{snr}\acrlong{snr}, \acrshort{snr}), time-information such as \ac{tof} and \ac{tdoa}, and angle information (\ac{aoa} and \ac{aod}) in order to obtain distance and direction estimates, which enable a device or group of devices to estimate either their own location, or the location of another device in their proximity, or both.
Fig.~\ref{fig:loc.flowdiag} offers a general view of this process considering the papers on \ac{mmw} localization surveyed in the literature. After a device has extracted location-dependent features from a received signal, such features are either used directly for localization, or further processed to extract additional information, or joined into a global map of the environment along with other measurements. The device then applies geometric or \ac{ml}/\ac{dl} algorithms to derive location information.

\begin{figure}[t]
    \centering
    \subfloat[Trilateration\label{fig:loc.trilaterate}]{\includegraphics[width=0.8\columnwidth, trim = 30mm 28mm 35mm 50mm, clip]{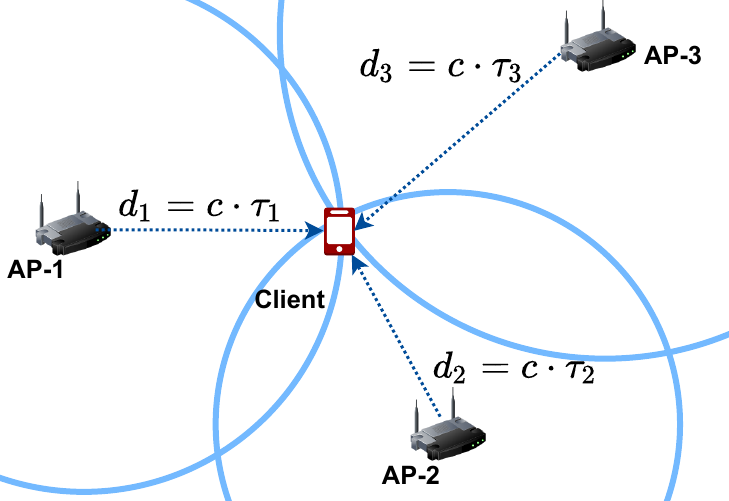}}
    \hspace{5mm}
    \subfloat[Triangulation\label{fig:loc.triangulate}]{\includegraphics[width=0.8\columnwidth]{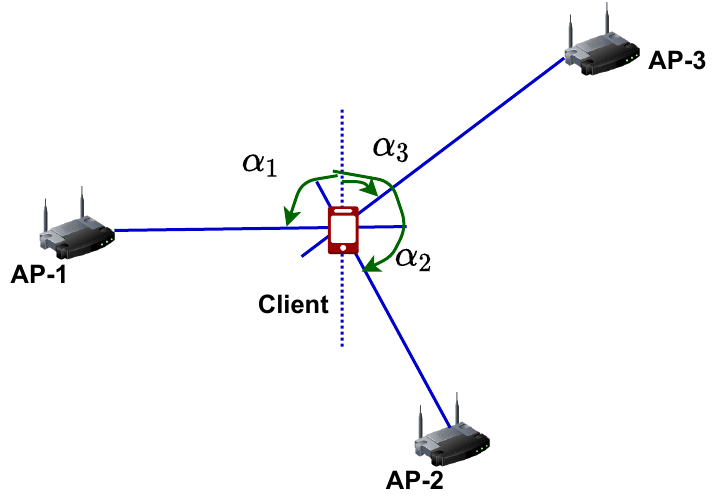}}
    \\
    \subfloat[Angle-difference of arrival \label{fig:loc.adoa}]{\includegraphics[width=0.8\columnwidth]{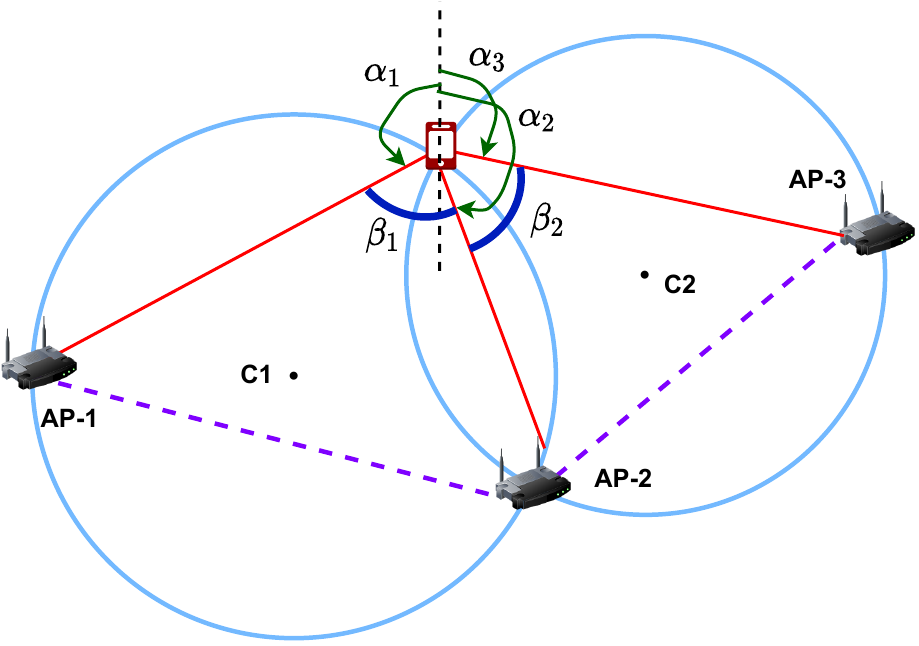}}
    \vspace{2mm}
    \caption{Illustration of the (a) trilateration, (b) triangulation, and (c) angle-difference of arrival processes using ToF, AoA, and ADoA localization geometries, respectively. Note that $d_i$ and $\tau_i$ respectively denote the distance and propagation delay between AP $i$ and the client, $c$ is the speed of light in air, $\alpha_i$ denotes the AoA of the signal from AP $i$, and $\beta_i$ is the ADoA, i.e., the difference of the AoAs from APs $i$ and $i+1$.}
    \label{fig:trilat_triang_adoa_concept}
\end{figure}

The most typical localization techniques rely on geometric algorithms. For example, \emph{trilateration} and \emph{triangulation} utilize distance and angle measurements from fixed reference points to compute an intersection, which yields the estimate of a device's location~\cite{indoorLocSurvey2019zafari}. The reference points are usually the location of the access points, and the localized device is typically a client. The distances between the \acp{ap} and the client are measured by exploiting either the \ac{tof} of the signal or by mapping the \ac{rssi} information to absolute distance using path-loss models. Fig.~\ref{fig:loc.trilaterate} shows an illustration of trilateration using \ac{tof} to estimate distances. 

\ac{aoa} (the angle at which the received signal strikes the receiver antenna or antenna array) and \ac{adoa} (the difference between two \acp{aoa}), are estimated by applying signal parameter estimation algorithms (like \ac{music}~\cite{music} and \ac{esprit}~\cite{ESPRIT}) on the received signal. The \acp{aoa} from different \acp{ap} are then triangulated to localize the client device. Fig.~\ref{fig:loc.triangulate} illustrates the triangulation-based technique, whereas Fig.~\ref{fig:loc.adoa} depicts \ac{adoa}-based localization. 

Wireless channel characteristics, e.g., in the form of the \ac{cir} between a transmitter and a receiver, also provide valuable information for localization purposes, including the \ac{tof} of the received signal. The \ac{csi} can also be extracted from the receiver antennas to obtain rich information about multipath signal components~\cite{Zheng2013rssiToCsi}. As a result, one can separate the \ac{los} propagation path from \ac{nlos} paths, or detect that only \ac{nlos} components reached the receiver, thus improving the accuracy of the signal parameter measurements.

The advent of bandwidth-hungry applications such as augmented reality, virtual reality, etc., and the ever-increasing demand for high data rates, has made \ac{mmw} communication technology a popular potential replacement for existing WLAN systems. This is mainly due to the availability of large bandwidth in the frequency range of 30-300~GHz, resulting in multi-Gbit/s data rates. \acp{mmw} propagate quasi-optically, thus reflecting crisply off indoor surfaces and obstacles with limited scattering just like light rays~\cite{rappaport2013mmw}. This makes finer measurements of signal parameters such as \ac{rssi}, \ac{aoa}, \ac{snr}, and \ac{tof}, more feasible and more accurate. Moreover, we remark that wireless devices typically collect location-dependent signal features through the interaction between a client and one or more \acp{ap}. Such interactions naturally take place in \ac{mmw} networks, e.g., during standard-compliant link establishment and beam refinement procedures (see also Section~\ref{sec:localgo.evtools}). Therefore, in principle the measurement of signal features does not require the devices to implement localization-specific message exchange protocols. This makes localization an almost-inherent feature of \ac{mmw} communication systems~\cite{yang2021integratedLocComm, fischione2014evaluation}.

As remarked in Section~\ref{sec:bfarch}, however, \ac{mmw} devices have peculiar characteristics that differentiate them from commonplace WiFi equipment. Specifically, \ac{mmw} arrays can incorporate a large number of antennas.
The presence of large arrays enable \ac{mmw} devices to output low-level physical layer measurements from each antenna separately. Once the device has locked onto a signal, each antenna receives the same signal with a different phase, corresponding to the delay incurred by the signal due to its spatial position in the array. These measurements can be made available as \ac{csi} and localization algorithms can exploit them to localize a device, either by converting them into \ac{aoa} estimates (e.g.,~\cite{joanCodebookOpt_MobiCom2018,mdTrack2019}) or by directly inferring the location of a device by exploiting the \ac{csi} as a location-dependent feature.

Whenever \ac{csi} measurements are not available, a device can still retrieve angle information by post-processing the output of standard-compliant beam training procedures. 
Typically, each \ac{mmw} has a number of pre-programmed beam patterns that provide it with the necessary flexibility to focus energy towards different directions.
Each beam pattern ideally covers a well-defined portion of the 3D space, so that observing each beam pattern separately makes it possible to implement a scan of all azimuthal and elevation angles that the \ac{mmw} array can cover. Therefore, measuring the power received through each beam pattern configuration would implement a sweep of lookout angles. By identifying the beam pattern that leads to the largest received power, a \ac{mmw} device could easily estimate angles of arrival.
We now proceed to discuss each type of location-dependent feature separately in the context of \ac{mmw} communications, highlighting the pros and cons of each feature.

\subsection{Pros and cons of location-dependent measurements for mmWave localization}
\label{sec:localgo.locdepmeas}

\noindent \textbf{Angles of arrival and departure, angle difference-of-arrival}~\cite{palacios2019single, olivier2016lightweight,henkAoD2020Icassp, Harvard2017Globecom, Yassin2017SimultaneousCI, palacios2017jade, clam2018Palacios} --- 
The term angle of arrival (\glsunset{aoa}\ac{aoa}) refers to the angle at which radio signals illuminate the antenna array at the receiver. The transmitter-based counterpart, the \ac{aod}, refers to the angle at which the radio signals emanate from the antenna array at the transmitter front-end in order to reach the receiver.  
In most cases, more than one antenna elements are required to compute angle information. Other methods to extract \ac{aoa} information from the receiver array involve the use of \ac{csi}, beamforming methods, or subspace approaches such as the well-known \ac{music}~\cite{music} and \ac{esprit}~\cite{ESPRIT} algorithms.
We cover angle-based approaches in Section~\ref{sec:localgo.aoa}.\\
\textbf{Pros}: Relatively accessible information in \ac{mmw} systems, thanks to the large number of antennas in transmitter and receiver arrays.\\
\textbf{Cons}: If not associated to some range information, can only yield location estimates in a relative coordinate system. Multipath propagation can distort angle estimates, if not properly modeled or compensated for.\\

\noindent \textbf{\Acrfull{csi}}~\cite{leap2019palacios, tensorGSCPD,ObjectDetAccurate2019Ajorloo} ---
\ac{csi} refers to the measurable properties of a received \ac{mmw} signal that relate to the propagation channel linking two devices, e.g., the \ac{ap} and the client. Different \ac{mmw} hardware may provide different forms of \ac{csi}. For example, patching TP-Link's Talon routers~\cite{talonRouter} with special firmware makes it possible to extract receiver-side \ac{csi} in the form of one complex gain coefficient per receiving antenna, expressing the attenuation and phase shift that affect the strongest propagation path at each antenna. Post-processing \ac{csi} yields different signal parameters, including path attenuation and angle information. If \ac{csi} values are sufficiently precise (e.g., no coarse quantization affects the amplitude or phase), collecting receiver-side \ac{csi} from multiple antennas also enables the estimation of \acp{aoa}.
We cover \ac{csi}-based approaches in Section~\ref{sec:localgo.csi}.\\
\textbf{Pros}: Rich information that can be readily used for ranging or as an input to learning-based approaches.\\
\textbf{Cons}: Typically not available straightforwardly on all devices. Different devices may provide different types of \ac{csi}.\\

\noindent \textbf{\Ac{rssi}}~\cite{rssi2014icc, fing2019mitsubishi1,wbfps2020Infocom,5gRssi2020beam,Vashist2020ml,fing2019mitsubishi2, deepL2020mitsubishi,wangMERLfingerprintingPart4} ---
\ac{rssi} is one of the simplest proxies for the range of a device in an environment. It is measured at a receiving device as the power or amplitude of the received RF signal. \ac{mmw} \ac{rss} measurements can be extracted from the physical or \ac{mac} layer of a device and used to measure the distance of a client from the \ac{ap}, based on the knowledge of a path loss model. The client is believed to lie on the circumference of the circle centered on the \ac{ap} and having the estimated range as the radius. Such estimates from more than two \acp{ap} can be trilaterated to approximate the location of the client.
We cover \ac{rssi}-based approaches in Section~\ref{sec:localgo.rssi-tof}.\\
\textbf{Pros}: Simple ranging method, typically available on communication devices. \\
\textbf{Cons}: Error-prone, typically requires an extensive tuning of the path loss model. \ac{rss} measurements are often affected by the losses in the front-end receiver architecture of the client and by the number of quantization bits in its \ac{adc} circuitry. \\

\input{taxonomySummary}

\noindent \textbf{Time information}~\cite{ToF2019Maletic} ---
Time information is another common proxy for the distance between two devices. Typical measurements used for this purpose involve \ac{tof} and \ac{tdoa} measurements.
\ac{tof} (also known as \ac{toa}) measurements exploit the time taken for a signal to propagate from the \ac{ap} to the client in order to estimate the distance between them. 
The client intuitively lies on the circumference of the circle with the \ac{ap} as the center and the distance estimate as the radius. Multilateration methods can be used to estimate the location of the client. It is important to note that \ac{tof} measurements require a tight synchronization between the \ac{ap} and the client. \ac{mmw} signals offer better \ac{tof} estimation accuracy (thus better ranging resolution), owing to the large bandwidth available, especially in the unlicensed bands.
We cover time-based approaches in Section~\ref{sec:localgo.rssi-tof}.\\
\textbf{Pros}: \ac{tof} information is usually accurate when directly extracted from a device's physical layer, which helps accurate localization. Such protocols as the \ac{ftm} protocol, when available on a device, can provide very accurate timing estimates.\\
\textbf{Cons}: Requires sub-nanosecond sampling times in a device's \ac{adc} in order to yield a sufficiently fine range resolution. \\

\noindent \textbf{Hybrid approaches}~\cite{accurate3D2018pefkianakis,bielsa2018indoor,kanhere2019map,sub6AoA2018Maletic,3DmmmMIMO2018Mathiopoulos,doaLf2017,pseudolateration2017,lms2018PIMRCmotion,losNlos2018msc,nlos2018SPAWCbrink, mmRanger2019,polar2020Infocom, Yassin2018GeometricAI, MOSAIC2018slam, 3dlocmap2018Yassin} ---
Several solutions propose to fuse information from multiple sources in order to improve localization accuracy. For example, several works merge \ac{aoa} and \ac{rssi}, or \ac{aoa} and \ac{tof} estimates.
We cover hybrid approaches in Section~\ref{sec:localgo.hybrid}.\\
\textbf{Pros}: Hybrid schemes usually achieve better accuracy. In some purely angle-based algorithms, side information such as \ac{rssi} and \ac{toa} can help resolve geometric translation, rotation, and scaling ambiguities.\\
\textbf{Cons}: The algorithms become more complex, and rely on the estimation of multiple quantities. In ill cases, errors compound and may make the location system more inaccurate than non-hybrid ones.\\

According to our survey of the literature on \ac{mmw} localization algorithms and to the above discussion, we identify two broad categories in the available literature:
\begin{enumerate}
    \item Algorithms \emph{tailored} to \ac{mmw} communication protocols and schemes, that exploit protocol operations to extract geometric scenario information and infer the location of the devices;
    \item \emph{General} algorithms that apply well-known range-based or range-free localization approaches to \ac{mmw} communications.
\end{enumerate}
The algorithms in the first category are mainly angle-based or \ac{csi}-based: they infer the angle of arrival structure by leveraging, e.g., sector measurements in communication protocols. Then, they use angle information to localize a device.
By way of contrast, the algorithms in the second category are not necessarily \ac{mmw}-specific. 
These works can be further subdivided by considering where the algorithm mainly runs:
\begin{enumerate}
    \item In \emph{client-centric} algorithms, the intelligence mainly resides on the client, which may collect location-dependent measurements by receiving signals from one or multiple \acp{ap}, and by estimating its own location locally. This approach is useful for systems that need to scale to up a large number of devices, as each device runs the algorithm independently.  Literature surveyed: \cite{5gRssi2020beam, Vashist2020ml,rssi2014icc,wbfps2020Infocom,olivier2016lightweight,palacios2019single,henkAoD2020Icassp,Yassin2017SimultaneousCI,clam2018Palacios,palacios2017jade,BeamAoD,polar2020Infocom,Yassin2018GeometricAI,MOSAIC2018slam, doaLf2017}
    
    \item In \emph{\ac{ap}-centric} algorithm, the intelligence resides in a computing entity connected to one or multiple APs, which coalesce their measurements from multiple clients in order to estimate the location of each client. These schemes are ideal for seamless network management purposes (e.g., to optimize client-\ac{ap} associations) but scale less than client-centric approaches when the number of clients increases.  Literature surveyed: \cite{fing2019mitsubishi2, deepL2020mitsubishi, wangMERLfingerprintingPart4, Harvard2017Globecom,leap2019palacios,tensorGSCPD,ObjectDetAccurate2019Ajorloo,bielsa2018indoor,accurate3D2018pefkianakis,3DmmmMIMO2018Mathiopoulos,lms2018PIMRCmotion,losNlos2018msc,mmRanger2019}
    \item Schemes based on \emph{\ac{ap}-client cooperation} are based on a shared intelligence, where both one or more \acp{ap} and the client run portions of the localization algorithm, and possibly exchange information to finally estimate the client location.  Literature surveyed:  \cite{fing2019mitsubishi1,ToF2019Maletic,nlos2018SPAWCbrink,pseudolateration2017, sub6AoA2018Maletic, kanhere2019map}
\end{enumerate}
In our scan of the literature, we observed a comparatively small number of works that employ a form of machine learning to compute location estimates. We believe this is due partly to localization being a somewhat understood problem (whereby the community prefers the use of understandable and optimizable signal processing algorithms rather than training black-box machine learning models) and partly to the sometimes daunting collection of training data. Yet, these prove a feasible solution in some cases, e.g., when a huge database of different location-dependent features is available, and the complexity of the considered indoor environment prevents straightforward modeling.

Table~\ref{tab:taxSummary2} summarizes the above preliminary subdivision pictorially, and conveys in what category most of the research efforts has concentrated so far. We observe that a few approaches have considered baseline \ac{rssi}, \ac{snr} and time measurements to localize \ac{mmw} devices. However, most of the research moved to exploit the fine angle resolution that large \ac{mmw} antenna arrays enable.
A significant number of works also consider hybrid approaches, which mix good angle resolution with the extra information yielded by time- or \ac{rssi}-based measurements, and thus achieve greater accuracy.
Finally, we observe that a few recent works (from 2017 to the time of writing) rely on \ac{ml} techniques, typically to process \ac{rssi} and \ac{snr} measurements and predict the location of a device. We highlight these works in green in Table~\ref{tab:taxSummary2}, in order to emphasize the emergence of this paradigm, previously unobserved in indoor \ac{mmw} localization. 

In \emph{client-centric} algorithms, the client collects signal measurements thanks to the interaction with different \acp{ap}. The client then trains an \ac{ml} model and employs it to estimate its own location. For example, in~\cite{Vashist2020ml}, the client collects \ac{snr} information to train \ac{ml} regression models. In~\cite{anish2022wcnc}, instead, the client resorts to \ac{aoa} information to train shallow neural networks and estimate its coordinates of the client. Other works in this survey that employ client-centric machine learning algorithms are~\cite{wbfps2020Infocom} and~\cite{doaLf2017}.

\emph{AP-centric} algorithms rely on \acp{ap} collecting location-dependent signal features that relate to the location of each client in a given environment. These radio fingerprints are then used to train models to localize the client. For example, in~\cite{fing2019mitsubishi2, deepL2020mitsubishi}, the \acp{ap} use the spatial beam \ac{snr} measurements collected during the beam training process in order to create a radio map of the environment. \Ac{dl} models are then trained to estimate the location and orientation of the client devices. Other works in this survey that employ \ac{ap}-centric machine learning algorithms are~\cite{wangMERLfingerprintingPart4} and~\cite{Harvard2017Globecom}.

Other algorithms rely on some form of \emph{\ac{ap}-client cooperation} to collect location-dependent signal features and train machine learning models. In these schemes, the features can be collected either by the \acp{ap} and the client separately and then exchanged, or through possibly multi-step procedures requiring \ac{ap}-client cooperation. The only work in the literature that uses this technique for \ac{ml} models is~\cite{fing2019mitsubishi1}. Here, \ac{rssi} and beam indices obtained both at the client and at the \acp{ap} after the beam alignment process are used to generate radio fingerprints at different client locations.

Notably, Table~\ref{tab:taxSummary2} clearly shows that \ac{ml}-based algorithms are mostly \ac{ap}-centric or hinge on a cooperation between \acp{ap} and clients. The main reason is most that \acp{ap} are infrastructured devices, and have easier access to compute power in local servers through fast cabled connections.

\subsection{Evaluation tools for mmWave localization}
\label{sec:localgo.evtools}

We now look into the tools that have been used so far to evaluate \ac{mmw} localization algorithms. From the surveyed literature, we observe both \emph{experimentation-based} and \emph{simulation-based} performance evaluation, depending on whether a proposed scheme is evaluated using \ac{mmw} hardware- or software-based setups.\\

\subsubsection{ Experimentation-based performance evaluation}
\label{sec:localgo.evtools.hw}

Localization experiments so far have been carried out using either laboratory-grade or commercial-grade equipment. 
Laboratory-grade equipment typically includes \acp{sdr} for signal generation and a \ac{mmw} up-converter, with a directional antenna to drive signal emission. 
For example, the above setup is used in~\cite{clam2018Palacios}, where the authors employ horn antennas to emulate narrow beam patterns. A similar setup is part of the work in~\cite{rttofMaletic} and~\cite{ToF2019Maletic}, where the authors employ the Zynq 7045-based \ac{sdr} and the \ac{usrp} X310-based \ac{sdr}, respectively, in addition to a 60-GHz analog front-end to emit the \ac{mmw} signals. The authors of~\cite{x60wlan,Shivang2020mwsim} have used an NI \ac{sdr} with a 60-GHz transceiver that enables the user to fully program of the \ac{phy}, \ac{mac}, and network layers, especially for \ac{wlan} applications. It also incorporates a 24-element Sibeam reconfigurable antenna array. A \ac{fpga}-based setup is discussed in~\cite{openMilli}, where the authors have used the XCKU040 Kintex UltraScale \ac{fpga} for the baseband processing of a 60-GHz reconfigurable phased antenna array. PEM-003 60~GHz transceivers were used as the RF front-end for the experimentation. Recently, the New York Uuniversity spin-off Pi-Radio~\cite{pi-radio} developed dedicated \ac{sdr} boards for \ac{mmw} wireless communications. The Pi-Radio~v1 \acp{sdr} consists of a 4-channel fully-digital transceiver board with a Xilinx's ZCU111 RF \ac{soc}~\cite{xilinxZcu111}, and operates over a bandwidth of about 2~GHz in the 57-64~GHz band.

Other platforms currently in use in experimental work include the open source \ac{mmw} experimentation platform proposed in~\cite{lacruz_mm-flex_2020}. It consists of a Xilinx Kintex Ultrascale \ac{fpga} with a 60-GHz front-end. The \ac{fpga} is integrated on an AMC599 board that implements hardware signal processing and storage for real-time frame processing. It can also provide antenna array reconfigurability for fast beam switching, e.g., for high mobility scenarios.
Moreover, Polese et al.~\cite{polese2019millimetera} propose a 60-GHz \ac{sdr}, fully digital experimentation platform. It uses a Xilinx KC705 and has 4 independent streams.

Alternatively, commercial-grade equipment can be leveraged for localization purposes, usually by substituting the provided operating system image with a custom build that embeds \acp{api} to access the output of the beam training procedure. 
For example, the work in~\cite{accurate3D2018pefkianakis} realizes a geometric 3D localization system using a 4$\times$8 phased array within a router that embeds a Qualcomm QCA9006 tri-band chipset for \ac{aoa} and \ac{tof} measurements. The work in~\cite{bielsa2018indoor}, instead, taps into the output made available by the Talon AD7200~\cite{talonRouter} routers' firmware. In the latter case, the hardware and the interface require significant adaptations of the angle estimation algorithms. For example, the firmware and operating system used in~\cite{bielsa2018indoor} returned coarsely quantized power measurements for each beam pattern and sometimes incomplete measurement outputs, which required to re-cast the angle estimation algorithm to be robust against quantization noise and missing values. 
The proprietary setup used in~\cite{accurate3D2018pefkianakis} returns the raw \ac{cir} measurements, which are then sanitised to extract the azimuth and elevation angles of arrival from the \ac{los} paths, and the \ac{tof} information for distance estimation.

%%%%%%%%%%%%%%%%%%%%%%%%%%%%%%%%%%%%%%%%%%%%%%%%%%%%%%%%%%%%
\input{localgo_evalTools.tex}
%%%%%%%%%%%%%%%%%%%%%%%%%%%%%%%%%%%%%%%%%%%%%%%%%%%%%%%%%%%%

Other works such as~\cite{x-array} also employ \ac{cots} devices like the 802.11ad-enabled Airfide \ac{ap}~\cite{airfide} to enhance the antenna array performance for omni-directional coverage and to improve link resilience in mobile and dynamic environments. 
Table~\ref{tab:evaltools} summarizes the above discussion by relating the works in our survey with the hardware platforms used to validate \ac{mmw} localization algorithms. We observe that software-defined platforms are still preferred, due to their greater versatility and to the availability of multiple digital receiver chains. \ac{cots} hardware is starting to appear in experimental evaluations, although this typically requires system management (and sometimes hacking) skills to flash the hardware with firmware and custom operating systems that give access to information from the radio receiver chain.

From a practical standpoint, the manufacturers of commercial-grade \ac{mmw} devices typically define a codebook of antenna weights that drive beam patterns to cover the largest set of lookout directions.
As a result, the corresponding beam patterns are not necessarily narrow, nor do they necessarily present a single direction where the gain is maximum~\cite{steinmetzer2017compressive}.

\begin{figure}[t]
    \centering
    \includegraphics[width=1\columnwidth]{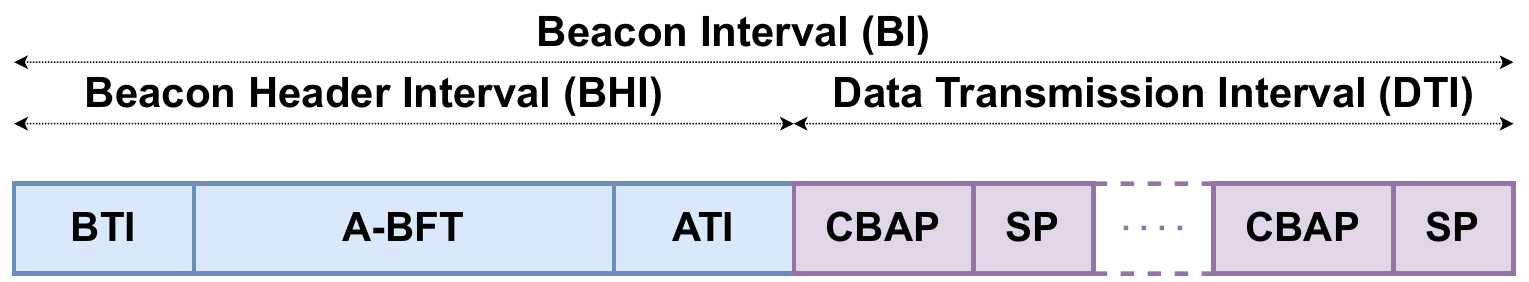}
    \caption{Beacon Interval frame of the IEEE 802.11ad standard~\cite{ieee80211adStandard}. It is important to note that after beam training process, \acp{sta} contend for the channel during the contention based access period (\ac{cbap}) and access it contention-free during the service period (\ac{sp}).}\vspace{2mm}
    \label{fig:loc.beaconInterval}
\end{figure}

\begin{figure}[t]
    \centering
    \includegraphics[width=1\columnwidth]{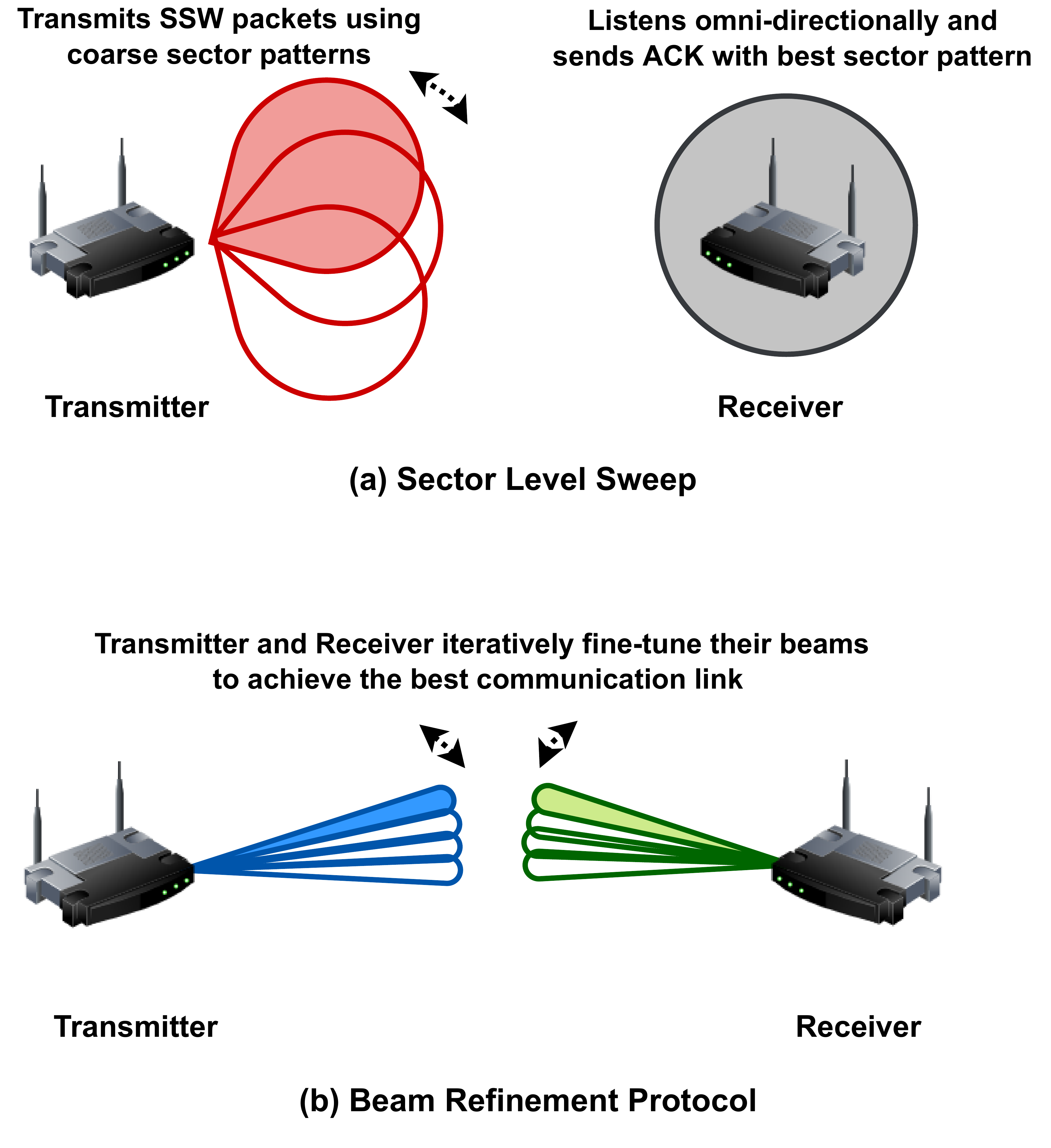}
    \caption{A simple illustration of the sector level sweep and beam refinement protocol as proposed in the IEEE 802.11ad standard~\cite{ieee80211adStandard}.}
    \label{fig:loc.slsbrp}
\end{figure}

Yet, standard-compliant beam training procedures still help retrieve location-dependent measurements through an automated process that is typically implemented in every device.
For example, the 802.11ad standard~\cite{ieee80211adStandard} presents a two-phase beam training process: 
\begin{itemize}
    \item \emph{\Ac{sls}}: During this phase, the transmitter (or \emph{beamformer}) periodically transmits \ac{ssw} frames using the different beam patterns defined in the sector codebook. The receiver (or \emph{beamformee}), receives these frames omnidirectionally and sends back an acknowledgment with the transmit sector yielding the highest signal quality. Subsequently, the two devices swap roles, and the receiver selects its best transmit sector. This phase provides coarse-grained beam patterns that are best suited for the two communicating devices.
    
    \item \emph{\Ac{brp}}: This optional phase can be used to refine the beam patterns chosen after the \ac{sls} phase. The \ac{brp} process is iterative. The two devices exchange special \ac{brp} packets requesting and acknowledging the \glsunset{tx}\glsunset{rx}transmit (\ac{tx}) and receive (\ac{rx}) training requests (TX-TRN and RX-TRN). The result is fine-grained beam patterns for the transmission and reception of the data, resulting in not just better directivity and therefore higher-throughput links, but also in a higher correlation between the beam pattern used and the \ac{aoa} of a signal.
\end{itemize}
These phases occur during the \ac{abft} subinterval of the \ac{bi}, as part of the channel access mechanism. The beamforming process during the \ac{dti} is to handle device mobility, blockage, etc. The \ac{bi} frame for channel access is shown in Fig.~\ref{fig:loc.beaconInterval} and the two beamforming phases are illustrated in Fig.~\ref{fig:loc.slsbrp}.

The more recent 802.11ay standard~\cite{claudioSilva-ITAW} formalized beam training procedures that enhance those of 802.11ad, namely the beam refinement protocol transmit sweep (\ac{brp} \glsunset{txss}\ac{txss}) and the \ac{abt}~\cite{802.11ay_claudioSilva}. These procedures rely on a channel reciprocity assumption to speed up beam training (through the \emph{\ac{brp} \ac{txss}} scheme) and slightly improve the process to compensate for the possibly different antenna gains at the \ac{ap} and at the client.

In addition, 802.11ay speeds up training in the presence of several clients through group beamforming, which extends beam training to manage multiple clients simultaneously.

When run with generic beam patterns, the above procedures do not yield a one-to-one relationship between the angle of arrival or departure of a \ac{mmw} signal and the antenna configuration that leads to the highest received power. Yet, if the beam patterns of the codebook are known, a \ac{mmw} device can still estimate angles of arrival via signal processing techniques involving compressive sensing~\cite{steinmetzer2017compressive}, or linear programming and Fourier analysis~\cite{bielsa2018indoor}.
Knowing angles of arrival enables angle-based localization techniques, with the additional advantage that angle estimation hinges on standard beam training procedures, with no need for external hardware components. In other words, localization becomes an embedded feature of \ac{mmw} communications.\\

% *************************************************

\subsubsection{ Simulation-based performance evaluation}
\label{sec:localgo.evtools.sw}

Simulation is the performance evaluation tool of choice if \ac{mmw} hardware is not available or if the available platforms do not offer sufficient flexibility to measure location-dependent features. A common practice observed in the literature is to employ ray tracers to mimic the propagation of \ac{mmw} signals. These ray tracing simulators are typically designed based on the channel models described in Section~\ref{sec:chan}. 
The main idea is to simulate the \ac{mmw} wireless channel characteristics at various indoor locations. Besides allowing the experimenter to measure channel features, ray tracers help create a radio map of the environment, and can thus substitute costly and time-consuming measurement campaigns~\cite{rappaport_Wireless6g}.

Two examples of such simulators are NYURay, a 3D \ac{mmw} ray tracer developed by New York University~\cite{rappaport_Wireless6g}, and S\_5GChannel, developed by Siradel.
NYURay was initially conceived as a geometry-based 2D ray tracer and was used in~\cite{locationNYUkanhere} to investigate indoor positioning algorithms based on \ac{aoa}, combined path-loss and \ac{aoa}, or \ac{rssi} values. NYURay was later extended in~\cite{kanhere2019map} to support 3D ray tracing by combining the shooting-and-bouncing rays (SBR) technique~\cite{sbr1997shin} and the geometry-based technique. NYURay found extensive use, not just in indoor environments, but also outdoors~\cite{kanhere2020millimeter, kanhere2021outdoor}.
Siradel developed the S\_5GChannel~\cite{s_5gchannel} \ac{5g} channel simulator to address the challenges of \ac{5g} signal propagation at \ac{mmw} frequencies indoors and outdoors. S\_5GChannel's ray model has been used in~\cite{MOSAIC2018slam} to develop a framework for joint localization and mapping.

A few additional works in the literature evaluate their proposed schemes using custom simulation software typically written in MATLAB or Python. The general purpose of such software is to generate synthetic datasets with realistic \ac{mmw} propagation characteristics, although typically restricted to the specific signal properties required for each study (e.g., \ac{aoa} values, \ac{toa} measurements, etc.).

In the following subsections, we explain the details of each surveyed work, and provide a synopsis of the main results of each paper and of the main enabling techniques in the form of summary tables at the end of the section.

\subsection{Angle-based algorithms}
\label{sec:localgo.aoa}

\ac{aoa} measurements, alongside the quasi-optical nature of \ac{mmw} signal propagation, facilitate high-accuracy localization based on triangulation. This is the simplest approach to localization using \ac{aoa}, wherein the angle information from the transmitting \acp{ap} and simple geometric principles are used to compute the client's position. In a 2-D plane, such position can be estimated using just two \acp{ap}~\cite{spatialContext2019Survey}.  

%%%%%%%%%%%%%%%%% Purely AoA based %%%%%%%%%%%%%%%%%%%%%%
%%%%%%%%%%%%%%%%%%%%%%%%%%%%%%%%%%%%%%%%%%%%%%%%%%%%%%%%%

Geometric methods are the simplest methods for localization when using \ac{aoa} estimates. In~\cite{olivier2016lightweight}, the authors present three lightweight single-anchor  algorithms based on the \ac{aoa} measurements. These algorithms are based on triangulation, \ac{adoa}, and fingerprinting, respectively. The algorithms have been simulated and also experimentally validated on pre-standard \ac{mmw} hardware operating at 60~GHz, showing that they achieve sub-meter accuracy with high probability, given the \ac{aoa} estimate errors are low.

The simplicity of these algorithms motivated the authors of~\cite{palacios2019single} to generalize the schemes in~\cite{olivier2016lightweight} for any number of \acp{ap}. These algorithms are extensively simulated as well as experimentally validated on 60~GHz \ac{cots} devices, in different indoor scenarios against two benchmark algorithms based on fingerprinting and \ac{aoa}. The two algorithms provide sub-meter accuracy in most indoor environments with multiple antennas. Triangulation-based scheme performs slightly better than the \ac{adoa}-based one in most scenarios, but independence of orientation and compass bias makes \ac{adoa} more preferable. 

The ideas proposed by~\cite{olivier2016lightweight} have also been used by the authors of~\cite{Yassin2017SimultaneousCI} for context inference and obstacle detection. They use the TV and \ac{adoa} algorithms for receiver localization using one \ac{ap}, estimate the locations of virtual anchor nodes, and thus infer the presence of obstacles.

\Ac{aoa} measurements have also been used for \ac{slam}. For example, in~\cite{palacios2017jade}, the authors propose a joint access point and device localization (JADE) algorithm that jointly maps the location of the client and of the physical and virtual \acp{ap}, while mapping the indoor environment, without any prior information (i.e. number of access points, boundaries of the room, etc.). The algorithm measures \acp{aoa} from the beam training procedure and leverages \acp{adoa} to estimate the location of the \acp{ap} and then of the client. Environment mapping follows by matching physical and virtual anchors and by predicting reflection points on surrounding surfaces. Simulation results show sub-meter accuracy in 90\% of the cases, even for erroneous \ac{aoa} estimates. JADE outperforms the approaches in~\cite{palacios2019single} in almost all scenarios. 

A similar algorithm that exploits \ac{aoa} information to derive \ac{adoa} estimates and fuses multiple measurements at different locations is CLAM~\cite{clam2018Palacios}. Like in~\cite{palacios2017jade}, the algorithm proceeds by first estimating the location of the anchor \acp{ap}, then of the client, and finally of the environment's boundaries. The algorithm is simulated and experimentally evaluated, showing sub-meter device localization errors in about 90\% of the cases.

A recent work explores deep learning-based localization scheme. The authors of~\cite{anish2022wcnc} propose a shallow neural network model to estimate the coordinates of the client device in an indoor environment, using \ac{adoa} measurements. The network is trained with imperfect location estimates from the JADE algorithm~\cite{palacios2017jade}, which jointly estimates the location of the \acp{ap} and the clients with zero knowledge of the environment. This relieves the burden of explicitly collecting the training dataset. The performance evaluation of the proposed scheme results in sub-meter client localization accuracy in $\approx 90\%$ of the scenarios, even with large \ac{aoa} errors. 
 
In~\cite{henkAoD2020Icassp}, the authors present mobile device positioning scheme in an indoor \ac{mmw} massive \ac{miso} scenario. The two-fold scheme utilizes coarse-grained \ac{aod} information from mobile clients with a single antenna to estimate the position of each client via downlink transmissions using adaptive beamforming. 

We can observe that angle-based algorithms usually rely on geometric approaches for device localization. However, \ac{ml} and neural network regression models can also be used to learn a non-linear mapping between \ac{aoa} measurements and client locations.
% \subsection{Angle of Departure}

\subsection{Channel information-based algorithms}
\label{sec:localgo.csi}

In recent works, \ac{mmw} \ac{csi} is also used to estimate the location of the client.
The definition of \ac{csi} varies from work to work. Typically, the term refers to the complex amplitude of the channel gain perceived at a receiving antenna, or to the vector of such gains measured by all elements of an antenna array.
A work exploiting \ac{csi} for localization is~\cite{tensorGSCPD}, where the authors present a channel parameter estimation method that transforms the \ac{mmw} uplink training signal into a higher-dimensional tensor using the canonical polyadic model. Tensor factorization using the proposed generalized structured canonical polyadic decomposition results in time delay, \ac{aoa}, and path fading coefficient estimates. These parameters are used to localize and track a mobile device.

A different way to exploit uplink \ac{csi} estimates~\cite{leap2019palacios} requires that the \acp{ap} convert the \ac{los} \ac{csi} measurements into angle information and then localize the client. The system is implemented on Talon AD7200 routers (without interfering with 802.11ad operations), and the authors propose to employ the location estimates to optimize \ac{ap}--client associations. The system achieves sub-meter localization accuracy in about 80\% of the cases.

With a focus on localizing passive objects, in~\cite{ObjectDetAccurate2019Ajorloo} the authors use the \ac{cir} captured after reflection from different objects and surfaces in an indoor environment to detect objects and also model the indoor environment in 2D. The proposed method has been evaluated using a testbed developed specifically for this purpose.

The use of \ac{csi} for localization is comparatively new for \ac{mmw} indoor device-based localization, most likely because retrieving full \ac{csi} or \ac{cir} data requires low-level hardware access, and only a few experimental firmware versions provide it. However, \ac{csi} and \ac{cir} can map to angle and time information, and therefore represent a promising and practical research direction, especially as feature-richer \ac{mmw} hardware and firmware emerges.

\subsection{RSSI and ToF}
\label{sec:localgo.rssi-tof}

%%%% Purely RSSI /SNR based %%%%%%%%%%%%%%
%%%%%%%%%%%%%%%%%%%%%%%%%%%%%%%%%%%%%%%%%%
\ac{rssi} and SNR based localization systems generally employ trilateration or fingerprinting-based techniques to localize the client. A number of works in the literature illustrate this concept. The authors in~\cite{rssi2014icc} investigate trilateration-based localization algorithm using \ac{rssi} measurements for 60-GHz IEEE 802.11ad \acp{wlan}. They modify the trilateration algorithm based on the concept of (weighted) center of mass. Simulations on randomly generated data points and the \ac{rssi} measured based on the IEEE 802.11ad channel model result in an average positioning error of about 1~m. This is among the earliest works on \ac{mmw}-based indoor localization that leverages \ac{rssi} measurements. 

%%%%% SNR %%%%%

\ac{rssi} is also the foundation of several fingerprinting-based localization schemes, especially in sub-6~GHz wireless networks. The authors of~\cite{fing2019mitsubishi1} propose a localization system that generates fingerprints of transmit beam indices and the corresponding \ac{rss} measurements between a pair of \ac{mmw} devices. Probabilistic location models are generated based on the fingerprint data and are leveraged for location estimation. The algorithm is experimentally evaluated using 60~GHz \ac{cots} devices. Many times, \ac{snr}-based fingerprinting is also at the core of some \ac{mmw} localization works, especially in combination with machine learning and deep learning techniques. The authors of~\cite{Vashist2020ml, kfloc2021vashisht} propose machine learning regression models for localization in warehouses. \ac{snr} information is collected from Talon AD7200 routers. The supervised regression models are trained offline and then deployed for localization at run time. The proposed method achieves sub-meter accuracy in 90\% of the cases. 

Similar machine learning regression models have been used for location estimation in~\cite{fing2019mitsubishi2}, where the authors use spatial beam \ac{snr} values, typically available during the beam training phase, in order to generate a location- and orientation-dependent fingerprint database. Deep learning techniques are also the main enablers for localization in~\cite{deepL2020mitsubishi} and~\cite{wangMERLfingerprintingPart4}, where the authors proposed ResNet-inspired models~\cite{he2016deep} for device localization in \ac{los} and \ac{nlos} scenarios.
To tackle the challenges imposed by \ac{nlos} conditions, the authors use spatial beam \ac{snr} values in~\cite{deepL2020mitsubishi}, whereas they employ multi-channel beam covariance matrix images in~\cite{wangMERLfingerprintingPart4}.

One example of how \ac{tof} measurements have been used in the \ac{mmw} context is presented in~\cite{ToF2019Maletic}. Here, the authors present a two-way ranging based on round-trip \ac{tof} (RTToF) information. The scheme estimates the distance between master and slave nodes, and then trilaterates the position of the slaves. The authors implement their algorithm on an \ac{sdr} with a 60~GHz \ac{soc}. The proposed system achieves an average distance estimation of 3~cm and an average positioning error below 5~cm.

Although conventional wireless localization schemes relying on \ac{rssi} or \ac{snr} measurements employed trilateration, machine learning-based fingerprinting algorithms are gaining more popularity for \ac{mmw}-based localization systems. This is due to the availability of mid-grained channel measurements from the beam training procedures of \ac{5g} and IEEE 802.11ad/ay systems~\cite{wangMERLfingerprintingPart4}. These techniques provide higher-accuracy location estimates compared to conventional techniques.

We also observe that \ac{mmw} systems do not rely on purely time-based measurements for localization. Even though the large bandwidth of \ac{mmw} signals can provide fine time measurements, such measurements tend to be fully available only on custom high-end \ac{mmw} transceivers. Therefore, many schemes tend to collect other signal measurements as well.

\subsection{Hybrid approaches}
\label{sec:localgo.hybrid}

A combination of two or more techniques mentioned above can be used to build systems that achieve better localization or mapping accuracy, with respect to stand-alone techniques. Coupling different sources of information is useful in challenging environments, where some \ac{mmw} parameter measurements may fail.

%%%%%%%%% Hybrid: AoA and RSSI based %%%%%%%%%%%%%%%%%%
%%%%%%%%%%%%%%%%%%%%%%%%%%%%%%%%%%%%%%%%%%%%%%%%%%%%%%%

Angle information along with \ac{rssi}-based ranging are the foundation of several \ac{mmw} localization approaches in the literature. The authors in~\cite{3DmmmMIMO2018Mathiopoulos} propose a positioning algorithm using \ac{rss} and \ac{aoa} measurements. These measurements are derived from a channel compression scheme designed for a \ac{mmw} mMIMO scenario with only one \ac{ap}. 
The \ac{rss} and \ac{aoa} estimates from the above methods are employed for position estimation. The system provides decimeter-level accuracy even at low \ac{snr}, and even lower errors as the \ac{snr} increases. 

%%%%%%%%%%%%%%%

As opposed to ranging, the algorithms proposed in~\cite{doaLf2017} and~\cite{wbfps2020Infocom} are based on location fingerprinting. In particular, the authors measure \ac{rssi} and \ac{aoa} information at various reference points in an indoor environment to generate location fingerprints. $K$~fingerprints nearest to the client measurement are selected from the dataset, and the location estimate corresponds to the weighted average of these $K$ reference points. The algorithm has been simulated with 2.4~GHz and 60~GHz, showing that  the average position error is much lower for \ac{mmw} signals than lower-frequency signals. To solve the problem of collecting a sufficiently large dataset,~\cite{wbfps2020Infocom} generates 3D beam fingerprints using \ac{rssi} and beam information. Weighted K-NN was used to localize an \ac{uav} in GPS-denied indoor environments. Particle filters were used along with the imperfect location estimates to track the motion \acp{uav}. The proposed scheme was experimentally validated, and the results showed sub-meter positioning accuracy on average.  

%%%%%%%%%%%%%%%%%

%%%%%%%%%%%%%%%%%% localization and mapping

\ac{rss} jointly with \ac{aoa} information enables \emph{mmRanger}~\cite{mmRanger2019} to autonomously map an indoor environment without infrastructure support. The mmRanger scheme senses the environment and uses time domain \ac{rss} sequences to reconstruct the path geometry via a path disentanglement algorithm. Then, \ac{aoa} and \ac{rss} information from the reflecting surfaces are exploited to reconstruct the geometries of each surface. Moreover, a robot pedometer assembles all estimated fragments to form a complete map of the environment. The results of the proposed system implementation show a mean estimation error of 16~cm for reflection points, and a maximum error of 1.72~m.

%%%%%%%%% Hybrid: AoD and RSSI based %%%%%%%%%%%%%%%%%%
%%%%%%%%%%%%%%%%%%%%%%%%%%%%%%%%%%%%%%%%%%%%%%%%%%%%%%%

In~\cite{bielsa2018indoor}, the authors leverage coarse-grained per-beam pattern \ac{snr} measurements provided by a modified operating system flashed on multiple TP-Link Talon AD7200 802.11ad-compliant \ac{cots} \ac{mmw} router. The \ac{aoa} estimation problem is formulated using linear programming, and the location is estimated using a modified particle filter and a Fourier analysis-based goodness function. The proposed scheme is experimentally validated and the system achieves sub-meter accuracy in 70\% of the cases. \ac{aod} and \ac{snr} information were used in~\cite{BeamAoD} to design beam-based midline intersection and beam scaling-based positioning algorithms. These were evaluated using both ray-tracing simulation and a WiGig \ac{soc} transceiver. The experiments, carried out under \ac{los} conditions, yielded centimeter-level location estimation errors.

%%%%%%%%% Hybrid: Angle and time based  localization%%%%%%%%%%%%%%%%%%
%%%%%%%%%%%%%%%%%%%%%%%%%%%%%%%%%%%%%%%%%%%%%%%%%%%%%%%

Time-based measurements are often enriched with angle information in order to achieve better positioning accuracy, especially for \ac{mmw} systems. For example, in~\cite{accurate3D2018pefkianakis}, the authors propose \emph{mWaveLoc}. The proposed system uses measured \acp{cir} to calculate \ac{aoa} and \ac{tof} data. The system is implemented on IEEE 802.11ad off-the-shelf devices leveraging the OpenWRT operating system, and achieves centimeter-level distance estimation and decimeter level 3D localization accuracy (median error 75~cm and sub-meter error in 73\% of the cases) in a realistic indoor environment. The system has also been evaluated in various experimental conditions. 

The author of~\cite{kanhere2019map} propose a map-assisted positioning technique using the fusion of \ac{tof} and \ac{aod}/\ac{aoa} information. A 3D map of the environment is either generated on the fly or assumed to be known a-priori. The scheme measures a set of possible user locations by fusing the estimated \ac{tof} values with angle information. These estimated locations are clustered, and the cluster centroid is output the final location estimate. The algorithm is simulated on the data collected at 28~GHz and 73~GHz by NYURay 3D ray tracer. The best-case and the worst-case mean localization error is found to be about 12~cm and 39~cm respectively.

Instead of explicitly fusing \ac{tof} and \ac{aoa} information, the authors of~\cite{pseudolateration2017} propose a pseudo-lateration protocol, that enacts the three following steps: i) sector sweeping for tracking \ac{los} and \ac{nlos} paths to compute physical and virtual anchors, respectively; ii) angular offsets measurements using extended sector sweeping; and iii) \ac{tof} measurements for distance estimation. A post-processing stage is employed for position estimation. The protocol has been simulated and implemented using a 60~GHz \ac{mmw} testbed. The protocol implementation achieves centimeter-level location estimation accuracy within 1.5~m and decimeter accuracy beyond 1.5~m.

The authors of~\cite{lms2018PIMRCmotion} explored adaptive filters for motion-assisted indoor positioning. An improved LMS filter estimates the \ac{aoa} of the client by using the client location, velocity and measured \ac{tof} as the inputs. \Ac{aoa} and \ac{toa} estimates are fed to an \ac{ukf} to track the client's position. The two-stage algorithm is simulated in an office environment with one \ac{ap}\ and achieves centimeter-level positioning accuracy.

Because mistaking \ac{los} for \ac{nlos} paths may offset location estimates significantly, the authors of~\cite{losNlos2018msc} propose a scheme to tell apart \ac{mmw} \ac{los} and \ac{nlos} \acp{mpc} having incurred up to one reflection. For this, they use \ac{tdoa} and \ac{aoa} information and apply the mean shift clustering technique. Then, they apply an \ac{aoa}-based localization scheme that computes least-squares estimates. The methods show a 98.87\% accuracy in path identification and positioning error of less than 75~cm in 90\% of the cases. \Ac{nlos} scenarios have also been exploited in~\cite{nlos2018SPAWCbrink}, where the authors propose a positioning scheme that relies on differential angle information, which is independent of angular reference. This scheme has been evaluated in an indoor environment with a geometric ray tracer based on an IEEE 802.11ay channel model, and achieves sub-30~cm position estimation errors in 90\% of the cases.

In~\cite{Yassin2018GeometricAI}, the authors present schemes for localization, mapping, obstacle detection and classification. Localization and mapping make use of \ac{aoa} and \ac{toa} measurements to estimate the location of the receiver and of virtual anchors. The latter are used to detect obstacles by estimating reflection points. Snell’s law and the relationship between the \ac{rss} and the reflection coefficient are used to classify the obstacles based on material composition. The presented algorithms have been simulated in an indoor environment.

Besides locating a client, the schemes presented in~\cite{Yassin2018GeometricAI} have been integrated into a \ac{slam} framework in~\cite{MOSAIC2018slam}. This framework involves algorithms for localization, obstacle mapping and tracking. \Ac{ekf}-based tracking helps improve obstacle detection and mapping. The framework has been simulated in an indoor environment, yielding sub-meter errors in 90\% of the cases. In the same context, the \ac{ekf} improves the obstacle mapping accuracy to sub-centimeter.

In~\cite{sub6AoA2018Maletic}, the authors present a device localization scheme, where the \ac{ap} and the client are equipped both with sub-6~GHz and with \ac{mmw} technology. Sub-6~GHz antennas are used for \ac{aoa} estimation and \ac{mmw} antennas are fed with the \ac{aoa} estimates for subsequent beam training and two-way ranging. The proposed method has been experimentally validated using SDR platforms, both in an anechoic chamber and in an office environment. Results show 2$^{\circ}$ \ac{aoa} errors and centimeter-level ranging accuracy in the anechoic chamber, and 5$^{\circ}$ \ac{aoa} error with an average 16-cm range error in the second one.

In~\cite{polar2020Infocom}, the authors propose to track the changes in the \ac{cir} measured at the station, that is equipped with an \ac{fpga}-based platform with IEEE 802.11ad, in order to localize a device-free object in an indoor industrial environment. The station uses the estimated \ac{cir} to measure the \ac{aod} and \ac{tof} of the signal reflecting off a moving object. Tracking \ac{cir} changes over time helps classify the reflections as static or mobile. Then, a Kalman filter smooths the trajectory of the mobile object. The results show sub-meter location errors in all scenarios, and a mean accuracy of 6.5~cm.

\label{sec:localgo.lastsecsurv}

From the literature surveyed above, we can observe that most localization schemes use angle information along with \ac{rssi}/time information, and often rely on geometric algorithms to compute high-accuracy location estimates. The use of adaptive filters such as \ac{lms} and Kalman filters helps mitigate location estimate errors, especially with mobile clients.

\subsection{Summary, highlights, and challenges}
\label{sec:localgo.hcls}

We now summarize the surveyed literature in order to highlight the main pitfalls and lessons learned from the methods.

\noindent\textbf{Geometry-based algorithms} -- The algorithms based on geometric techniques mostly rely on angle information (\ac{aoa}/\ac{aod}) for localization. As \ac{mmw} signals propagate quasi-optically, angle information becomes a reliable means to estimate the direction of the source. \ac{rssi} and \ac{tof} information help estimate the distance between a \ac{mmw} source and its receiver; thus, applying geometric methods such as triangulation and trilateration can help localize a client. However, the accuracy of such algorithms depends upon the accuracy of angle and time measurements, and most of them require accurate indoor floor plan information to work reliably. 

From the perspective of \ac{cots} devices, angle measurements are obtained either by decomposing \ac{csi} measurements using parameter estimation techniques or from the beam patterns chosen after beam training. However, imperfect beam patterns with broad main lobes and non-negligible sidelobes can lead to angle estimation errors.

Similar issues affect the estimation of time information through \ac{cir} or packet exchange means. We can obtain fine time measurements thanks to the large bandwidth of the \ac{mmw} signals. However, this requires a very tight synchronization between the \ac{ap} and the client devices.

\noindent\textbf{\ac{ml}-based algorithms --}
Owing to recent contributions, we observe a paradigm shift towards self-learning location systems that exploit the information from \ac{mmw} signals. In these works, \ac{ml} and \ac{dl}-based models use information extracted from \ac{mmw} signals at different locations to form a dataset and eventually learn an accurate model to estimate client positions. However, most of these models are specific for the location the training data comes from, and do not translate well to other locations.
Most of the algorithms in the literature train \ac{ml} and \ac{dl} models through \ac{rssi} or \ac{snr} fingerprint maps. Recent works have showed how \ac{aoa} and \ac{csi} information also help \ac{ml} models learn a non-linear function either to estimate the location of the client or to associate a client to the best \acp{ap}.  

Although these systems provide good localization accuracy, they also face several challenges: the collection of large training dataset; the computational complexity which limits their application to \ac{cots} or embedded devices; their dependence on the training environment. \Ac{ml} methods have thus found comparatively limited application to date. A valuable contribution to the community would be a collaborative effort towards a public benchmark dataset, that different authors would use to feed different machine learning approaches.

\noindent\textbf{Error mitigation in \ac{mmw} localization --} Errors in signal measurements due to imperfect signal parameter estimation limit the performance of localization systems~\cite{yang2021integratedLocComm}. These errors are often due to the unpredictable interference between multiple propagation paths and the fading that results, or to \ac{nlos} arrivals reaching a device~\cite{kanhere2021position}. 
A detailed mathematical analysis for error mitigation is presented in~\cite{yang2021integratedLocComm}. In the case of \acp{mmw}, measurement errors may affect angle and time measurements. The works in the literature resort to adaptive filter-based techniques mostly to mitigate the location estimation errors resulting from localization algorithms fed with error-prone data~\cite{xiao2020overview6G}. For example, the approaches in~\cite{wbfps2020Infocom, bielsa2018indoor} resort to particle filters to mitigate client location errors. Different types of Kalman filters are another typical choice to smooth out client location estimates and trajectories~\cite{scalingmmWave2019Fiandrino}. The authors of~\cite{losNlos2018msc} use \ac{lms} filters to mitigate large errors in \ac{aoa} and \ac{tof} measurements. A detailed account of the tools and techniques employed in each surveyed paper is provided in Table~\ref{tab:summaryLocAlgos} at the end of this section.

%%%%%%%%%%%%%%%%%%%%%%%%%%%%%%%%%%%%%%%%%%%%%%%%%%%%%%%%%%%%
\input{localgo_mainTechniques}
%%%%%%%%%%%%%%%%%%%%%%%%%%%%%%%%%%%%%%%%%%%%%%%%%%%%%%%%%%%%

%%%%%%%%%%%%%%%%%%%%%%%%%%%%%%%%%%%%%%%%%%%%%%%%%%%%%%%%%%%%
\input{localgo_evaluationMethod.tex}
%%%%%%%%%%%%%%%%%%%%%%%%%%%%%%%%%%%%%%%%%%%%%%%%%%%%%%%%%%%%

Table~\ref{tab:toolsUsed} supports the above discussion by summarizing the techniques employed in each of the surveyed works. We observe a preference for geometry-based localization approaches, with different supporting signal processing algorithms.

\subsection{Discussion and future research directions}
\label{sec:localgo.summary}

We summarize the findings of our survey in Table~\ref{tab:summaryLocAlgos}. The table succinctly conveys the main proposition of each paper, the main techniques used among those outlined in Sections~\ref{sec:localgo.aoa} to \ref{sec:localgo.lastsecsurv}, the tools employed, and a highlight of the performance attained.
We observe a number of \ac{mmw} localization approaches exploiting different features of \ac{mmw} signals as well as different properties of \ac{mmw} propagation. While some schemes rely on well-known techniques, e.g., based on \ac{tof} and \ac{rssi} measurements, even these techniques have been further developed to leverage the sparsity of \ac{mmw} multipath patterns in order to collect more precise measurements with a finer time resolution. In some environments, typically with special-purpose lab-grade \ac{mmw} hardware, the corresponding localization schemes often yield decimeter- or sub-decimeter-level accuracy, but require specific protocols to exchange the data the algorithms need.

With respect to such approaches, angle-based localization schemes relying on \ac{aoa} and \ac{adoa} measurements still prove accurate, and yield the additional benefit that \ac{aoa} measurements can directly result from beam training operations at link setup time. Hybrid solutions that leverage both angle information and time/\ac{rssi} information tend to show even better performance, although only a minority of them has been tested in operational environments with \ac{cots} equipment.

Finally, Table~\ref{tab:evaluationMethod} summarizes the performance evaluation approach taken in each surveyed paper. We observe a majority of evaluations based on experiments with real hardware, although simulation is still used in several contexts, e.g., as a tool to quickly and affordably generate large datasets.

\ac{mmw} technologies are expected to keep gaining momentum as part of the \ac{5g}-and-beyond ecosystem, and there exist a wealth of promising research directions to realize the vision of embedding localization as a feature of \ac{mmw} communications. According to our analysis, we identify the following key research directions. The community needs more low-cost \ac{mmw} platforms that implement standard-compliant operations while still providing \acp{api} for researcher and developers to access low-level physical layer measurements, such as per-beam pattern \ac{csi}, or even better, full \acp{cir}. This would democratize the research on practical algorithms that fully integrate localization as part of standard communications operations. In particular, such platforms would help research better algorithms to manage scenarios featuring multiple \acp{ap}, which are expected to be common in indoor \ac{mmw} deployments.
Moreover, there is ample space for the design of zero-initial knowledge algorithms that require no input data from the user, and autonomously bootstrap the algorithm by finding the location of all anchors (e.g., all \acp{ap}), localizing the clients, and using the joint location information of all clients and \acp{ap} to estimate the floor plan of indoor environments in a \ac{slam} fashion, both as a stand-alone solution and as a complement to the device-free radar-based approaches described in Section~\ref{sec:radarsurv}.

From the point of view of \ac{ml} schemes, we observe that most approaches still require lengthy training data collection operations before achieving practical accuracy levels. Moreover, a trained \ac{ml} algorithm remains specific to the area where training data was collected. Therefore, further research is needed on machine learning approaches that work with less training data, federate training results from different clients and \acp{ap} in order to speed up the training phase, and can be transferred across different, even previously unseen environments.

\input{localgo_paperSummaryLocAlgos.tex}

%% file: taxonomySummary.tex
\newcommand{\bwDoc}{\includegraphics[width=1.05em]{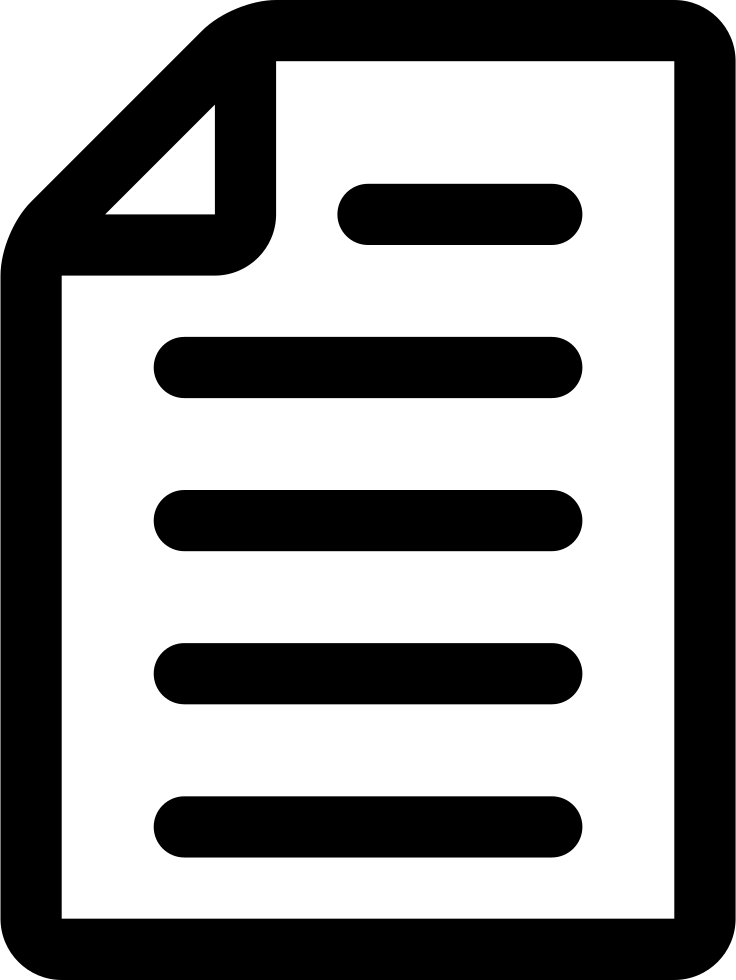}}
\newcommand{\grDoc}{\includegraphics[width=1.05em]{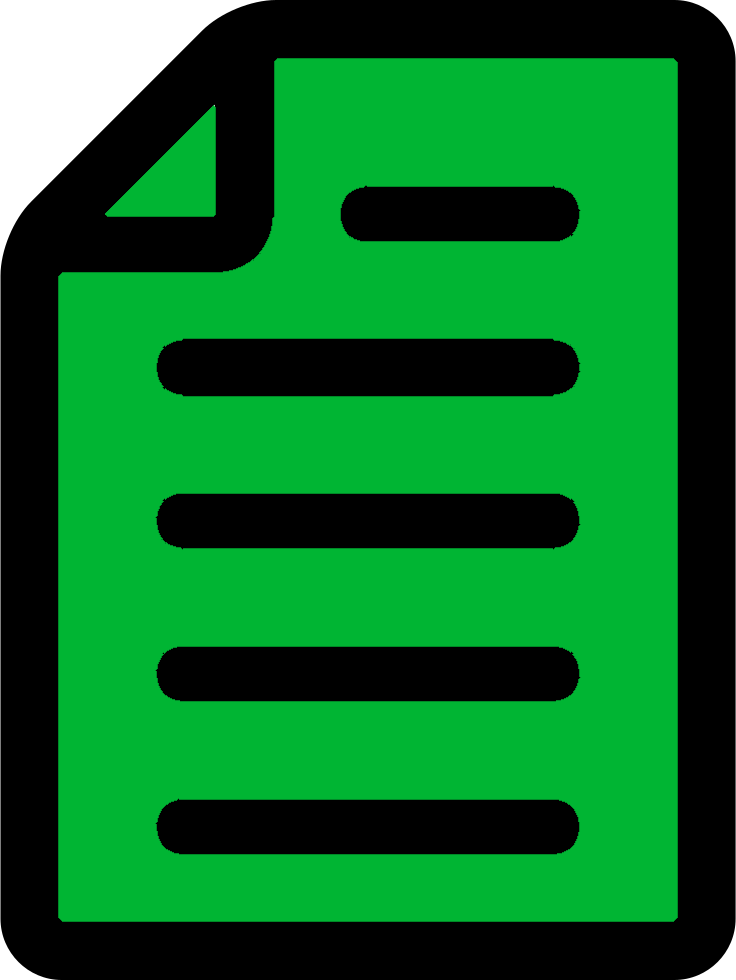}}

%%%%%%%%%%%%%%%%%%%%%%%%%%%%%%%%%%%%%%%%%%%%%%%%%%%%%%%%%%%%%%%%%%%%%%%%%%%%%%%%%%%%%%%%%%%%%%%%%%%%%%%%%%%%%%%

\begin{table*}[!ht] % << PC: make it a "p" so you have it in a separate page
\centering
\caption{Visual representation of the distribution of research efforts for device-based mmWave localization.\protect\linebreak Green icons represent recent papers that employ some form of machine learning.}
\label{tab:taxSummary2}

\resizebox{0.8\linewidth}{!}{ \centering \renewcommand{\baselinestretch}{1.0}\small 
\renewcommand{\arraystretch}{1.25}
\begin{tabular}[h!]{@{\hspace{1mm}}L{1.75cm}C{2.5cm}C{1.75cm}l C{2.75cm}C{1.5cm}C{3.5cm}@{\hspace{1mm}}}

% \cmidrule[.08em]{2-7}
\multicolumn{1}{c}{\phantom{M}} & \multicolumn{6}{c}{\textbf{\textsc{\large Indoor mmWave localization}}} \\
\cmidrule[.08em]{2-7}

\multicolumn{1}{c}{\phantom{M}} 
& \multicolumn{2}{c}{\textbf{Traditional methods}}
& \phantom{,}
& \multicolumn{3}{c}{\textbf{Tailored methods}} \\
\cmidrule{2-3} \cmidrule{5-7}

& \textbf{RSSI and SNR}  
& \textbf{Time information} 
& \phantom{,}
& \textbf{Angle information}
& \textbf{CSI-based}
& \textbf{Hybrid approaches} \\

& {(e.g.~\cite{fing2019mitsubishi1},~\cite{Vashist2020ml})}
& {(e.g.~\cite{ToF2019Maletic})}
& \phantom{,}
& {(e.g.~\cite{palacios2019single,henkAoD2020Icassp})}
& {(e.g.~\cite{leap2019palacios,tensorGSCPD})}
& {(e.g.~\cite{doaLf2017,BeamAoD})} \\

\midrule

\textbf{Client-centric} 
&  \bwDoc{}\,\bwDoc{}\,\grDoc{}\,\grDoc{}   
&    
& \phantom{,}
& \bwDoc{}\,\bwDoc{}\,\bwDoc{}\,\bwDoc{}\,\bwDoc{}\,\bwDoc{}\,\grDoc{}\,
& 
&  \bwDoc{}\,\bwDoc{}\,\bwDoc{}\,\bwDoc{}\,\bwDoc{}\,\grDoc{}   \\ 
\midrule

\textbf{AP-centric} 
&  \grDoc{}\,\grDoc{}\,\grDoc{}   
& 
& \phantom{,}
&  \grDoc{} 
&  \bwDoc{}\,\bwDoc{}\,\bwDoc{}\,
& \bwDoc{}\,\bwDoc{}\,\bwDoc{}\,\bwDoc{}\,\bwDoc{}\,\bwDoc{}\\
\midrule

\textbf{AP-client cooperation} 
&  \grDoc{}
& \bwDoc{}
& \phantom{,}
& 
&   
& \bwDoc{}\,\bwDoc{}\,\bwDoc{}\,\bwDoc{}\,\\
\bottomrule

\end{tabular}
}

\end{table*}

%% file: localgo_evalTools.tex
%%%%%%%%%%%%%%%%%%%%%%%%%%%%%%%%%%%%%%%%%%%%%%%%%%%%%%%%%
%%%% Table for evaluation tools/strategy
%%%%%%%%%%%%%%%%%%%%%%%%%%%%%%%%%%%%%%%%%%%%%%%%%%%%%%%%%

\begin{table*}[t!]
\centering
\caption{Summary of the hardware and software platforms used in mmWave localization algorithms}
\label{tab:evaltools}
\renewcommand{\baselinestretch}{1.0}\footnotesize 
\renewcommand{\arraystretch}{1.25}
\begin{tabular}{ll}

  \toprule
  \textbf{Hardware Platform} & \textbf{Related Literature}  \\
  \midrule
  Vubiq 60~GHz development system & 
   \cite{palacios2019single},
   \cite{olivier2016lightweight},
   \cite{clam2018Palacios},
   \cite{bielsa2018indoor},
    \cite{pseudolateration2017} \\
  Zynq 7045 based SDR with 60~GHz analog front-end & \cite{ToF2019Maletic} \\
  4$\times$8 phased array AP with QCA9006 triband chipset & \cite{accurate3D2018pefkianakis} \\
  TP-Link Talon AD7200  &   \cite{leap2019palacios}, \cite{fing2019mitsubishi1}, \cite{Vashist2020ml}, \cite{fing2019mitsubishi2}, \cite{deepL2020mitsubishi}, \cite{wangMERLfingerprintingPart4} \\
  QCA6320 baseband module with QCA6310 RF front-end & \cite{mmRanger2019} \\
  %NYURay 3-D mmWave ray tracer & \cite{kanhere2019map} \\ \hline
  USRP X310 and TwinRX daughterboard with 60~GHz analog front-end & \cite{sub6AoA2018Maletic} \\
  MicroTik wAP 60G & \cite{polar2020Infocom} \\
  
  \bottomrule\\[-3.4mm]
  \textbf{Software Platform} & \textbf{Related Literature}  \\
  \midrule
  NYURay Ray tracer & \cite{kanhere2019map},\cite{locationNYUkanhere} \\ 
  S\_5GCHANNEL simulator & \cite{MOSAIC2018slam} \\
  \bottomrule
\end{tabular}
\end{table*}

%% file: localgo_mainTechniques.tex
\begin{table*}[!tbh]
\centering
\caption{Summary of the main techniques used in the surveyed papers}
\label{tab:toolsUsed}

\renewcommand{\baselinestretch}{1.0}\footnotesize
\renewcommand{\arraystretch}{1.25}

\begin{tabular}{ll}
  \toprule
  \textbf{Analytical Tools} & \textbf{Related Literature}  \\
  \midrule
  
  Beamforming techniques & \cite{henkAoD2020Icassp} \\ 
  Clustering methods &  \cite{tensorGSCPD}, \cite{losNlos2018msc},\cite{mmRanger2019}   \\  
  Deep learning &  
  \cite{Harvard2017Globecom},
  \cite{deepL2020mitsubishi}, \cite{wangMERLfingerprintingPart4},  \cite{anish2022wcnc}\\ 
  Fourier analysis &  \cite{bielsa2018indoor}\\ 
  Geometry &  \cite{palacios2019single},\cite{olivier2016lightweight},\cite{Yassin2017SimultaneousCI}, \cite{rssi2014icc},\cite{5gRssi2020beam},\cite{BeamAoD},\cite{accurate3D2018pefkianakis},  \cite{kanhere2019map},  \cite{Yassin2018GeometricAI},\cite{3dlocmap2018Yassin}      \\ 
  Kalman filters &  \cite{leap2019palacios}, \cite{lms2018PIMRCmotion}, \cite{polar2020Infocom}, \cite{MOSAIC2018slam} \\ 
  Least mean square filters & \cite{lms2018PIMRCmotion}\\ 
  Levenberg-Marquardt (LM) method & \cite{nlos2018SPAWCbrink}\\ 
  Linear programming & \cite{bielsa2018indoor} \\ 
  Machine learning models & \cite{Harvard2017Globecom}, \cite{wbfps2020Infocom}, \cite{Vashist2020ml},  \cite{fing2019mitsubishi2}, \cite{doaLf2017} \\ 
  MUSIC & \cite{sub6AoA2018Maletic} \\ 
  Particle filters &  \cite{wbfps2020Infocom}, \cite{bielsa2018indoor} \\ 
  Probabilistic data modelling & \cite{fing2019mitsubishi1} \\ 
%   Extended Kalman filtering & \cite{MOSAIC2018slam} \\ 
  Tensor analysis & \cite{tensorGSCPD} \\ 
  \bottomrule
   
\end{tabular}
\end{table*}

%% file: localgo_evaluationMethod.tex
%%%%%%%%%%%%%%%%%%%%%%%%%%%%%%%%%%%%%%%%%%%%%%%%%%%%%%%%%%%%
%%%%% Table for Evaluation method
%% This should be in order [1] - [end]
%%%%%%%%%%%%%%%%%%%%%%%%%%%%%%%%%%%%%%%%%%%%%%%%%%%%%%%%%%%%

\begin{table*}[t]
\caption{Summary of the evaluation methods used in the mmWave localization algorithms}
\label{tab:evaluationMethod}

\renewcommand{\baselinestretch}{1.0}\footnotesize 
\renewcommand{\arraystretch}{1.25}

\centering
\begin{tabular}{ll}
  \toprule
  \textbf{Evaluation} & \textbf{Related Literature}  \\
  \midrule
  Experimentation &    \cite{olivier2016lightweight, palacios2019single},\cite{clam2018Palacios}, \cite{ObjectDetAccurate2019Ajorloo},  
 \cite{fing2019mitsubishi1},
     \cite{wbfps2020Infocom},\cite{Vashist2020ml},\cite{deepL2020mitsubishi}, \cite{wangMERLfingerprintingPart4},\cite{ToF2019Maletic},\cite{BeamAoD},\cite{accurate3D2018pefkianakis},
\cite{bielsa2018indoor}, \cite{sub6AoA2018Maletic},  \cite{pseudolateration2017}, \cite{mmRanger2019},  \cite{polar2020Infocom}
       \\ 
  Simulation &  \cite{Yassin2017SimultaneousCI, henkAoD2020Icassp}, \cite{tensorGSCPD}, \cite{rssi2014icc},\cite{5gRssi2020beam},\cite{doaLf2017},\cite{kanhere2019map},   \cite{3DmmmMIMO2018Mathiopoulos},  \cite{lms2018PIMRCmotion},  \cite{losNlos2018msc}, \cite{nlos2018SPAWCbrink}, \cite{Yassin2018GeometricAI, MOSAIC2018slam},  \cite{3dlocmap2018Yassin}, \cite{anish2022wcnc} \\ 
  \bottomrule

\end{tabular}
\end{table*}

%% file: localgo_paperSummaryLocAlgos.tex
%%%%%%%%%%%%%%%%%%%%%%%%%%%%%%%%%%%%%%%%%%%%%%%%%%%
%%%%% Table for tools of the work
%%%%%%%%%%%%%%%%%%%%%%%%%%%%%%%%%%%%%%%%%%%%%%%%%%%

\onecolumn
{ \centering \renewcommand{\baselinestretch}{1.05}\small 
\renewcommand{\arraystretch}{1.25}

% \begin{table*}[p]
% \caption{Summary of the literature on indoor mmWave localization}\label{tab:summaryLocAlgos}%\vspace{-10mm}
% \end{table*}

\setlength{\LTcapwidth}{\textwidth}
\begin{longtable}[c]{|m{4cm}|C{1.75cm}|m{4cm}|m{5.5cm}|}

\caption{\textsc{Summary of the literature on indoor mmWave localization}}\label{tab:summaryLocAlgos}\\ %\vspace{-10mm}
\hline 
\multicolumn{1}{|c|}{\textbf{Proposition}} &
  \textbf{Techniques} &
  \multicolumn{1}{c|}{\textbf{Tools Used}} &
  \multicolumn{1}{c|}{\textbf{Performance}} \\ \hline \hline
\endfirsthead
\caption[]{\textsc{Summary of the literature on indoor mmWave localization (continued)}}\\ %\vspace{-10mm}
\hline 
\multicolumn{1}{|c|}{\textbf{Proposition}} &
  \textbf{Techniques} &
  \multicolumn{1}{c|}{\textbf{Tools Used}} &
  \multicolumn{1}{c|}{\textbf{Performance}} \\ \hline \hline
\endhead

\multicolumn{4}{|c|}{\textbf{Localization Algorithms}}\\ \hline \hline

%%%%%%%%%%%%%%%%%%%%%%%%%%%%%%%%%%%%%%%%%%%%%

Round-trip time based localization~\cite{ToF2019Maletic}  & \centering
 ToF & Geometry & Distance estimation error within 3~cm and location estimation error within 5~cm. \\ \hline

%%%%%%%%%%%%%%%%%%%%%%%%%%%%%%%%%%%%%%%%%%%%%%%%%%%%%%%%%%%%%

Accurate 3D indoor localization using a single \ac{ap}~\cite{accurate3D2018pefkianakis}  & 
AoA-ToF &
Geometry & 3-D localization with median accuracy of 75~cm with sub-meter accuracy in 73\% of the cases. \\ \hline

%%%%%%%%%%%%%%%%%%%%%%%%%%%%%%%%%%%%%
Improving localization accuracy~\cite{bielsa2018indoor} & AoD, RSSI & Linear programming, Fourier analysis, Particle filters & Sub-meter accuracy in 70\% of the cases. \\ \hline

%%%%%%%%%%%%%%%%%%%%%%%%%%%%%%%%%%%%%%%%%%%%%%%%%%%%%%%%%%%

Improving the accuracy of device localization~\cite{palacios2019single, olivier2016lightweight} &
  AoA &
  Geometry &
  Sub-meter localization accuracy in 70\% of the cases.  \\ \hline

%%%%%%%%%%%%%%%%%%%%%%%%%%%%%%%%%%%%%%%%%%%%%

Improving location estimation accuracy and network performance~\cite{leap2019palacios} & 
  CSI & Regression trees, Extended Kalman filter &
  Sub-meter localization accuracy in 80\% cases and throughput improvement between 8.5\% and 57\%, lesser outage probability, SNR within 3 dB of optimum. \\ \hline

%%%%%%%%%%%%%%%%%%%%%%%%%

Fingerprinting based indoor localization~\cite{fing2019mitsubishi1} & RSSI & Probabilistic models & Mean and median localization error of $\approx$30~cm. \\ \hline

Indoor localization for intelligent material handling~\cite{Vashist2020ml}  & 
 SNR & Multi layer perceptron regression, Support vector regression, Logistic regression
& Centimeter-level accuracy with \ac{rmse} of 0.84~m and MAE of 0.37~m. \\ \hline

%%%%%%%%%%%%%%%%%%%%%%%%%

Fingerprinting based indoor localization~\cite{fing2019mitsubishi2, deepL2020mitsubishi}  &
  SNR &
Machine learning algorithms, Deep learning & Avg. \ac{rmse} is 17.5~cm with coordinate estimates within 26.9~cm in 95\% of the cases. Median and mean \ac{rmse} of 9.5~cm and 11.1~cm respectively.   \\ \hline

%%%%%%%%%%%%%%%%%%%%%%%%%%%%%%%%%%%%%%%%%%%%%%

Fingerprinting based indoor localization in \ac{nlos} environments~\cite{wangMERLfingerprintingPart4} & SNR & Deep learning, Machine learning algorithms & Location classification accuracy greater than 80\%. Median location estimation error of about 11~cm. \\ \hline

%%%%%%%%%%%%%%%%%%%%%%%%%%%%%%%%%%%%%%%%%%%%%%%%
Map-assisted indoor localization~\cite{kanhere2019map} & AoA, AoD, ToA & Geometry & Mean localization accuracy of 12.6~cm and 16.3~cm in \ac{los} and \ac{nlos} respectively.  \\ \hline

%%%%%%%%%%%%%%%%%%%%%%%%%%%%%%%%%%%%%%%%%%%%%%%
Sub-6 GHz-assisted device localization~\cite{sub6AoA2018Maletic}  & \centering
 ToF-AoA & MUSIC, Geometry & \ac{aoa} estimation error less than 5$^{\circ}$ and about 16~cm distance estimation error. \\ \hline

%%%%%%%%%%%%%%%%%%%%%%%%%%%%%%%%%%%%%%%%%%%%%%%%

Single-antenna client localization using downlink transmissions~\cite{henkAoD2020Icassp} & \ac{aod} & Adaptive beamforming & 60\% improvement in the accuracy in the downlink scenario as compared to in the uplink scenario. \\ \hline

%%%%%%%%%%%%%%%%%%%%%%%%%%%%%%%

Indoor positioning for wideband multiuser millimeter wave systems~\cite{tensorGSCPD} & CSI & Tensor decomposition, Clustering methods & Decimeter-level position estimation errors. \\ \hline  

%%%%%%%%%%%%%%%%%%%%%%%%%%%%%%%%%%%%%%%%%%%%%%% 

Indoor network localization~\cite{rssi2014icc} & RSSI & Geometry & Mean positioning error around 1~m. \\ \hline

%%%%%%%%%%%%%%%%%%%%%%%%%%%%%%%%%%%%%%%%%%%%%%%%%%%%%%%%%%%%%

3D indoor positioning for \ac{mmw} massive \ac{mimo} systems~\cite{3DmmmMIMO2018Mathiopoulos}  & \centering
AoA-RSSI &
Geometry & Low complexity channel compression and beamspace estimation developed. Decimeter-level positioning errors achieved in \ac{nlos} scenarios. \\ \hline

%%%%%%%%%%%%%%%%%%%%%%%%%%%%

 Location fingerprint-based localization~\cite{doaLf2017}  &
AoA-RSSI &
K-nearest neighbours & Average positioning error for mmWave is 4 times less compared to lower frequency signals. \\ \hline

%%%%%%%%%%%%%%%%%%%%%%%%%%%

UAV positioning in GPS-denied environments~\cite{wbfps2020Infocom} & RSSI & Weighted K-NN, Particle filters & Sub-meter 90th-percentile location errors in different cases. \\ \hline

%%%%%%%%%%%%%%%%%%%%%%%%%%%%%%%%%%%%%%%%%%%%%%

Beam-based UE positioning in indoor environment ~\cite{doaLf2017} & SNR, AoD & Geometry & Centimeter-level estimation error in all cases. Experimental results approach the simulations results with MSE difference of 0.1~m. \\ \hline

%%%%%%%%%%%%%%%%%%%%%%%%%%%%%%%%%%%%%%%%%%%%%%%%%%%%%%%%

Single RF chain-based localization~\cite{pseudolateration2017}  & \centering
 ToF-AoA & Geometry & Centimeter accuracy in location estimation within 1.5~m and decimeter accuracy beyond 1.5~m. \\ \hline

%%%%%%%%%%%%%%%%%%%%%%%%%%%%%%%%%%%%%

Motion feature-based 3D indoor positioning~\cite{lms2018PIMRCmotion}  & \centering
AoA-ToA &
\Ac{lms} and Kalman filters & Centimeter-level positioning accuracy. \\ \hline

%%%%%%%%%%%%%%%%%%%%%%%%%%%%%%%%%%%%%%%%%%%%%%

\ac{los} and \ac{nlos} path identification and localization~\cite{losNlos2018msc}
& TDOA, AOA & Mean shift clustering, Geometry & 98.87\% accuracy in path identification and positioning accuracy $\leq$ 0.753~m in 90\% of the cases. \\ \hline

%%%%%%%%%%%%%%%%%%%%%%%%%%%%%%%%%%%%%%%%%%%%%%%%%

 \ac{nlos} mmWave indoor positioning~\cite{nlos2018SPAWCbrink}  & 
 AoA, AoD, ToA &
Geometry,  Levenberg-Marquardt (LM) method & Positioning accuracy within 30~cm in 90\% of the observations with differential angle information along with time information.  \\ \hline

%%%%%%%%%%%%%%%%%%%%%%%%%%%%%%%%%%%%%%%%%%%%%%

 5G mmWave indoor positioning~\cite{5gRssi2020beam}  & 
 RSSI &
Geometry, Beamforming & Single-beam geometric model for indoor positioning. Mean error of 0.7~m for stationary in \ac{los} and 2.4~m for a mobile user in \ac{los}/\ac{nlos} scenario. \\ \hline

%%%%%%%%%%%%%%%%%%%%%%%%%%%%%%%%%%%%%%%%%%%%%%%%%
 
 Data-driven indoor localization~\cite{Harvard2017Globecom}  & \centering
 AoA & Multi-layer perceptron, Deep learning & Sub-meter localization accuracy.  \\  \hline
 
 %%%%%%%%%%%%%%%%%%%%%%%%%%%%%%%%%%%%%%%%%%%%%%%

 Indoor localization with imperfect training data~\cite{anish2022wcnc}  & \centering
 AoA & Deep learning &Sub-meter localization accuracy in $\approx 90\%$ of the cases, when trained with client location estimates from JADE algorithm.\\ 

%
%%%%%%%%%%%%%%%%%%%%%%%%%%%%%%%%%%%%%%%%%%%%%
%
%%%%%%% SLAM based algos
%
%%%%%%%%%%%%%%%%%%%%%%%%%%%%%%%%%%%%%%%%%%%%%
%
\hline  \hline
%
%%%%%%%%%%%%%%%%%%%%%%%%%%%%%%%%%%%%%%%%%%%%%
\multicolumn{4}{|c|}{\textbf{Localization and Mapping Algorithms}} \\ \hline \hline

%%%%%%%%%%%%%%%%%%%%%%%%%%%%%%%%%%%%%%%%%%%%%%%%

Autonomous environment mapping~\cite{mmRanger2019} &
  \centering RSSI, AoA &
   Geometry, K-means clustering &
  Reflection point mean estimation error of 16~cm with a max error of 1.72~m. \\ \hline 
 
%%%%%%%%%%%%%%%%%%%%%%%%%%%%%%%%%%%%%%%%%%%%%%%%

Passive object localization~\cite{polar2020Infocom} & AoD, ToF & Kalman filters & Sub-meter accuracy in all cases with 6.5~cm mean error accuracy. \\ \hline

%%%%%%%%%%%%%%%%%%%%%%%%%%%%%%%%%%%%%%%%%%%%

Localization and obstacle detection~\cite{Yassin2017SimultaneousCI} & AoA & Geometry & Sub-meter accuracy in 70\% of the cases. High accuracy obstacle detection and obstacle limits estimation.  \\ \hline 

%%%%%%%%%%%%%%%%%%%%%%%%%%%%%%%%%%%%%%%%%%%%%%%%%

Localization and mapping~\cite{palacios2017jade, clam2018Palacios} &
  AoA &
   Geometry &
  Sub-meter localization accuracy in 90\% of the cases. SLAM without any prior knowledge. \\ \hline
%%%%%%%%%%%%%%%%%%%%%%%%%%%%%%%%%%%%%%%%%%%%%%%%

%%%%%%%%%%%%%%%%%%%%%%%%%%%%%%%%%%%%%%%%%%%%%%

Accurate object detection ~\cite{ObjectDetAccurate2019Ajorloo} & CIR & Geometry & Accuracy of about 2~cm achieved in most experiments. \\ \hline
%%%%%%%%%%%%%%%%%%%%%%%%%%%%%%%%%%%%%%%%%%%%%%%%%

Simultaneous localization and mapping without a-priori knowledge~\cite{Yassin2018GeometricAI, MOSAIC2018slam} &
  \centering AoA, ToF, RSSI &
   Geometry, Extended Kalman filters &
  Sub-meter device localization accuracy in 90\% of the cases. Sub-centimeter obstacle mapping accuracy. \\ \hline

%%%%%%%%%%%%%%%%%%%%%%%%%%%%%%%%%%%%%%%%%%%%%%%%%%

3-D localization and mapping~\cite{3dlocmap2018Yassin} &
  \centering RSSI, AoA, ToA &
   Geometry &
  Perfectly maps the environment for AoA errors $\leq$ 5$^{\circ}$. \\ \hline %\hline

%%%%%%%%%%%%%%%%%%%%%%%%%%%%%%%%%%%%%%%%%%%%%%%%%
%\as{*not related to our section} Radars with mmWave massive antenna arrays~\cite{mmRadar2014dardari, pmmRadar2015dardariJ} &
%  \centering mmWave Sensing &
%   Bayesian State Space Analysis, Extended Kalman filter &
%  Map reconstruction accuracy increase with number of antennas. Sensitive to radar orientation and position errors.\\
%%%%%%%%%%%%%%%%%%%%%%%%%%%%%%%%%%%%%%%%%%%%%%%%%
%
%
%
%%%%%%%%%%%%%%%%%%%%%%%%%%%%%%%%%%%%%%%%%%%%%%%%%
% \multicolumn{4}{|c|}{\textbf{Object and Obstacle Detection for Localization}} \\ \hline \hline
%%%%%%%%%%%%%%%%%%%%%%%%%%%%%%%%%%%%%%%%%%%%%%%%%

\end{longtable}
\addtocounter{table}{-1}
}

\twocolumn

%% file: 06_RADAR_survey.tex
\section{mmWave radar-enabled device-free localization and sensing}   \label{sec:radarsurv}

\subsection{Introduction}    
\label{sec:radarsurv.intro}

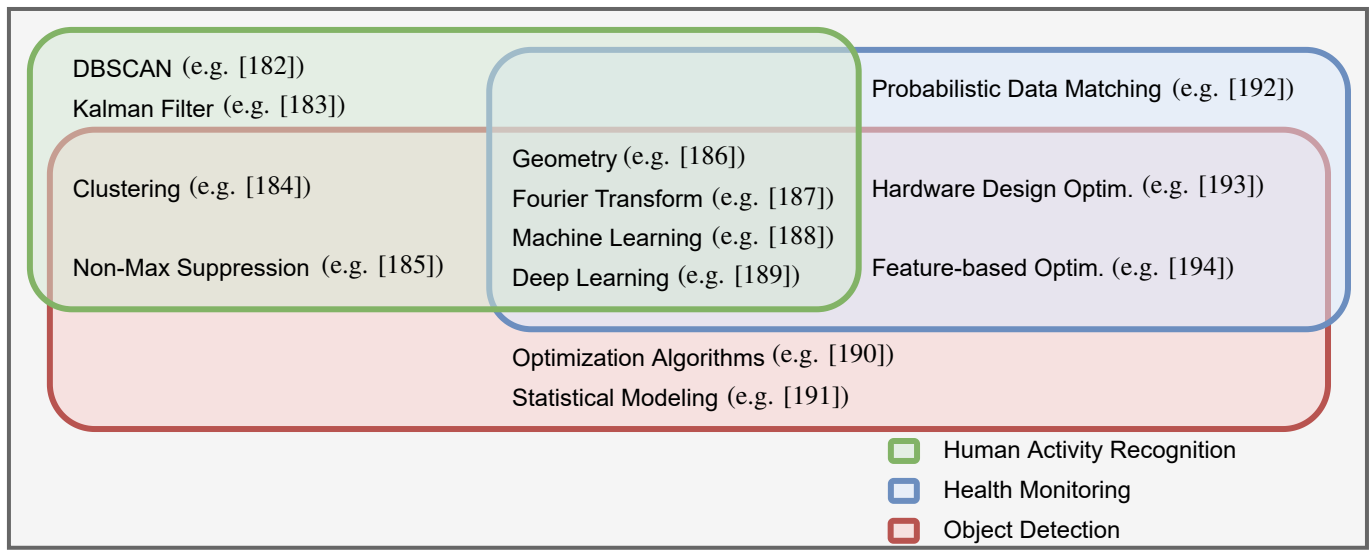
\begin{figure*}[b]
    \centering
    \begin{tikzpicture}
     
        % Include the image in a node
        \node [inner sep=0] (image) at (0,0) {\includegraphics[width=1\linewidth]{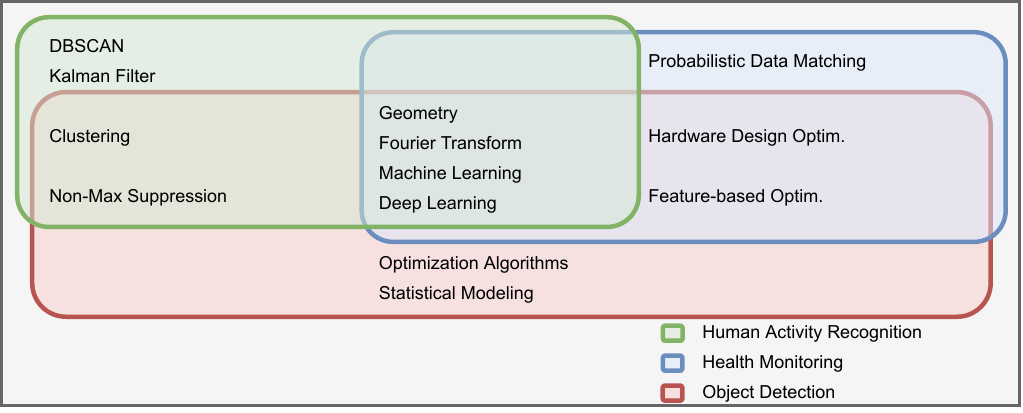}};
         
        % Create scope with normalized axes
        \begin{scope}[shift=(image.south west), % origin and scale
            x={(image.south east)},y={(image.north west)}]

            \node[anchor=west] at (0.122,0.888) {(e.g.~\cite{Wang2021})}; % DBSCAN
            \node[anchor=west] at (0.152,0.818) {(e.g.~\cite{pegoraro2020multiperson})}; % Kalman
            \node[anchor=west] at (0.127,0.67) {(e.g.~\cite{pegoraro2021-point-clouds})}; % Clust
            \node[anchor=west] at (0.224,0.527) {(e.g.~\cite{li2020capturing})}; % Nonmax
            
            \node[anchor=west] at (0.445,0.725) {(e.g.~\cite{jiang2020mmvib})}; % Geo
            \node[anchor=west] at (0.51,0.65) {(e.g.~\cite{Regani2021})}; % Fourier
            \node[anchor=west] at (0.51,0.580) {(e.g.~\cite{Nickalls2021})}; % ML
            \node[anchor=west] at (0.485,0.509) {(e.g.~\cite{Ozturk2021})}; % DL
            \node[anchor=west] at (0.554,0.365) {(e.g.~\cite{zhao2020heart})}; % Optim
            \node[anchor=west] at (0.521,0.29) {(e.g.~\cite{Cui2021})}; % Stat
            
            \node[anchor=west] at (0.847,0.849) {(e.g.~\cite{wang2020precise})}; % Prob
            \node[anchor=west] at (0.825,0.673) {(e.g.~\cite{Vodai2021})}; % Hard
            \node[anchor=west] at (0.803,0.524) {(e.g.~\cite{lu2020see})}; % Feature

        \end{scope}
     
    \end{tikzpicture}
    \caption{Overview of techniques used for \ac{mmw} radar sensing applications.}
    \label{fig:radarsurv.overview}
\end{figure*}

In this section, we focus on \ac{mmw}-based radar systems that operate over short distances (a few tens of meters), which have recently emerged as a low-power and viable technology for environment sensing. These devices are expected to find extensive use in a number of relevant applications, by replacing standard camera-based systems, either fully or in part. Survey papers have been recently published on \ac{mmw} sensing, with a focus on signal processing~\cite{davoli2021machine} (both with traditional and machine learning-based approaches) and applications~\cite{van2021millimeter}. In the following review, we emphasize the main signal processing algorithms that are being successfully exploited for indoor sensing, discussing their pros and cons. In doing this, we especially focus on \ac{nn} algorithms, discussing their different flavors, and exposing the most promising directions for research and development. We also comment on the level of maturity of this technology, i.e., about whether the proposed techniques are robust and work without requiring environment-specific and manual calibration. Our analysis will also discuss on the role of the supporting architecture, which should provide communication and computing/processing capabilities, and on the opportunity of implementing networks of radar devices. This would extend current systems, which often involve a single radar, to the large-scale monitoring of physical spaces.
An illustrative overview of the main techniques used for sensing applications that exploit \ac{mmw} radars is provided in Fig.~\ref{fig:radarsurv.overview}.

Two main components of a radar device are \ac{tx} and \ac{rx} \ac{rf} antennas, which are combined with an \ac{adc}, \acp{mcu}, \acp{dsp} and a clock. The main idea behind such a system is to transmit a properly shaped radio wave (e.g., pulses or continuous waves) and estimate the modifications that occur in the back-scattered copy of such wave, i.e., which is returned as a reflected signal from the surrounding environment. Through some signal processing algorithms (usually, Fourier transform-based), it is then possible to estimate the distance, angle, velocity, and, to some extent, the shape of the targets. \ac{tx} and \ac{rx} are usually co-located within the same device: the transmitter sends a first version of the modulated signal and the receiver detects its back-scattered copy from the surrounding environment, after a very short time delay.

Modern radar systems utilize two main {\it wave functions}; \ac{pw} and \ac{fmcw}. While radars are traditionally used to detect and track objects that move in the far field, such as vehicles and airplanes, here we are concerned with indoor or city environments where the objects to be tracked may be cars or humans. Moreover, besides the tracking, vital signs can also be monitored such as the breathing rate and the heart rate. Although these recent applications share common features with traditional (far field) approaches, they also exhibit major differences due to the short distance of the radar from the targets (near field).

\vspace{-1mm}

\subsection{Pulsed Wave Radar}
With \ac{pw} radars the electromagnetic waves from an antenna are emitted in short bursts. The logic behind \ac{pw} is to wait for the reflections from the previously transmitted signal to reach to the antenna before sending the next burst. Thus, the reflected signal from the initial emitted sequence of pulses are sampled via a secondary sequence of pulses with a different repetition time. In \ac{pw} radars, energy of the transmitted pulse is relatively small due to the limited peak amplitude. This limitation in amplitude together with sequential sampling limits the dynamic range and results in a relatively poor \ac{snr} at larger distances. For these reasons, \mbox{\ac{pw}-type} radars have fallen out of favor, and are not used for the applications that will be discussed next, which are mainly about object and people detection, tracking/identification, and vital sign monitoring. From the next section onward, we thus concentrate on \ac{fmcw} systems, which typically are the technology of choice for medium and larger ranges. Still, for short range applications, such as gesture tracking, \mbox{\ac{pw}-type} radars might still be a viable alternative.

\subsection{Frequency-Modulated Continuous Wave Radar}

As the name implies, \ac{fmcw} radars transmit a frequency-modulated signal in a continuous fashion. Due to the larger temporal duration of continuous-wave signals, \ac{fmcw} yields a much larger energy on the emitted signal as compared to \ac{pw}. In order to cover the desired frequency band, the signal is linearly modulated over time starting from the lower frequency within the band to the higher frequency (or vice-versa). This type of signal is most frequently referred to as a chirp, and the linear modulation of the signal is called frequency sweep. An analogue continuous-wave signal can be generated with a \ac{vco}, providing flexible adjustments to the sweep duration independent of the bandwidth. A frequency synthesizer together with a \ac{vco} can be used to provide a digital alternative. This technique also provides a higher spectral purity which makes it possible to avoid accidental emission of frequencies adjacent to the desired band, and thus to comply with given regulations. In \ac{fmcw} radars, the received signal is multiplied by the \ac{tx} signal. The intermediate-frequency signal component that results is then isolated via low-pass filtering. Additionally, a low-cost \ac{adc} can be used to convert the received signal into the digital domain. Due to the recent developments on radar hardware, the wider operating frequency range and the above mentioned advantages, \ac{fmcw} radars are currently preferred over \ac{pw} ones, especially in millimeter-wave band applications.

\noindent \textbf{\ac{fmcw} Signal:} As previously mentioned, a chirp is a linearly modulated \ac{fmcw} signal: it is a sinusoidal function formulated as $\mathbf{x_{\mathrm{tx}}} = \sin(\omega_{\mathrm{tx}} t+\phi_{\mathrm{tx}})$, where the frequency $f_{\mathrm{tx}}=\omega_{\mathrm{tx}}/(2 \pi)$ increases linearly over time. After transmission, the reflected chirp signal from an object is collected at the \ac{rx} antenna and can be written as $\mathbf{x_{\mathrm{rx}}} = \sin(\omega_{\mathrm{rx}} t+\phi_{\mathrm{rx}})$. The \ac{if} signal is produced by mixing \ac{rx} and \ac{tx} signals in the mixer component of the radar as $\mathbf{x_{\mathrm{if}}}=\sin\big((\omega_{\mathrm{tx}}-\omega_{\mathrm{rx}})t + (\phi_{\mathrm{tx}}-\phi_{\mathrm{rx}})\big)$. The time delay between the \ac{rx} and the \ac{tx} signals is $\tau = 2d/c$, where \textit{d} is the distance to the objects and \textit{c} is the speed of light in air. The start of the \ac{if} is at $\tau$, which is also when \ac{rx} chirp is realized and ends when the \ac{tx} chirp is entirely transmitted. Time delay is the foundation for computing the range and velocity of a target in an environment. While the given introductory \ac{fmcw} concepts are sufficient for the purpose of this paper, further details on \ac{fmcw} radars can be found in~\cite{jankiraman2018fmcw}.

\noindent \textbf{Range Measurement and Resolution:} Range resolution is defined as the ability of a radar to identify closely packed objects. When the distance separating two objects is smaller than the resolution, the radar becomes unable to distinguish between them, returning a single range reading. %However, when the range resolution is increased, it is possible to identify the two objects in which they would not be observed at the same distance but rather two or more readings will be captured. 
The range measurement is carried out by computing the phase difference between \ac{tx} and \ac{rx} chirps, yielding the initial phase of an \ac{if} signal, that is formulated as $\phi_0 = 2{\pi}f_c{\tau}$, where $f_c = c / \lambda$ stands for the frequency, $c$ is the speed of light and $\lambda$ is the wavelength. Hence, the distance $d$ to an object, the so called range $d=\tau c / 2$, can be retrieved as $d = \phi_0 c / (2 \pi f_c) = \phi_0 \lambda / (4 \pi)$. When multiple objects are present, a single \ac{tx} chirp results in the reception of multiple reflected (\ac{rx}) signal copies. According to the different time delays ($\tau$) between the \ac{tx} and each of the \ac{rx} chirps, multiple \ac{if} signals are computed, and range measurements for each corresponding object are derived. The range resolution $d_{\rm res} = c / (2 B)$, highly depends on the bandwidth $B$ of the radar~\cite{iovescu2017fundamentals}: it can be improved by increasing the bandwidth swept by the chirp, yielding a longer \ac{if} signal and, in turn, leading to a more precise reading of the environment.  

\noindent \textbf{Velocity Measurement and Resolution:} In an \ac{fmcw} radar the velocity computation (commonly referred to as Doppler) can be achieved using two \ac{tx} chirps. Initially, the object range is calculated by applying a \ac{fft} to the \ac{rx} chirps. This range calculation is commonly called \mbox{range-\ac{fft}}. The \mbox{range-\ac{fft}} of separate chirps at the same location will yield different phases. The object velocity is then derived according to the phase difference between the two chirps as $\mathbf{v} = \lambda\Delta\phi/(4{\pi}T_c)$, where $\Delta\phi$ is the phase difference and $T_c$ is the chirp duration. However, in the case of multiple moving objects having the same distance from the radar, the above method no longer works. To overcome this, the radar needs to transmit $N$ chirps with equal separation, i.e., a so called \textit{chirp frame}. When the chirp frame is passed through the \mbox{range-\ac{fft}}, it yields a phase difference containing combined phase differences of all the moving objects. The result of the \mbox{range-\ac{fft}} is passed through a second \ac{fft} called \mbox{Doppler-\ac{fft}} to identify specific phase differences $\omega$ of each object. In the case of two objects, the corresponding phase differences, $\omega_1$ and $\omega_2$, can be used to derive two velocities as $\mathbf{v_1} = \lambda\omega_1 / (4{\pi}T_c)$ and $\mathbf{v_2} = \lambda\omega_2/(4{\pi}T_c)$. The velocity resolution, $\mathbf{v_{\rm res}}$, of the radar is inversely proportional to the duration of a single frame, $T_f = N T_c$. By knowing the frame duration $T_f$, the velocity resolution is $\mathbf{v_{\rm res}} = \lambda/(2 T_f)$~\cite{iovescu2017fundamentals}. 

\noindent \textbf{Angle Measurement and Resolution:} In radar sensing applications, most often the ``angle'' refers to the horizontal-plane \ac{aoa} at the receiver (or \textit{azimuth} in a spherical coordinate system). It is calculated by observing the phase changes occurring on the \mbox{range-\ac{fft}} or \mbox{Doppler-\ac{fft}} peaks. In order to observe these changes, there have to be at least two \ac{rx} antennas. The difference between the readings of each antenna is what produces the phase change in the \ac{fft} peaks. The phase change is formulated as $\Delta \phi = 2 \pi \Delta d/\lambda$ s.t. $\Delta d = \ell \sin(\theta)$, where $\ell$ is the distance between the antennas. Accordingly, the angle can be estimated as $\theta = \sin ^ {-1} \big(\lambda \Delta \phi / (2 \pi \ell)\big)$. The closer $\theta$ is to zero, the more accurate the angle estimation becomes. In fact, the angle resolution $\theta_{\rm res} = \lambda / N d \cos(\theta)$ is usually given assuming $\theta = 0$ and $d = \lambda / 2$ which simplifies it to $\theta_{\rm res} = 2 / N$. The field of view of the radar depends on the maximum AoA that can be measured~\cite{iovescu2017fundamentals}. 

We remark that the distance and angle resolution of a \ac{mmw} radar device are especially important as they characterise the density and the minimum separation of the points that are detected in the radar maps (see next section). This, in turn, has a major impact on the resolution of the clustering algorithms that are used to separate signals reflected by different subjects and objects in the radar maps (see, e.g., \ac{dbscan} in the following sections) and, on the final tracking performance of any signal processing pipeline.

\input{passiveRadar_hardware}

In Table~\ref{tab:passiveHardware}, we summarize the types of passive \ac{mmw} radars employed in the literature covered by our survey. We observe that the availability of commercial evaluation boards from Texas Instruments (TI) and of software interfaces enabling the retrieval of raw radar data has made TI devices the platforms of choice in many of the works. However, others still prefer powerful but less commercial devices or come up with custom boards when commercial platforms are not sufficient to satisfy the requirements of the application.

\subsection{Key Processing Techniques}    
\label{sec:radarsurv.proc}

Next, we describe some key processing techniques that are utilized in modern \ac{mmw} based radar systems. As detailed below, these are used for various purposes such as noise removal, object/people tracking, people detection and identification, vital signal estimation, etc. Note also that multiple techniques are often used concurrently as part of the same solution. By processing distance, velocity and angle information, it is possible to get two or three dimensional data points, such as \mbox{\textit{\acrlong{rd}}} (\glsunset{rd}\ac{rd}), \mbox{\textit{\acrlong{ra}}} (\glsunset{ra}\ac{ra}) or \mbox{\textit{\acrlong{rda}}} (\glsunset{rda}\ac{rda}) maps. This type of data shape, with temporal information between the data frames, can be further processed to provide valuable information about objects and users in positioning, tracking and identification applications.

\noindent \textbf{micro-Doppler} --- In addition to the main (bulk) movement of an object, it is possible to have mechanical vibrations within the object body as well. These internal vibrations are called {\it micro motions}. The \mbox{micro-Doppler} phenomenon is observed when these micro motions from the object cause a frequency modulation on the returned signal~\cite{chen2006micro}. An example for this would be the individual movements of the legs and the arms of a person while walking, or rotations of the propellers of a fix-winged aircraft while flying. Assuming that the scalar range from the radar to an object is $r(t)$ and that the carrier frequency is $f_c$, then the phase of the baseband signal is defined as $\phi(t) = 2 \pi f_c \frac{2 r(t)}{c}$. With this, it is possible to obtain the \mbox{micro-Doppler} frequency shift caused by the motion of an object. Taking the time derivative of the phase yields $ \frac{d}{dt}\phi (t) = 2 \pi f_c \frac{2}{c} \frac{d}{dt} r(t)$. We manipulate this equation by introducing the Doppler shift induced by the rotation of the object and referring to vector $\mathbf{p}$ as the location of an arbitrary point on it. Thus, the \mbox{micro-Doppler} frequency shift equation is obtained as $ f_D = \frac{2f}{c} (\mathbf{v} + \mathbf{\Omega} \times \mathbf{p})^T \cdot \mathbf{n}$. The first term of the equation is the Doppler shift due to the object's translation $ f_{\rm trans} = \frac{2f}{c} \mathbf{v} \cdot \mathbf{n}$, where $\mathbf{v}$ is the bulk velocity of the object and $\mathbf{n}$ is the radar's line of sight direction. The second term is the Doppler shift due to the object's rotation $ f_{\rm rot} = \frac{2f}{c} (\mathbf{\Omega} \times \mathbf{p})^T \cdot \mathbf{n}$, where $\mathbf{\Omega}$ is the angular velocity of the object. In order to get \mbox{time-varying} frequency distribution of \mbox{micro-Doppler} modulation, the \ac{stft}~\cite{allen1977unified} is used. \ac{stft} is a moving window Fourier transform where the signal is examined for each window interval in order to generate a time-frequency distribution. This process can be pictured as a \ac{dft} multiplied by the sliding window's spectrum, which yields a spectrogram of time-varying \mbox{micro-Doppler} modulation \cite{cohen1995time, chen2014radar}. %An \mbox{in-depth} theoretical analysis and practical discussion about \mbox{micro-Doppler} signature is provided in~\cite{chen2014radar}.
Due to the different characteristics in micro-Doppler, it is possible to detect a moving body and even to identify it, by capturing the particular modes of motions of the body parts. 

\noindent \textbf{\acrfull{kf}\glsunset{kf}} --- It is a key tool for the analysis of \mbox{time-series} containing noise or inaccuracies, providing a precise understanding on how the signal changes with time. The \ac{kf} estimates the {\it state} of the monitored process through time, by removing random noise. It is commonly used in movement control, navigation and activity recognition, and it is as well widely employed in radar applications. The \ac{dkf} was initially developed in~\cite{kalman1960new}. It is a \mbox{two-step} recursive algorithm. The first step of the recursive loop is the \textit{prediction} step, where a projection from the current state of the model and corresponding uncertainties into the next \mbox{time-step} is made. Second, the \textit{correction} (or update) step is where adjustment of the projection is made by taking the weighted average of the projected state with the measured information. In linear systems with additive Gaussian noise, \ac{dkf} works as an optimal \mbox{least-square} error estimator. While for non-linear systems, the most common \ac{kf} variants are the \ac{ekf} and the \ac{ukf}. One of the possible ways of obtaining state estimations in non-linear models is converting the system into a linear one. At each time step, the \ac{ekf} uses a \mbox{first-order} partial derivative matrix for the evaluation of the next predicted state starting from the current one. This essentially forces the system to use {\it linearized versions} of the model in the correction step~\cite{welch1995introduction}. However, when the model is highly non-linear, the \ac{ekf} could experience very slow convergence to the correct solution. In such non-linear models, the \ac{kf} is used with an unscented transformation and hence the derivation of \ac{ukf}. In order to carry out predictions, the \ac{ukf} picks a finite set of points (called sigma points~\cite{julier2002reduced}) around the mean and generates a new mean by passing this set through the non-linear function that describes the system. Thus, the new estimate is obtained. While the computational complexity of both filters are same, for most cases the \ac{ukf} practically runs faster as compared to the \ac{ekf}, as it does not calculate  partial derivatives~\cite{wan2000unscented}. 
%Nonetheless, it should be noted that for specific systems, it is possible to adapt the \ac{ukf} solving these problems~\cite{menegaz2015systematization}

In radar systems, he \ac{kf} makes it possible to reliably estimate the trajectory of the targets, which is achieved by filtering the temporal sequences of points in the RD, RA or RDA maps, by identifying the \ac{com} of the moving target(s) and estimating its (their) trajectory (trajectories) over time. \ac{kf} allows coping with random noise, obtaining robust trajectories, and to also estimate tracks for the targets in those cases where some temporal \ac{rd}/\ac{ra}/\ac{rda} frames are lost due to occlusions see, e.g.,~\cite{pegoraro2020multiperson}. Given the sampling time of radar applications and the typical speed of movement of people, linear \ac{kf} models are usually appropriate for human trajectory tracking. Also, most prior works use \ac{kf} to track the \ac{com} of an object or person, treating it as an idealized \mbox{point-shaped} reflector.

A recent solution for \ac{mmw} indoor radars~\cite{pegoraro2021-point-clouds} uses an extended object tracking \ac{kf}, which makes it possible to jointly estimate the \ac{com} and the {\it extension} of the target around it. In~\cite{pegoraro2021-point-clouds}, such extension is mapped onto an ellipse around the \ac{com}, whose shape and orientation matches those of the target. This \ac{kf} technology has similar performance as standard \ac{kf} assuming \mbox{point-shaped} reflectors in terms of tracking accuracy for the \ac{com}, but {\it additionally it makes it possible to track the object extension over time}. In the case of human sensing, the ellipse represents the way the torso is oriented within the monitored environment. This information, combined with the target trajectory, reveals where the target is steering at, which may be a valuable information for some applications, e.g., for the retail industry.

\subsection{Main learning techniques}
\label{sec:radarsurv.learning}

Nowadays, \ac{ml} and especially \ac{dl} is successfully being applied to many different fields and applications. Although most of these techniques have been developed for a long time, they are recently emerging due to hardware advancements. \ac{ml} methods are used for regression, classification and clustering tasks. A more comprehensive analysis and discussion of \ac{ml} and \ac{dl} techniques can be found in~\cite{mitchell1997machine} and~\cite{goodfellow2016deep}, respectively. Just like many other fields, these techniques are being successfully and abundantly exploited within \ac{mmw} radar sensing systems. %One particular technique is seen more often than the others and we are going to go through this technique in the rest of this section.

In some cases, it is required to group sets of objects into categories, i.e., to perform \textit{clustering}. This technique is widely used in such areas as pattern recognition, image analysis and machine learning. This is particularly relevant when there are scattered data points in the observed space, and the information about which point belongs to what category is non trivial. In our setup, it is used for the analysis of radar images. After the cluster analysis, if the results are good, the clustering method could be exploited to compute labels on the dataset, and it could even be used as a part of a more sophisticated system, e.g., for a subsequent identification of the subject or of the human that has generated each data cluster within an image. Often, the clustering task is carried out in an {\it unsupervised} fashion. Over the years, many researchers have designed clustering algorithms tailored for a variety of models. Some of the well known of clustering algorithms are \textit{k-means} (based on partitioning), \textit{AUTOCLASS} (based on Bayesian statistics), \textit{\acrlong{em}} (\glsunset{em}\acrshort{em}) (based on parametric statistics) and also unsupervised neural networks and \glsunset{dbscan}\acrshort{dbscan}~\cite{ester1996density} (density based). More on the existing clustering models and algorithms, their categorization and discussion can be found in~\cite{estivill2002so}.\\

\noindent \textbf{\ac{dbscan}} --- Considering the data gathered by mapping the radar signal on the environment are tightly packed points in range, angle and velocity dimensions, one algorithm stands out in the field of radar sensing, \acrfull{dbscan}~\cite{ester1996density}. \ac{dbscan} is a \mbox{density-based} clustering technique where the points belonging to a high density region are grouped discarding those that are recognized to be isolated, in accordance with precise definitions of the neighborhood of a point and of its local density. The algorithm starts at an initial point featuring a dense neighborhood and tags it as a \textit{core} point. The remaining points within the \textit{core} point's neighborhood, i.e., within a preset radius from it, are referred to as \textit{reachable}. %While a \textit{core} point can reach surrounding points, it is not possible to reach any other point from a \textit{reachable} point. For this reason, the notion of \textit{connectedness} is introduced, where the two points are \mbox{\textit{density-reachable}} if there exists a third point that can be reached from the first two points. This proves that while \textit{connectedness} is a symmetric relation, \textit{reachability} is not. 
Upon initializing the first \textit{core} point, DBSCAN evaluates the neighborhood density of each \textit{reachable} points within its neighborhood, and the ones residing within a dense neighborhood are chosen as the new \textit{core} points. The {\it density connected} region of the cluster is thus extended by connecting dense neighborhoods, constructing clusters of generic shapes and only containing densely connected points. This process is continued in a recurrent fashion until there are no more \textit{reachable} points whose neighborhood exceeds the minimum density. Finally, a cluster is defined as the collection of all points that are either \mbox{density-connected} or \mbox{density-reachable}. Multiple clusters are possible and represents density-connected regions of points. Points that are not contained within a high-density cluster are referred to as {\it outliers} (these are termed noise points, and are rejected).

In \ac{mmw} based radar applications, \ac{dbscan} has been extensively and successfully used to extract the clusters of data points in the \ac{rd}, \ac{ra} or \ac{rda} maps associated with the tracked humans and/or objects (e.g., vehicles) in the monitored environment~\cite{kellner2012grid}. This technique was found to be very robust and efficient due to the following reasons: i) most importantly, \ac{dbscan} is an unsupervised method, its simplest version only needs two parameters to work (a density threshold and a radius for the density neighborhood), while it does not need one to know in advance the number of clusters (objects/humans) to be tracked. The \ac{dbscan} parameters are to be set at training time and, for given hardware (mainly, working frequency, distance and angle resolution) and environment choices (e.g., indoor vs outdoor), their set values remain rather effective across a large number of scenarios~\cite{pegoraro2020multiperson}, ii) \ac{dbscan} runs fast, with a time complexity of $\mathcal{O}(n \log n)$, where $n$ is the number of points to be evaluated, iii) the clusters do not have to be spherical, \ac{dbscan} works well with clusters of any shape and it is very effective in rejecting random noise, which is quite common in radar maps. Further discussion on how \ac{dbscan} is used in radar systems and applications from the state-of-the-art is presented in Section~\ref{sec:radarsurv.detloc}. \\

% \textcolor{red}{tbd type of neural networks that were used, e.g., CNN, RNN, autoencoders, CNN with attention, ResNets, PointNets, Gants. What they are used for (e.g., identity recognition by processing the micro-doppler signal, and explain why you do not do it with standard ML techniques) and pros and cons} \\

\textbf{Neural networks} --- The term \textit{neural network} \glsunset{nn}(\acrshort{nn}) comes from biological processes where a collection of neurons create a network. In the modern sense, \acp{nn} are the technology counterpart of the brain. They try to achieve learning by identifying the relationships in a set of data similarly to how brain does~\cite{hasson2020direct}. The most basic \ac{nn} is the perceptron originally devised by Rosenblatt~\cite{rosenblatt1958perceptron}. It only has a single layer and performs a classification task based on taking the input and multiplying it by given weights, summing the resulting signals, and passing the result through a non-linear decision function. Essentially, this is the idea behind the whole \ac{dl} field. Below, we talk briefly about some state-of-the-art \ac{dl} architectures, which have captured the attention of researchers working on radar sensing applications.\\

\noindent \textbf{\Acp{cnn}} --- One of the most common \ac{nn} models is the \ac{cnn} \cite{lecun1998gradient}. \acp{cnn} usually consist of back to back convolutional and pooling layers with a final fully-connected layer. The convolutional layers take the input and process it via a kernel function (a filter) where the feature detection is carried out. These feature maps are then fed to the pooling layer where dimension reduction of the domain is performed. This process is continued until a fully-connected layer, where a flattened feature map is computed and used to obtain the classification output (either via a single non-linearity or a softmax layer). \ac{cnn} is a feedforward \ac{nn} where information can only move in the forward direction from the input to the output layer, without cycles nor loops. While the convolution operation is naturally invariant with respect to rigid translations of input patterns, it does not work with other types of transformations, such as rotations. For this reasons, in the radar sensing field dedicated \ac{cnn}-based approaches have been specifically proposed for radar point clouds, which are discussed shortly below.\\
%\textbf{Pros}: Very good at classification of image data. \\
%\textbf{Cons}: Affected by in-image visual characteristics such as position, color and composition details of an object \eb{\cite{rehman2018optimization}}. \\
\textbf{Use with \ac{mmw} radar signals:} due to the lack of mathematical models to describe \ac{rd}/\ac{ra} and \ac{rda} maps from \ac{mmw} radars and to the presence of strong noise components (e.g., from ghost reflections and metallic objects), \ac{cnn} have been extensively used to automatically obtain meaningful feature representations from radar readings. Usually, \acp{cnn} are applied to the \ac{rd}/\ac{ra}/\ac{rda} clusters found by a preceding clustering algorithm, e.g., \ac{dbscan}, assuming that each cluster represents a target object within the monitored space. These representations can be then utilized to detect objects within an environment \cite{chang2020spatial}, to assess the type of activity a person is carrying out \cite{zhang2018real}, or to even track their identity~\cite{pegoraro2020multiperson}. \\

\noindent \textbf{\Acp{rnn}} --- Unlike feedforward \acp{nn}, \acp{rnn} \cite{williams1990efficient} utilize their internal memory to retain information from previous input samples. This allows temporal sequences to be used as input and thus the learning process can extract temporal correlation. Hence, \acp{rnn} remember the information during the learning process, while feedforward \acp{nn} cannot. This is especially relevant for radar data, as it makes it possible to extract temporal features from a sequence of radar maps (i.e., a trajectory). For example, such features can describe the way a person moves his/her limbs while performing a certain activity. 
%Due to its short-term memory capability, these type of networks are preferred in streaming applications such as person tracking \eb{survey citations} \\
%\textbf{Pros}: Works over a time period instead of a time instance and can take arbitrary length of input series without affecting the network size. \\
%\textbf{Cons}: Due to the recurrent nature, computation is slower and is prone to vanishing or exploding gradient problems. Also, cannot remember longer time period information. \\
Vanishing or exploding gradients are commonly seen during the \mbox{back-propagation} \cite{goodfellow2016deep} based training of an \ac{rnn}. This prevents the \ac{nn} to effectively learn, leading to a premature stopping of the training process. \Ac{lstm} cells, or alternatively \ac{gru} cells~\cite{hochreiter1997long}, extend the original \ac{rnn} neuron to effectively cope with vanishing or exploding gradients~\cite{hochreiter1998vanishing}, by intelligently redefining the structure of a memory unit. This solves the gradient vanishing problems at the cost of a greater complexity.\\
%by making \mbox{back-propagated} errors flow backward in unlimited number of virtual layers unfolded in space. \eb{survey citations} \\
%\textbf{Pros}: Reduces the gradient problems and can remember information from past sequence of data. \\
%\textbf{Cons}: It is computationally demanding for \ac{lstm} networks to be used. It is not easy to use dropout functionality \eb{\cite{?}}. \\
\textbf{Use with \ac{mmw} radar signals:} activity recognition usually cannot be achieved on data coming from a single time step, e.g., from a single RD/RA or RDA map. For an activity to be determined, analysis of a sequence of such radar maps should be carried out. Combining this with the \mbox{micro-Doppler} effect observed in humans, it is possible to estimate the identity of a person based on the specific micro motions of their body parts~\cite{dcoss2019, pegoraro2020multiperson, pegoraro2021-point-clouds}.\\

\noindent \textbf{\Acp{ae}} --- Autoencoders \cite{vincent2010stacked} encode the input and then decode it to generate the output. While an AE is trained to copy the input onto the output, the main rationale behind this is to learn internal representations (features) that describe the manifold where the high-dimensional input signal resides. That is, the AE features should be highly representative of the input and can be used to classify it with high accuracy. For a proper training of the AE, the encode/decode functionalities are constrained, e.g., by limiting the number of neurons in the inner layer or forcing some sparsity for the inner representation. This forces the AE network to approximate the output by preserving the {\it most} relevant features. Denoising autoencoders~\cite{vincent2010stacked} are trained to denoise the input signal and reconstruct, at their ouput, the original (noise-free) signal version. This was found to produce better features in the AE inner layers. In addition, the denoising capability of such \ac{nn} architectures is valuable for RD/RA/RDA radar maps due to their noisy nature.

%\textbf{Pros}: Very good at feature extraction and can even converge to the PCA \eb{acronym} representation of the data. Can also be used for unsupervised learning \\
%\textbf{Cons}: If the dataset used for training an AE does not represent the problem well enough, model will not clarify but obscure the information. \\
\noindent \textbf{Use with \ac{mmw} radar signals:} radar system are prone to noisy data and can be significantly affected by unwanted or fake reflections (e.g., ghost reflections). Due to this, many radar applications use the AE encode/decode functionalities as a middle ground for the reconstruction of the desired observation such as anomaly analysis for human fall detection~\cite{jin2020mmfall}, person detection for surveillance systems~\cite{wagner2019target} and indoor person identification~\cite{pegoraro2020multiperson}.\\

\noindent \textbf{\Acp{gan}} --- In general, \acp{gan}~\cite{goodfellow2014generative} are divided into two sub-models called the \textit{generator} and the \textit{discriminator}. In the \textit{generator} network the expected outcome is a newly generated sample which should reflect the features found in the input data/domain. Conversely the task of the \textit{discriminator} network is to classify an input to detect whether it is a fake (generated) or a real example. Learning proceeds as a game, where the generator becomes progressively better in generating fakes, and the discriminator improves at detecting them. The final goal is again to learn meaningful representations (features) of a (usually) high-dimensional input signal.

% \textbf{Pros}: Markov chain is not required, for gradient calculations only back propagation is used, no inference is needed during training, and a wide range of functions can be incorporated into the model. \\
% \textbf{Cons}: Highly depends on the minimization of the gradient function and it is hard to generate discrete data such as text. \\

\noindent \textbf{Use with \ac{mmw} radar signals:} Because of the competitive nature of the generator and discriminator networks, jumping from one to another during training makes them better at their respective tasks. Most often, algorithms exploit this fact to generate the required data and use this newly generated input whenever it fits. For instance, \acp{gan} have been used in~\cite{lu2020see} to generate dense maps from sparse inputs (also referred to as super resolution imaging) for the purpose of environment mapping in a \mbox{low-visibility} environment. In this case, the generator network is used to improve the image resolution and the discriminator to train the generator better.\\

\noindent \textbf{\Acp{resnet}} --- ResNets~\cite{he2016deep} use shortcuts to skip layers. Typically, the skipped layers include activation and batch normalization~\cite{ioffe2015batch}. The reason behind using shortcuts is to overcome vanishing gradients and/or gradient degradation problems. Despite the seemingly simple architectural change, this leads to a major change in terms of learning paradigm, which preserves the correct propagation of the error gradients across the whole network, allowing one to build very deep networks with hundreds of layers and with a remarkable representation (feature extraction) effectiveness.

%\textbf{Pros}: Large networks can be trained without increased training error percentage. Vanishing gradient problem is tackled by identity mapping \\
%\textbf{Cons}: With deeper networks, if layer dropping was not used, training time can become very large \\
\noindent \textbf{Use with \ac{mmw} radar signals:} Due to the large number of layers that can be stacked, ResNets are exploited in complex scenarios where the input signal contains a high number of patterns to be concurrently classified. Examples include human skeletal posture estimation~\cite{sengupta2020mm}, where the detection of more than 15~joints and the subsequent tracking of the person are carried out, or \mbox{real-time} object detection for autonomous driving~\cite{chang2020spatial}, where real time obstacle detection is performed.\\

\noindent \textbf{PointNet and PointConv} --- Images are represented through dense regular grids of points, whereas point clouds are irregular and also unordered. For these reasons, using the convolution operation with them can be difficult. Pointnet~\cite{qi2017pointnet} is a deep neural network which uses unordered 3D (graph) point clouds as input. The applications of PointNets are object classification and semantic segmentation. An extension of this network is Pointnet++~\cite{qi2017pointnet++}, where the PointNet architecture is recursively applied on a nested partitioning of the given point cloud. PointNet++ can identify local features on a greater contextual scale. The key reason of using such architecture is to make the extracted features {\it permutation invariant} with respect to the input signal. Along the same lines, in~\cite{pointconv2019} the convolution filter is extended into a new operation, called PointConv, which can be effectively applied to point clouds to build convolutional neural networks. These new network layers can be used to perform translation-invariant and permutation-invariant convolutions (and obtain invariant features) on any point set in the 3D space. These qualities are especially important for radar point clouds. When tracking people or objects from radar data, being rotation invariant is relevant as the traits that we want to capture about the target (movement of limbs, body shape, etc.) do not depend on their orientation in space.

\noindent \textbf{Use with \ac{mmw} radar signals:} In the recent papers~\cite{pegoraro2021-point-clouds} and~\cite{mmgait2020}, novel Pointnet based \acp{nn} are presented to track and identify people from point clouds obtained by \ac{mmw} radars. We remark that \ac{mmw} systems can either operate on dense radar Doppler maps, or on point clouds which can be derived from these dense maps by only retaining the most significant (strongest) reflections. Point clouds are less informative, as some information is lost when moving from dense to sparse representations, but are on the other hand easier to store, transmit and their processing is also lighter. For these reasons, algorithms that operate on sparse point clouds are particularly appealing and are gaining momentum.

%paper from us:
% https://ieeexplore.ieee.org/abstract/document/9440989

\subsection{Selected Applications}
\label{sec:radarsurv.detloc}

Some of the works that we review in this section adopt a custom design for the whole sensing system, from the radar hardware to the implementation of the software. Others, instead, use off-the-shelf radar devices and present new algorithms. Most of the applications deal with human activity recognition, object detection and health monitoring, but other use cases are emerging such as vibration detection, environment mapping and even speech recognition. \\

\noindent \textbf{Human Activity Recognition}
\label{sec:radarsurv.detloc.har}

For the purpose of tracking and identity recognition of humans moving in a room, the authors of~\cite{pegoraro2020multiperson} use \mbox{micro-Doppler} signatures obtained from \mbox{back-scattered} \ac{mmw} radio signals. An \mbox{off-the-shelf} radar is used to gather \ac{rda} maps and noise removal is carried out. Hence, \ac{dbscan} is applied to the \mbox{pre-processed} \ac{rda} maps to detect the data points (signal reflections) generated by each of the human targets in the monitored environment. With \ac{dbscan}, a cluster of \ac{rda} points is obtained for each subject and updated as the targets move, across subsequent time steps. Trajectory detection is carried out by applying a \ac{kf} to the clusters and, as a final step, identity recognition is carried out using a \ac{cnn} with inception layers. The average accuracy is of $90.69$\% for single targets, $97.96$\% for two targets, $95.26$\% for three targets and $98.27$\% for four targets. Similarly, authors in~\cite{pegoraro2021rapid} have designed RAPID in order to use off-the-shelf IEEE 802.11ay devices for person detection and activity recognition. Underlying techniques for human activity recognition are similar to the previous work (e.g \mbox{micro-Doppler} signatures, \ac{kf}, \ac{cnn}). However, RAPID uses CIR estimation and TRN fields to expose targets movement information from the radio signals. As a result, the authors have achieved person detection accuracy between 97.8\% (for 2 subjects being the highest) and 90\% (for 7 subjects being the lowest). In addition, activity recognition rates for walking, running, sitting, and waving hands are 92.9\%, 71.6\%, 99.8\%, and 89.9\% respectively.

Similarly, in~\cite{zhang2018real} micro-Doppler signatures are extracted and exploited for human motion detection, where both \ac{rd} data cubes as a whole, and RD point clouds are considered. The \mbox{real-time} information is received by passing RD data through \mbox{Doppler-time} extraction. \ac{dbscan} is used to group the RD point cloud data from each of the tracked users in the monitored space. The movements of arms, torso and legs of a walking person are then identified via a \ac{cnn} model. Tests were carried out for walking and leaving, waving hands, sitting to walking transition, walking back and forth, and combining all behaviors. An average accuracy of $96.32$\% (walking), $99.59$\% (leaving), $64$\% (waving hands), $91.18$\% (sitting to walking), $97.84$\% (walking back and forth) and $95.19$\% (all) was observed for each scenario, respectively.

In the same vein, movement pattern detection for of one or two patients is the key result in~\cite{jin2019multiple}. Together with \ac{dbscan}, Kalman filtering has been applied to track the trajectory of each patient. Walking, falling, swinging, seizure and restless movements are the movement patterns which are classified by the proposed \ac{cnn} model. For these movement types, the authors have obtained accuracy values ranging between $82.77$\% and $95.74$\%.

The authors of~\cite{gu2019mmsense} have proposed a framework called ``mmSense''. It uses an \mbox{\ac{lstm}-based} classification model for localization. Initially, environment fingerprinting is carried out with and without human presence. Hence, the presence of people and their location within the environment are estimated using an \ac{lstm} model. Moreover, an approach combining human outline profile and vital sign measurements gathered from $60$~GHz reflected signal strength series is devised to identify the targets. mmSense was tested on five people concurrently sharing the same physical space, achieving an accuracy of $97.73$\% for classification and of $93$\% for identification tasks, respectively.

With the purpose of preventive decision making in autonomous driving applications, the authors of~\cite{sengupta2020mm} propose \mbox{``mm-Pose''}, a model for estimating the posture of a person in real-time. To achieve this, \ac{rda} data is used to obtain 3D cloud point representations and \ac{rgb} projections of \mbox{depth-Azimuth} and \mbox{depth-Elevation} are used. \ac{cnn} is used to cope with noise and unwanted reflections and also to detect skeletal joint coordinates. The final model was able to locate $17$ human skeletal joints with errors of $3.2$~cm, $2.7$~cm and $7.5$~cm on the depth, elevation and azimuth axes, respectively.

A similar application is presented in~\cite{li2020capturing} for human skeletal pose estimation. In this model, \mbox{range-angle} heatmaps are initially fed to a \ac{cnn} followed by a \ac{fscn}. To exactly locate the target points, the \mbox{non-max} suppression algorithm was used and the obtained key points were combined, implementing and testing the proposed solution on single user scenarios. The evaluation metrics used in this work are \ac{oks} and Mean Average Precision (AP) over different \ac{oks} thresholds. The authors obtained an average \ac{oks} of $70.5$ over eight different body parts. As a comparison, camera based pose estimators achieve higher performance, i.e., Openpose (avg.~\ac{oks}:~$93.3$) and Leave One Out (avg.~\ac{oks}:~$66.6$).

In~\cite{jin2020mmfall}, a fall detection framework, called ``mmFall'' is presented. 4D cloud points are used, i.e., range, azimuth angle, elevation angle and Doppler. To perform fall detection, the authors exploit a \mbox{sequence-2-sequence} \ac{hvrae} model that utilizes an encoder/decoder logic. They use a tailored loss function along with a simplified version (HVRAE\_SL) for testing purposes. They also test the model on vanilla Recurrent Autoencoders (RAE). Overall, HVRAE achieved $98$\%, HVRAE\_SL had $60$\% and vanilla RAE had $38$\% probability of detecting a fall.

The authors of~\cite{Smith2021} designed a system to classify static hand gestures, namely, palm facing the radar, hand perpendicular to radar and thumbs-up gesture. In addition to the real data, artificial reconstructions of the gestures were used to gather synthetic data. Tests were performed both on range and \ac{ra} maps and, $85$\% and $90$\% accuracy were respectively achieved  with them, while with the addition of synthetic data, the accuracy increased up to $93.1$\% and $95.4$\% for range and RA maps, respectively.

A framework for human detection and tracking by using radar fusion is presented in~\cite{Cui2021}. Ground truth data is obtained via a camera system. \ac{dbscan} is used for clustering and temporal relationships between clusters are exploited to obtain the probability distribution of the new positions to perform tracking, similar to Kalman filtering. The concurrent use of two radars allows improving the accuracy from $46.9$\% to $98.6$\%.

The ``GaitCube'' algorithm was proposed in~\cite{Ozturk2021}. It utilizes so called gait data cubes, i.e., 3D joint-feature representations of micro-Doppler and micro-Range signatures for human recognition purposes. The idea behind this algorithm is to exploit the radar's \mbox{multi-channel} capabilities to improve the recognition accuracy. Their proposed system achieves $96.1$\% accuracy with a single antenna, $98.3$\% when using all antennas and an average accuracy of $79.1$\% when tested in an environment not seen at training time.

Akin to the objectives of the above paper,~\cite{Nickalls2021} develops a posture estimation algorithm using \ac{dbscan} to cluster and single out real targets. The authors generate their dataset by installing the radar on the ceiling, and receiving data at $10$~frames per second. The data processing model is based on \acp{cnn}, and the \ac{cnn} network is trained on lying, seated and upright moving postures. Classification results demonstrate a mean accuracy of $99.1$\% and an average processing time of $0.13$s.

Another work in~\cite{Tiwari2021} performs the classification of 7~fitness exercises. \ac{cnn} and \ac{lstm} neural network models are utilized for the classification task, by training them on \ac{rd}, \ac{ra}, \ac{ad} and joint-image data. For these data types, a classification accuracy of $92.08$\%, $98.65$\%, $97.7$\% and $99.27$\% is attained, respectively. In~\cite{Wang2021}, fitness activities were tested both in offline and also in online scenarios. Classification was performed on $5$ human activities achieving an accuracy of $93.25$\% and $91.52$\% for the offline and online operation modes, respectively. The system was also tested on multiple locations and the obtained average accuracy is $88.83$\%.

Authors of~\cite{twoRadar2021Dahnoun, DahnounMECODetection} have created a human detection and tracking algorithm by using two radars simultaneously. In both of these works, Kalman filter and \ac{dbscan} were used for tracking and identifying the location of the person, and the synchronization of the radars were carried out in an offline fashion. The results in~\cite{DahnounMECODetection} show a significant improvement when a \mbox{two-radar} setup was used with an accuracy of $98.6$\% compared to $46.9$\% from \mbox{single-radar} setup in human detection. In~\cite{twoRadar2021Dahnoun}, radars were placed so that one had a \mbox{top-view} and the other had \mbox{side-view} angle. This work in addition to prior work proposes an alarm system and a posture estimation method. An alarm system is triggered upon a positive evaluation in the change of cluster number, number of points in the cluster or the center point of the cluster. The posture estimation is done for standing, sitting and lying poses by analysing the height of a person at a particular instance and the accuracy of estimation is from $92.5$\% to $93.7$\%. 

Towards performing human activity recognition, any combination of range and Doppler (or in some cases of range, Doppler and angle) is used. \ac{rda} is typically used with DBSCAN and/or Kalman filtering to identify the clusters within the environment. After extracting micro-Doppler data, a \ac{nn} architecture (i.e., \ac{cnn}, \ac{rnn}, \ac{ae} etc.) is employed to perform activity/sensing applications. If properly designed, \ac{dl} models are generally the preferred way to identify movement patterns of \ac{rda} clusters, as this is the common theme among most of the surveyed material above. Deterministic algorithms often fail to provide good performance due to the need of a careful parameter tuning (which is very sensitive to the monitored scenario) and to the lack of mathematical models that accurately represent the signals at the receivers.
\\

\noindent \textbf{Object Detection}
\label{sec:radarsurv.detloc.object}

The authors of~\cite{chang2020spatial} propose a method called \ac{saf} for obstacle detection with \ac{mmw} radar and vision sensors. A \ac{fcos} \ac{nn} is used for the detection of objects. For the training of this neural network, radar data is converted into radar maps (images) and during the feature extraction step, the \ac{saf} block within the \ac{fcos} network is used for combining radar with vision features. The proposed \mbox{SAF-FCOS} model is trained and tested on the nuScenes dataset with a \mbox{ResNet-50} backbone, achieving an average precision of $90.0$\% with an \mbox{intersection-over-union} of $0.50$ or higher. 

The detection of concealed objects implies additional challenges, as it becomes necessary to single out hidden objects from rest of the scenery. In~\cite{lee2010automatic}, the authors employ \ac{em} to fit a Gaussian mixture model of the image acquired: through a \mbox{two-step} image segmentation procedure, they first extract the body area from the image and then detect concealed objects. The model is evaluated in terms of average probability of error and the authors report that \mbox{multi-level} \ac{em} has an increased performance of up to $90.0$\% when compared to conventional \ac{em}.

A \mbox{real-time} outdoor hidden object detection model is proposed in~\cite{yeom2011real}. This work also utilizes \ac{em}, Bayesian decision making and Gaussian mixture model for image segmentation, with an architecture similar to that of~\cite{lee2010automatic}. However, vector quantization is adopted for the first segmentation level to achieve faster computation times. The authors also state that \ac{em} can be avoided as a whole to reduce computation time (and complexity) significantly, but this causes an error increase as well. As a result,~\cite{yeom2011real} achieves a computation time of $1.11$~s (with \ac{em}) and $0.134$~s (without \ac{em}) per frame.

Along a similar line, the authors of~\cite{wei2015mtrack} propose a writing object (e.g., a pen) tracking system called ``mTrack,'' that uses dedicated mechanisms to suppress interference from background reflections. After this, the \ac{rx} antenna is steered according to the peak response observed on the reflected received signal. In other words, the antenna is adaptively steered to face and track the direction of the pen. Finally, the target movement tendency is evaluated by the trend and amount of phase shifting. The system can detect the location of the pen at a $94$\% accuracy, with a tracking error smaller than $10$~mm across $90$\% of the trajectory.

In~\cite{kapilevich2011fmcw}, a \mbox{non-imaging} sensor for hidden object detection is developed. The authors test their device both in an outdoor scenario with a gun and in an indoor scenario where plastic sticks of diameter equal to 2~cm are covered by a fabric. Final results of the model are evaluated by applying the Fourier transform to \ac{if} chirps to get the range map on horizontal and vertical scans of the environment compared with a captured image. In~\cite{kapilevich2013passive}, an improved version of the sensor is proposed, using a horn antenna integrated with a focusing dielectric lens operating in the 80--100~GHz frequency range. This sensor can be operated with any preferred movement (e.g., \mbox{up-down}) and the authors claim that the probability of detection can go up to $100$\%.

In~\cite{Regani2021}, an IEEE~802.11ad device is used as a pulsed wave radar to perform passive handwriting tracking. Slow- and \mbox{fast-time} dimension analysis of the complex \ac{cir}, \ac{cacfar} and \ac{spi} are the underlying techniques used in their algorithm. After applying digital beamforming, the authors could extract Doppler maps and by choosing the bins with higher Doppler power, could localize the writing tool (a pen). Finally, the pen is tracked by picking the lowest elevation angle of its lower part at each time-step. With this, they achieved a tracking accuracy between $3$~mm (at a distance of $20$~cm) and $40$~mm (at a distance of $3$~m) and a character recognition accuracy ranging from $72$\% to $82$\%.

The authors of~\cite{Bhatia2021} perform object classification considering three classes: humans, drones and cars. The feature set used in their algorithm consists of radial range, area under the peak, width of the peak, height and standard deviation of the peak in the \mbox{range-\ac{fft}} domain. Logistic regression and Naive Bayes led to a classification accuracy of $86.9$\% and $73.9$\%, respectively.

In~\cite{wang2020precise}, authors have developed a new deep learning model called \ac{hdc} for concealed object detection. \ac{hdc} uses two-class semantic segmentation network for keeping a high resolution in order to detect small objects. As a design rule and assignment strategy, \ac{ecd} assignment is applied. In this assignment stage, the first dilation rate group forms the ``rising edge'' (increasing dilation) and the rest forms the ``falling edge'' (decreasing dilation) of dilation rates. As a result, their average precision with intersection over union of $0.5$ is at $0.69$\% which outperforms rest of the existing techniques.

As it may be apparent from our discussion, a wide variety of algorithms have been used for object detection. Initially, signal processing with \ac{dft} or \ac{fft} is performed to distill signal features. Next, such features are either converted into images such as radar maps, or further data processing is applied, e.g., \ac{cir} analysis. \ac{ml} and \ac{dl} methods, or decision making algorithms such as \ac{em}, are then applied to obtain the final results. In general, there is not a single winning methodology. Rather, the optimal approach depends on hardware, software, environment and application limitations.

\noindent \textbf{Health Monitoring}
\label{sec:radarsurv.detloc.health}

The authors of~\cite{bakhtiari2011real} propose a model for remote \ac{hr} monitoring and analysis. They use the \ac{lm} algorithm for the extraction of \mbox{heart-rate} information. The sum of heartbeat complex, respiration, body movement, background noise, and electronic system noise is gathered by expressing the received in-phase and quadrature-phase components from \ac{lm} as the cosine and sine of the received signal. One distinctive advantage of this method is that there is no phase unwrapping as the fitting of the \ac{hr} signal is directly carried out on the cosine and sine of the received phase modulated signal, which is important for low \ac{snr} scenarios. The method is able to estimate beat-to-beat \ac{hr} and individual heartbeat amplitude, both having a critical role in the diagnosis of heart diseases.

The authors of~\cite{alizadeh2020remote} demonstrate a remote breathing and sleep position monitoring system over multiple people at the same time. High resolution \ac{aoa} detection is used to identify closely located targets. A \ac{svm} is used for finding the sleep position, and an optimal filter to estimate the breathing rate. The designed system achieved an accuracy of $97$\% for breathing rate estimation and of $83$\% for sleep position detection. 

% \eb{?} In~\cite{kao2013design}, a design and analysis of 60GHz radar system is given for \mbox{vital-sign} and vibration detection.
In the same vein, the work in~\cite{yang2017vital} proposes vital sign and sleep monitoring system. Initially, the location of the person is detected by using the reflection loss as the classification parameter, performing a $360$\textdegree\ sweep of the environment. After localizing the human, reflected signal strength samples from the reflected signal directed at the human are gathered. For heart rate detection, \ac{fft}, customized \mbox{band-pass} filter, \ac{ifft} and peak detection are applied, while for breathing rate detection only \ac{ifft} and peak detection were sufficient. The achieved accuracy was $98.4$\% and the mean estimation error in breathing rate and heart rate estimation for an incident angle of $70$\textdegree\ was smaller than $0.5$~bpm and $2.5$~bpm, respectively. 

A similar purpose is found in~\cite{zhao2020heart}, which designs a robot for human detection and heart rate monitoring. The Hungarian Algorithm and Kalman filtering are used to detect and track the user position. Once the person is located, the robot approaches him or her and starts the scanning process. The biquad cascade \ac{iir} filter is used to extract the heartbeat waveforms from the signal, whereas a \ac{nn} is used for predicting the heart rate. The proposed system achieved an accuracy between $91.08$\% and $97.89$\% across eight different poses.

For the purpose of remote glucose level monitoring, the authors of~\cite{omer2018glucose, omer2020blood} observe reflected \mbox{multi-channel} signal signatures collected through the SOLI \ac{mmw} sensor~\cite{lien2016soli}. The signal is analyzed by obtaining average \ac{psd} of each gated signal vector by applying \ac{dft} and \ac{fft}. With this, they were able to sense the change in dielectric constant due to a varying glucose level in the blood. 

The authors of~\cite{Vodai2021} use the radars' multi-channel capabilities to improve the estimation of vital signs (heart rate). Experiments are performed on $4$ different scenarios by changing the location of the radar and the posture of the subject. Authors claim that using \mbox{multi-channel} Maximal Ratio Combining (MRC) outperforms single channel estimates in most cases, quantifying the benefits for each scenario.

Although an increasing number of articles is appearing on health monitoring via \ac{mmw} radars, this field of application deserves significant additional work. In fact, prior art presents results for rather ad-hoc and artificial scenarios, where people are still, positioned at known locations, etc. A fully automated monitoring system should instead operate in free living conditions, where users are free to move and no prior location information is available. Further research is thus needed to enable multi-user tracking of vital signs, by also compensating for people movement, which has a detrimental effect on the estimation of breathing and heart rate.\\

\noindent \textbf{Other Applications}
\label{sec:radarsurv.detloc.other}

In~\cite{jiang2020mmvib}, a system namely ``mmVib'' for micrometer level vibration detection is presented. The authors propose a \mbox{multi-signal} consolidation model to understand \mbox{In-phase} and Quadrature (IQ) domain and in turn exploit the consistency among the two obtained signals to estimate the vibration characteristics of an object. With this, they can detect vibrations at micrometer level.

The authors of~\cite{lu2020see} propose an indoor mapping system called ``milliMap'', designed for \mbox{low-visibility} environments. A lidar is used for environment mapping as a ground truth data collector. A \ac{gan} is used to construct the grid map by recognizing obstacles, free spaces and unknown areas. Finally, semantic mapping is applied for the classification of obstacles. 

A \mbox{noise-resistant} speech sensing framework, ``WaveEar,'' is proposed in~\cite{xu2019waveear}. Directional beamforming is used to make the system robust to noise. After localizing the throat and receiving the data, voice reconstruction is achieved by a neural network based on an \mbox{encoder-decoder} (autoencoder) architecture. As a result, WaveEar achieves a stable $5.5$\% word error rate at a distance of about $1.5$~m from the user. The authors also point out that joint optimization speech-to-text and WaveEar would further enhance the capabilities of their system.

In~\cite{Albuquerque2019}, a \ac{mmw} radar device is mounted on a robot to estimate its position. This is achieved by exploiting the interference produced by other \ac{mmw} radars located in the same environment (with known positions), and by only estimating the angle of arrival of each other radar interference. The proposed system attains position errors for the robot ranging from $14$~cm (with three radars) down to $6$~cm (ten radars).

The applications presented in this section vary from micrometer-level activity recognition to speech recognition. We observe that radio sensing enables new and unforeseen use cases, such as vibration detection~\cite{jiang2020mmvib}, indoor navigation~\cite{lu2020see} and speech reconstruction~\cite{xu2019waveear}. However, it is still unclear whether these signals can be reliably detected in an environment with mobility and other sources of noise. Additional experimental data would be required to check the performance of these solutions in general settings and to possibly improve their robustness.

%%%%%%%%%%%%%%%%%%%%%%%%%%%%%%%%%%%%%%%%%%%%%%%%%%%%%%%%%%%%
%\vspace{-2mm}
\input{passiveRadar_environment}

%\vspace{-2mm}
\input{passiveRadar_tools}
%\vspace{-2mm}

%%%%%%%%%%%%%%%%%%%%%%%%%%%%%%%%%%%%%%%%%%%%%%%%%%%%%%%%%%%%

\subsection{Summary}
\label{sec:radarsurv.summary}

In this section, we have summarized the recent advances and trends in signal processing for passive \ac{mmw} radar systems for indoor spaces. These systems are rapidly gaining momentum as radar devices become commercially available, at a low cost. A number of applications are emerging, targeting diverse scenarios such as people detection, tracking and identification, estimation of biosignals such as respiration and heart rate, detection of gestures/activities/falls, vibrations, speech or environmental mapping. Table~\ref{tab:passiveSummary} summarizes these application-oriented propositions, while Table~\ref{tab:passiveEnvironment} categorizes them based on the environment where the experiments were carried out. While early works used standard machine learning algorithms such as expectation maximization and support vector machines, latest developments have been dominated by neural networks. This is clearly evident from Table~\ref{tab:passiveTools}, which presents a summary of the analytical tools discussed in the survey. These are being implemented in their many flavors, and are allowing researchers to obtain good results in scenarios where no analytical models are available. As far as human data monitoring is concerned (e.g., people tracking, activity monitoring, etc.), the key processing algorithms are \ac{dbscan} clustering for the separation of user data in the radar \ac{rd}/\ac{ra}/\ac{rda} maps and Kalman filtering to reliably track their trajectories. Neural network architectures are evolving from standard \acp{cnn} to more advanced convolutions (PointConv and PointNets) that were specifically designed for radar point clouds. Some solutions then use \acp{rnn} to capture and exploit the temporal correlation of radar signals. Advanced architectures, such as \ac{gan} based, are also being exploited to extract features from radar images. 

Although many applications and uses of this technology have emerged lately, a lot of research and implementation work is still required. As far as research is concerned, vital sign monitoring is still in its infancy as more robust algorithms are to be developed, capable of working in free living conditions, i.e., in the presence of user mobility and other noise sources. In addition, while advanced user tracking and positioning techniques are available for single radar systems, no substantial work can be found for multi-radar setups, i.e., {\it radar networks}. With multiple radar devices, many additional problems have to be tackled, including time synchronization, data fusion among radar signals, distributed calibration and means to quantify whether and to which extent radar devices share a common portion of their field of view. For what concerns implementation, much work still has to be performed architecturally, e.g., where to place the \ac{ml} based intelligence, which messages are to be exchanged between the radars and the computing units, which protocols are to be exploited to synchronize multiple devices along time and data dimensions, etc. Lastly, experimental work is key to the development of robust algorithms, as analytical or simulated models often fail to accurately represent all the noise sources. Hence, the collection of experimental data and its publication along with the code of the developed solutions are vital to make progress.

%%%%%%%%%%%%%%%%%%%%%%%%%%%%%%%%%%%%%%%%%%%%%%%%%%%%%%%%%%%%

\input{passiveRadar_summary}

%% file: passiveRadar_hardware.tex
%%%%%%%%%%%%%%%%%%%%%%%%%%%%%%%%%%%%%%%%%%%%%%%%%%%%%%%%%
%%%% Table for evaluation tools/strategy
%%%%%%%%%%%%%%%%%%%%%%%%%%%%%%%%%%%%%%%%%%%%%%%%%%%%%%%%%

\begin{table}[t]
\caption{Summary of the hardware platforms used in the literature}
\label{tab:passiveHardware}
\centering
\begin{tabular}{ll}
  \toprule
  \textbf{Hardware platform} & \textbf{Related literature}  \\  
  \midrule
  Google SOLI
  & ~\cite{omer2020blood}
  \\ 
  Infineon SiGe
  & ~\cite{omer2018glucose}
  \\ 
  INRAS RadarLog
  & ~\cite{pegoraro2020multiperson}
  \\ 
  Keysight EXG N5172B
  & ~\cite{gu2019mmsense, yang2017vital}
  \\ 
  Qualcomm 802.11ad device
  & ~\cite{Regani2021}
  \\ 
Xilinx Kintex Ultrascale
  & \color{blue}~\cite{pegoraro2021rapid}
  \\ 
  TI AWR1243
  & ~\cite{alizadeh2020remote}
  \\ 
  TI AWR1443
  & ~\cite{lu2020see, Ozturk2021, Cui2021, Wang2021}
  \\ 
  TI AWR1443BOOST
  & ~\cite{Smith2021,Cui2021}
  \\ 
  
  TI AWR1642
  & ~\cite{zhang2018real, Tiwari2021}
  \\ 
  TI AWR1642BOOST
  & ~\cite{jin2019multiple, sengupta2020mm, jiang2020mmvib,Albuquerque2019}
  \\ 
  TI AWR1643BOOST
  & ~\cite{li2020capturing}
  \\ 
  TI AWR1843BOOST
  & ~\cite{jin2020mmfall}
  \\ 
  TI AWR2243
  & ~\cite{Vodai2021}
  \\ 
  TI AWR6843
  & ~\cite{zhao2020heart,Nickalls2021}
  \\ 
  TI IWR1443
  & ~\cite{twoRadar2021Dahnoun, DahnounMECODetection}
  \\ 
  \bottomrule

\end{tabular}
\end{table}

%% file: passiveRadar_environment.tex
%%%%%%%%%%%%%%%%%%%%%%%%%%%%%%%%%%%%%%%%%%%%%%%%%%%%%%%%%%%%
%%%%% Table for Evaluation method
%% This should be in order [1] - [end]
%%%%%%%%%%%%%%%%%%%%%%%%%%%%%%%%%%%%%%%%%%%%%%%%%%%%%%%%%%%%

\begin{table*}[t]
\caption{Summary of the environments in which the experiments have been carried out}
\label{tab:passiveEnvironment}
\centering
\begin{tabular}{ll}
  \toprule
  \textbf{Evaluation} & \textbf{Related literature}  \\  
  \midrule
  Indoors 
  & \cite{pegoraro2020multiperson,pegoraro2021rapid,zhang2018real,gu2019mmsense,jin2019multiple,jin2020mmfall,li2020capturing,lee2010automatic,wei2015mtrack,kapilevich2011fmcw,kapilevich2013passive,bakhtiari2011real,alizadeh2020remote,kao2013design,omer2018glucose,omer2020blood,yang2017vital,zhao2020heart, jiang2020mmvib, lu2020see, xu2019waveear, Wang2021,Ozturk2021,twoRadar2021Dahnoun, Smith2021, Albuquerque2019, Regani2021, Vodai2021, Cui2021, Nickalls2021, wang2020precise, DahnounMECODetection}
  \\ 
  Outdoors 
  & \cite{sengupta2020mm,li2020capturing,chang2020spatial,yeom2011real,kapilevich2011fmcw,kapilevich2013passive, Bhatia2021, Tiwari2021}
  \\ 
%   Both 
%   & ~\cite{},~\cite{}
%   \\ 
  \bottomrule
\end{tabular}
\end{table*}

%% file: passiveRadar_tools.tex
\begin{table*}[t]
\centering
\caption{Summary of the main techniques used in the surveyed papers}
\label{tab:passiveTools}

\begin{tabular}{ll}
  \toprule
  \textbf{Analytical Tools} & \textbf{Related Literature}  \\  
  \midrule
    % micro-Doppler signature analysis
    % & ~\cite{pegoraro2020multiperson,zhang2018real,jin2019multiple}
    % \\  
     DBSCAN
    & ~\cite{pegoraro2020multiperson, zhang2018real,twoRadar2021Dahnoun, Cui2021, Nickalls2021, Wang2021, DahnounMECODetection}
    \\  
    Deep learning
    & ~\cite{pegoraro2020multiperson,pegoraro2021rapid,zhang2018real,gu2019mmsense,jin2019multiple,sengupta2020mm,jin2020mmfall,li2020capturing,chang2020spatial,zhao2020heart,lu2020see,xu2019waveear,Wang2021,Ozturk2021, Smith2021, Cui2021, Nickalls2021, Tiwari2021, Wang2021}
    \\ 
    Fourier transform
    & ~\cite{kapilevich2011fmcw, omer2018glucose, omer2020blood, yang2017vital, jiang2020mmvib, Smith2021, Albuquerque2019, Regani2021, Bhatia2021}
    \\ 
    Hungarian algorithm
    & ~\cite{pegoraro2020multiperson,zhao2020heart}
    \\ 
    Kalman filter 
    & ~\cite{pegoraro2020multiperson,pegoraro2021rapid,zhao2020heart,twoRadar2021Dahnoun, DahnounMECODetection}
    \\ 
     Levenberg-Marquard method
    & ~\cite{bakhtiari2011real}
    \\ 
    Machine learning
    & ~\cite{alizadeh2020remote,Nickalls2021, Bhatia2021}
    \\ 
    Non-max suppression
    & ~\cite{li2020capturing}
    \\ 
    Signal processing
    & ~\cite{pegoraro2021rapid, gu2019mmsense, wei2015mtrack,yang2017vital, xu2019waveear, Regani2021, Vodai2021}
    \\ 
    Statistical modeling
    & ~\cite{lee2010automatic,yeom2011real, Cui2021}
    \\ 
    \bottomrule
   
\end{tabular}
\end{table*}

%% file: passiveRadar_summary.tex
%%%%%%%%%%%%%%%%%%%%%%%%%%%%%%%%%%%%%%%%%%%%%%%%%%%
%%%%% Table for tools of the work
%%%%%%%%%%%%%%%%%%%%%%%%%%%%%%%%%%%%%%%%%%%%%%%%%%%

% \newcolumntype{L}[1]{>{\raggedright\let\newline\\\arraybackslash\hspace{0pt}}m{#1}}
% \newcolumntype{C}[1]{>{\centering\let\newline\\\arraybackslash\hspace{0pt}}m{#1}}
% \newcolumntype{R}[1]{>{\raggedleft\let\newline\\\arraybackslash\hspace{0pt}}m{#1}}

\onecolumn
{ \centering \renewcommand{\baselinestretch}{1.05}\small 
\renewcommand{\arraystretch}{1.25}

%\begin{table}[t]
%    \caption{Summary of the mmWave radar sensing works in the literature}\label{tab:passiveSummary}
%    \vspace{-5mm}
%\end{table}

\setlength{\LTcapwidth}{\textwidth}
\begin{longtable}[t]{|m{4cm}|m{4.25cm}|c|m{5.5cm}|}

\caption{\textsc{Summary of the mmWave radar sensing works in the literature}}\label{tab:passiveSummary} \\

\hline 

\multicolumn{1}{|c|}{\textbf{Proposition}} & 
  \multicolumn{1}{c|}{\textbf{Tools Used}} &
  \multicolumn{1}{c|}{\textbf{Band (GHz)}} &
  \multicolumn{1}{c|}{\textbf{Performance}} 
  \\ \hline \hline
\endfirsthead  
  
\caption[]{\textsc{Summary of the mmWave radar sensing works in the literature (continued)}} \\

\hline

\multicolumn{1}{|c|}{\textbf{Proposition}} & 
  \multicolumn{1}{c|}{\textbf{Tools Used}} &
  \multicolumn{1}{c|}{\textbf{Band (GHz)}} &
  \multicolumn{1}{c|}{\textbf{Performance}} 
  \\ \hline \hline
\endhead 
% \endfirsthead
% %
% \endhead

%%%%%%% Human Detection %%%%%%%
\multicolumn{4}{|c|}{\textbf{Human Activity Recognition Algorithms}} \\ 
\hline \hline
%%%%%%%%%%%%%%%%%%%%%%%%%%%%%%%%%%%%%%%%%%%%%%%%
Multi-person tracking and identification~\cite{pegoraro2020multiperson} 
& Micro-Doppler, DBSCAN, Kalman filter, Hungarian algorithm, CNN
& 77
& Continuous identification of multiple persons with up to 98\% accuracy.
\\ \hline
%%%%%%%%%%%%%%%%%%%%%%%%%%%%%%%%%%%%%%%%%%%%%%%%
Indoor human detection and sensing~\cite{pegoraro2021rapid} 
& CIR, micro-Doppler, Kalman filter, CNN %, TRN fields
& 60
& Person detection accuracy of 97.8\% to 90\%. Walking, running, sitting and waving hands accuracy of 92.9\%, 71.6\%, 99.8\% and 89.9\% respectively.
\\ \hline
%%%%%%%%%%%%%%%%%%%%%%%%%%%%%%%%%%%%%%%%%%%%%%%%
Multi-person detection and identification~\cite{gu2019mmsense} 
& LSTM-based model, RSS series analysis 
& 60 
& Multi person classification and identification accuracy of 97.73\% and 93\% respectively.
\\ \hline
%%%%%%%%%%%%%%%%%%%%%%%%%%%%%%%%%%%%%%%%%%%%%%%%

Gait-based human recognition ~\cite{Ozturk2021}
& CNN
& 77
& Classification accuracy of 96.1\% and 98.3\% with single gait cycle, when using single and all receive antenna respectively.
\\ \hline
%%%%%%%%%%%%%%%%%%%%%%%%%%%%%%%%%%%%%%%%%%%%%%%%

Human detection and tracking ~\cite{Cui2021}
& DBSCAN, probability distribution matching, Kalman filter-like algorithm
& 77-81
& Human detection sensitivity and precision of 90\% and 98.6\% respectively.
\\ \hline
%%%%%%%%%%%%%%%%%%%%%%%%%%%%%%%%%%%%%%%%%%%%%%%%

Real-time human activity recognition~\cite{Wang2021} 
& DBSCAN, CNN, RNN 
& 77 & Offline and real-time activity recognition accuracy of 93.25\% and 91.52\% respectively, over five different human activities. \\ \hline

%%%%%%%%%%%%%%%%%%%%%%%%%%%%%%%%%%%%%%%%%%%%%%%
Hand gesture classification \cite{Smith2021}
& Deep learning, Signal processing (FFT)
& 77-81
& Hand gesture classification accuracy of 93\% and 95\% on range and range-angle data respectively.
\\ \hline
%%%%%%%%%%%%%%%%%%%%%%%%%%%%%%%%%%%%%%%%%%%%%%%%

Human motion
behavior detection~\cite{zhang2018real} 
& Micro-Doppler, DBSCAN, CNN
& 77 
 & Accuracy of over 90\% in detecting various human motion behaviours.%  in walking and leaving 96.32\%, waving hands 99.59\%, sitting to walking 64\%, walking 91.18\%, without micro-Doppler 97.84\%, all behaviors 95.19\%
\\ \hline

%%%%%%%%%%%%%%%%%%%%%%%%%%%%%%%%%%%%%%%%%%%%%%%%
Activity recognition and fitness tracker ~\cite{Tiwari2021}
& Deep learning, CNN
& 77-81
& Classification accuracy of 92.08\%, 98.65\%, 97.7\%, and 99.27\% for RD, RA, Angle-Doppler (AD), and joint-image evaluation respectively.
\\ \hline
%%%%%%%%%%%%%%%%%%%%%%%%%%%%%%%%%%%%%%%%%%%%%%%%

Real-time patient behaviour detection~\cite{jin2019multiple} 
& Micro-Doppler, STFT, CNN
& 77 
& Over 84.31\% prediction accuracy for different behaviors for a single patient. Around 80\% prediction accuracy for different behaviors for two patients.%classification between 97.77\% and 98.94\%, behaviour analysis other 95.74\%, walking 94.13\%, falling 84.49\%, swing 82.77\%, seizure 86.36\%, restless movement 84.31\%
\\ \hline

%%%%%%%%%%%%%%%%%%%%%%%%%%%%%%%%%%%%%%%%%%%%%%%%
Human skeletal pose estimation~\cite{sengupta2020mm} 
& CNN
& 77
& Detection of 17 human skeletal joints with 3.2~cm, 2.7~cm and 7.5~cm localization error on depth, elevation, and azimuth axes respectively.
\\ \hline

%%%%%%%%%%%%%%%%%%%%%%%%%%%%%%%%%%%%%%%%%%%%%%%%

Human pose estimation~\cite{li2020capturing} 
& CNN, Fractionally strided CNN
& 77
& Average object keypoints similarity of 70.5 over 8 different parts.% while they compare their results with camera based pose estimators, Openpose (avg. OKS 93.3) and Leave One Out (avg. OKS 66.6)
\\ \hline

%%%%%%%%%%%%%%%%%%%%%%%%%%%%%%%%%%%%%%%%%%%%%%%
Fall detection system~\cite{jin2020mmfall} 
& LSTM, RNN
& 77
& Proposed scheme achieves 98\% fall detection rate and outperforms the baseline techniques.
\\ \hline

%%%%%%%%%%%%%%%%%%%%%%%%%%%%%%%%%%%%%%%%%%%%%%%%

Real-time posture estimation system~\cite{Nickalls2021}
& DBSCAN, CNN, LSTM, Decision trees
& 77
& Posture estimation with an accuracy of 99.1\% at a processing time of 0.13s
\\ \hline

%%%%%%%%%%%%%%%%%%%%%%%%%%%%%%%%%%%%%%%%%%%%%%%%
Human detection and tracking~\cite{twoRadar2021Dahnoun},~\cite{DahnounMECODetection} 
& DBSCAN, Kalman filters 
& 76-81
& Human detection sensitivity of over 90\%. Two-radar setup improves precision from 46.9\% to 98.6\%. Posture estimation precision from $92.5$\% to $93.7$\% \\ \hline\hline
%%%%%%%%%%%%%%%%%%%%%%%%%%%%%%%%%%%%%%%%%%%%%%%%

\multicolumn{4}{|c|}{\textbf{Object Detection Algorithms}} \\ \hline \hline
%%%%%%%%%%%%%%%%%%%%%%%%%%%%%%%%%%%%%%%%%%%%%%%%%

%%%%%%%%%%%%%%%%%%%%%%%%%%%%%%%%%%%%%%%%%%%%%%%%%
Handwriting tracking ~\cite{Regani2021}
& STFT, CIR, Cell averaging-constant false alarm rate (CA-CFAR)
& 60
& Tracking accuracy of 3~mm to 40~mm and character recognition accuracy of 72\% to 82\%.
\\ \hline
%%%%%%%%%%%%%%%%%%%%%%%%%%%%%%%%%%%%%%%%%%%%%%%%
Obstacle detection for autonomous driving~\cite{chang2020spatial} 
& Deep learning, CNN
& 77
& Average precision of 90\% with intersection of unions greater than 0.5.
\\ \hline

%%%%%%%%%%%%%%%%%%%%%%%%%%%%%%%%%%%%%%%%%%%%%%%%
Concealed object detection~\cite{lee2010automatic} 
& Gaussian smoothing filter, expectation-maximization, Bayesian
& 37.47
& Usage of multi-level EM increased performance up to 90\% compared to conventional EM.
\\ \hline

%%%%%%%%%%%%%%%%%%%%%%%%%%%%%%%%%%%%%%%%%%%%%%%%
Real-time concealed object detection~\cite{yeom2011real} 
& Expectation-Maximization, Bayesian decision making, Gaussian mixture model
& 94
& Computation time of 1.11~s and 0.134~s with reduced processing.
\\ \hline 

%%%%%%%%%%%%%%%%%%%%%%%%%%%%%%%%%%%%%%%%%%%%%%%%
Writing object tracking (mTrack)~\cite{wei2015mtrack} 
& RSS, phase change analysis
& 60
& The system tracks/locates a pen with sub-centimeter accuracy in 90\% of the cases.
\\ \hline

%%%%%%%%%%%%%%%%%%%%%%%%%%%%%%%%%%%%%%%%%%%%%%%%
Concealed object detection ~\cite{kapilevich2011fmcw, kapilevich2013passive}
& FFT
& 80-100
& Object detection accuracy up to $100$\%.
\\ \hline

%%%%%%%%%%%%%%%%%%%%%%%%%%%%%%%%%%%%%%%%%%%%%%%%

Object classification ~\cite{Bhatia2021}
& FFT, Logistic regression, Naive Bayes
& 77-81
& 86.9\% and 73.9\% classification accuracy using Logistic Regression and Naive Bayes respectively.
\\ \hline

%%%%%%%%%%%%%%%%%%%%%%%%%%%%%%%%%%%%%%%%%%%%%%%%
Hidden object detection~\cite{wang2020precise}
& Semantic segmentation, CNN
& 60
& The proposed expand-contract dilation (ECD) scheme has an average precision (AP@0.5) of 0.69, and outperforms all the existing techniques.
%Obtained performance with ECD is at 0.69 on AP@0.5 which is 10\% over hybrid dilated convolution (HDC) network
\\
\hline
%%%%%%%%%%%%%%%%%%%%%%%%%%%%%%%%%%%%%%%%%%%%%%%%
\hline

%%%%%%%%%%%%%%%%%%%%%%%%%%%%%%%%%%%%%%%%%%%%%%%%
%%%%%%% Health Monitoring %%%%%%%
\multicolumn{4}{|c|}{\textbf{Health Monitoring Algorithms}}
\\ \hline \hline

%%%%%%%%%%%%%%%%%%%%%%%%%%%%%%%

% Heart rate sensing \cite{alizadeh2018remote}
% & FFT, Doppler FFT, FIR bandpass filter
% & Vital-sign, remote heart rate detection
% & \eb{?} / Indoor / 77-81
% \\ \hline
%%%%%%%%%%%%%%%%%%%%%%%%%%%%%%%
Blood glucose level monitoring \cite{omer2020blood} 
& DFT, FFT
& 57-64 
& Remote detection of blood glucose levels by sensing the change in dielectric constant and loss tangent.
\\ \hline

%%%%%%%%%%%%%%%%%%%%%%%%%%%%%%%

Glucose level detection \cite{omer2018glucose}
& Energy-density comparison, DTFT
& 57-64
& Demonstrates accurate identification of blood glucose levels.
\\ \hline

%%%%%%%%%%%%%%%%%%%%%%%%%%%%%%%
Vital sign and sleep monitoring \cite{yang2017vital}
& RSS, IFFT
& 60
& Human finding accuracy of 98.4\% and the mean estimation error in breathing rate and heart rate is less then 0.43~Bpm and 2.15~Bpm.
\\ \hline
%%%%%%%%%%%%%%%%%%%%%%%%%%%%%%%

Breathing and sleep position monitoring~\cite{alizadeh2020remote}
& FFT, DOA, optimum filter, SVM
& 77-81
& Accuracy of 97\% and 83\% for breathing rate estimation and sleep position detection respectively.
\\ \hline

%%%%%%%%%%%%%%%%%%%%%%%%%%%%%%%

Vital sign monitoring ~\cite{Vodai2021}
& Arctangent demodulation (AD), Maximal ratio combining (MRC)
& 77-81
& Proposed signal processing chain significantly improves the heart rate estimation accuracy in all cases.
\\ \hline
%%%%%%%%%%%%%%%%%%%%%%%%%%%%%%%
Heart rate sensing~\cite{zhao2020heart}
& Neural networks, Hungarian algorithm, Kalman filter
& 60-64
& Accuracy of 91.08--97.89\% over 8 different human poses.
\\ \hline

%%%%%%%%%%%%%%%%%%%%%%%%%%%%%%%%%%%%%%%%%%%%%%%%
Heart rate analysis~\cite{bakhtiari2011real}
& Non-linear Levenberg-Marquardt
& 94
& Capability of estimating beat-to-beat heart rate and individual heartbeat amplitude.
\\ \hline
%%%%%%%%%%%%%%%%%%%%%%%%%%%%%%%
%
% \eb{Check} Vital-sign and vibration detection~\cite{kao2013design}
% & FFT
% & 60
% & Detection of vibrations with displacement of 0.2~mm and up to 2~m away.
% \\ \hline
% %%%%%%%%%%%%%%%%%%%%%%%%%%%%%%%%%%%%%%%%%%%%%%%%
%
\hline
%%%%%%%%%%%%%%%%%%%%%%%%%%%%%%%%%%%%%%%%%%%%%%%%

%%%%%%% Others Monitoring %%%%%%%
\multicolumn{4}{|c|}{\textbf{Other Algorithms}}
\\ \hline \hline

%%%%%%%%%%%%%%%%%%%%%%%%%%%%%%%%%%%%%%%%%%%%%%%%
Indoor mapping ~\cite{lu2020see}
& GAN
& 77
& Map reconstruction error within 0.2~m. Obstacle classification accuracy of ~90\%.
\\ \hline

%%%%%%%%%%%%%%%%%%%%%%%%%%%%%%%
Vibration detection ~\cite{jiang2020mmvib}
& FFT, AoA
& 77
& Median amplitude error of 3.4~$\mu$m for the 100~$\mu$m amplitude vibration.
\\ \hline

%%%%%%%%%%%%%%%%%%%%%%%%%%%%%%%%%%%%%%%%%%%%%%%%

Robot position estimation ~\cite{Albuquerque2019}
& AoA, range and doppler FFT
& 77
& Position estimation of the robot with an error below 20~cm.
\\ \hline
%%%%%%%%%%%%%%%%%%%%%%%%%%%%%%%%%%%%%%%%%%%%%%%%
Speech sensing ~\cite{xu2019waveear}
& SSNR, Neural network
& 77
& 5.5\% word error rate around 1.5~m distance
\\ \hline
%%%%%%%%%%%%%%%%%%%%%%%%%%%%%%%%%%%%%%%%%%%%%%%%
% \eb{add to survey sec.} ~\cite{santhalingam2020mmasl}
% &
% &
% &
% \\ \hline
%%%%%%%%%%%%%%%%%%%%%%%%%%%%%%%%%%%%%%%%%%%%%%%%

%
%\hline
%
%%%%%%%%%%%%%%%%%%%%%%%%%%%%%%%%%%%%%%%%%%%%%%%%

\end{longtable}
\twocolumn
}

%% file: 07_Discussion_open_research_dir.tex
\section{Discussion and open research directions}
\label{sec:disc}

% \oureditor{Lead: PC; ALL to help for their sub-parts}

% \ourtodonotes{This is a way to draw some summaries of the material discussed thus far and to also answer the question: ``What can be improved? What's a promising research direction? What's left to be done in this field?''}

Our comprehensive review of the state of the art in \ac{mmw} localization and sensing shows that a sizeable set of contributions have already covered significant work in this research area. Such works show that current \ac{mmw} equipment, even \ac{cots} devices, already offer sufficient opportunities to incorporate localization as part of communication processes. Moreover, commercial implementations of \ac{mmw} radars are currently very compact, and cater for precise device-free localization and sensing.
However, additional efforts are required to democratize these tasks and make them natively available to vertical applications that rely on \ac{mmw} connectivity. 

At the current stage of hardware development, fully-custom signal processing algorithms only apply to software-defined radio platforms, where fully-digital transceiver architectures can be available upfront. Conversely, commercial-grade hardware does not give full access to internal signal samples and measurements, requiring more complex processing and often yielding limited performance. For example, while theoretical analysis predicts millimeter-level device localization accuracy and fully digital architectures achieve centimeter-level accuracy, algorithms for commercial-grade \ac{mmw} devices typically achieve decimeter-level 3rd-quartile localization errors.\sloppy

In this perspective, we conclude that promising research directions in the above field would greatly benefit from new-generation standard-compliant \ac{mmw} transceiver hardware that also exposes channel state information to external algorithms. While some efforts in this direction have been announced, there is still no such platform available on the market. The same observation holds for hybrid beamforming architectures. While preliminary works exist that exploit hybrid beamforming to improve beam pattern directivity and adaptivity, or to make the 802.11ay \ac{sls} operations faster, these architectures could also help localize \ac{mmw} devices faster, e.g., by enabling faster angular spectrum scanning.
Moreover, the field still needs scalable algorithms that flexibly manage the presence of multiple \acp{ap} or of several clients in the same area. These algorithms should work, if possible, with zero initial knowledge of the floor plan and surrounding area, and ideally estimate the whole environment, including the location of the \acp{ap} and of all reflective surfaces automatically in a \ac{slam} fashion, in order to relieve the need of input from the user.
Significant research opportunities also exist for integrating \ac{ml} algorithms into location systems. Here, the main challenges relate to: relieving the need for extensive training datasets, whose collection requires important efforts; creating models that transfer well across different environments, especially indoors; speeding up the convergence of the trained models, e.g., through federated learning, particularly when involving heterogenous clients.

All of the above would be important enablers of a fully integrated device-based sensing and localization system, for which significant research is still needed. The benefits of such a system would be enormous, as the seamless integration of device-based localization and communications would enable advanced location-based services in multiple domains (including but not limited to healthcare, massive \ac{iot}, industrial scenarios, safety, and mission-critical applications), as well as multiple network optimizations (such as optimal client-\ac{ap} associations, predictive re-association before link breakage due to movement or obstacles, or location-aided beam training and tracking).

Regarding passive radar sensing, a number of major advancements are envisaged. First, most commercial low-cost radars incorporate linear antenna arrays, which have limited detection and tracking capabilities. Bi-dimensional antenna arrays would make it possible to detect higher resolution radar images in the 3D space, enabling new uses of this technology (e.g., human gait analysis). Even though commercial \ac{mmw} radars with enhanced capabilities and 2D antenna arrays are becoming available, e.g., TI AWR/IWR radars~\cite{ti-all-radars} and TI cascaded imaging radar MMWCAS~\cite{ti-mmwcas-rf-evm} with relatively large antenna array size, very little work is available to date exploiting massive \ac{mimo} radars. These would allow high resolution sensing, which makes it possible to detect finer movements and shapes. Also, most of the available research only involves a single radar sensor, whereas radars could be as well \mbox{co-deployed}, allowing for large-scale monitoring applications. This will give rise to new opportunities and technical challenges to face, such as new techniques to perform sensor fusion from multiple radar views, self-calibration algorithms for the distributed radar sensors, transmission and compression of radar features from multiple sensing units. Architecturally, no clear approach was found on where the supporting computing facilities are to be located, which messages should be sent to them and what is the preferred interaction model between the field sensors and the computing units. All of this is of key importance especially for large deployments involving multiple sensors. Additional opportunities concern the combination of \ac{mmw} radar systems with camera-based ones (including thermal cameras), to perform data/feature extraction and fusion across different sensing domains. Finally, a promising research avenue is to modify commercial off-the-shelf communication technology, such as the forthcoming IEEE~802.11ay, so that it can double as a passive \ac{mmw} radar. This would enable joint communications and passive sensing, potentially without having to deploy a dedicated \ac{mmw} radar network. The recent creation of the TGbf task group (working on research and standardization of \ac{wlan} sensing towards the IEEE 802.11bf amendment) testifies the interest of the community on these emerging topics.

As a general observation, the research on machine learning methods applied to device-based localization remains limited compared to device-free radar-based sensing. For device-based localization, machine learning methods find their typical application in fingerprinting approaches. Yet, these schemes require a typically lengthy preliminary measurement effort, which is often deemed excessive or impractical. Conversely, modern \ac{mmw} radar systems are both compact and affordable, and expose a number of features that can be more easily passed on to complex learning and clustering algorithms to map environments, track movement, or estimate the occurrence of some events of interest. The applicability of machine learning algorithms to to either field could change if more features become available, e.g, from multiple digital transceiver architectures integrated in the same client. For example, this would make it possible to use machine learning to increase the speed of intermediate localization algorithm steps (e.g., angle computations, ranging and simultaneous distance estimation among multiple \ac{mmw} devices, or joint angle/distance estimates based on radio features).

%% file: 08_Conclusions.tex
\section{Conclusions}   \label{sec:concl}

\Acrfull{mmw} communication devices will soon become a fundamental component of \ac{5g}-and-beyond communication networks. This survey put the lens on recent research advances in localization and sensing algorithms for indoor \ac{mmw} communication and radar devices. After introducing the most important properties of \ac{mmw} signal propagation and communication chain architectures that enable \ac{mmw} channel measurements, we presented a thorough account of localization algorithms for \ac{mmw} devices. These are based on a broad range of techniques, that include both traditional methods based, e.g., on timing and \acrfull{rssi} information, and more specific methods that exploit the properties of \ac{mmw} devices and signal propagation, e.g., by processing \acrfull{csi}.

Then, we turned our attention to consumer-grade \ac{mmw} radar devices, which are becoming extremely cost-effective sensing platforms. After introducing the basic structure of such radar architectures, we discussed different approaches that tackle applications such as human activity recognition, object detection and health monitoring. We unveiled that several research directions remain open in both fields, including better algorithms for localization and sensing with consumer-grade devices, data fusion methods for dense deployments, as well as an educated application of machine learning methods to both device-based localization and device-free sensing.